\documentclass[12pt,english]{article}
\pdfoutput=1
\usepackage[T1]{fontenc}
\usepackage{amssymb}
\usepackage{amsmath}
\usepackage{cite}
\usepackage{epsfig}
\usepackage{scalerel}
\usepackage[unicode=true,bookmarks=true,bookmarksnumbered=false,bookmarksopen=false,
 breaklinks=false,pdfborder={0 0 1},backref=false,colorlinks=true]
 {hyperref}
\usepackage[hang,flushmargin]{footmisc}
\usepackage{color}
\usepackage{shuffle}
\usepackage{cleveref}
\usepackage{nicematrix}
\usepackage{todonotes}
\usepackage{booktabs} 				
\usepackage[latin1]{inputenc}
\usepackage{amsmath}
\usepackage{amsfonts}
\usepackage{amssymb}

\usepackage{graphicx}
\usepackage{subcaption}
\usepackage[export]{adjustbox}
\usepackage[most]{tcolorbox}

\usepackage{cleveref}
\renewcommand\thesubfigure{(\alph{subfigure})}

\usepackage{bm}

\usepackage[bbgreekl]{mathbbol}
\DeclareMathSymbol\bbDelta  \mathord{bbold}{"01}
\DeclareMathOperator{\Res}{\text{Res}}

\newcommand{\preprint}[1]{\begin{table}[t]  
    \begin{flushright}      
             {#1}        
             \end{flushright}                 
             \end{table}}  
\renewcommand{\title}[1]{\vbox{\center\LARGE{#1}}\vspace{5mm}}
\renewcommand{\author}[1]{\vbox{\center#1}\vspace{5mm}}
\newcommand{\address}[1]{\vbox{\center\em#1}}

\numberwithin{equation}{section}

\newcommand{\tred}[1]{\textcolor{red}{#1}}

\newcommand{\bigger}[1]{\raisebox{-0.95pt}{\scalebox{1.25}{$#1$}}}
\newcommand{\be}{\begin{equation}}
\newcommand{\ee}{\end{equation}}
\newcommand{\bea}{\begin{eqnarray}}
\newcommand{\eea}{\end{eqnarray}}
\newcommand{\eq}[2]{\be\begin{aligned}#1 \label{#2}\end{aligned}\ee}

\newcommand{\fwbox}[2]{\text{\makebox[#1][c]{$\hspace{-150pt}\displaystyle#2\hspace{-150pt}$}}}
\newcommand{\fwboxL}[2]{\text{\makebox[#1][l]{$#2$}}}
\newcommand{\fwboxR}[2]{\text{\makebox[#1][r]{$#2$}}}
\newcommand{\equivR}{\fwbox{14.5pt}{\hspace{-0pt}\fwboxR{0pt}{\raisebox{0.47pt}{\hspace{1.25pt}:\hspace{-4pt}}}=\fwboxL{0pt}{}}}

\Crefname{equation}{Eq.}{Eqs.}

\ifdefined \Ref
\renewcommand{\Ref}[1]{Ref.~\cite{#1}}
\else
\newcommand{\Ref}[1]{Ref.~\cite{#1}}
\fi

\newcommand{\vev}[1]{\langle #1 \rangle}
\newcommand{\bra}[1]{\langle #1 |}
\newcommand{\ket}[1]{| #1 \rangle}

\newcommand{\edgeA}{\text{{\normalsize${\color{hblue}a}$}}}
\newcommand{\edgeB}{\text{{\normalsize${\color{hblue}b}$}}}
\newcommand{\edgeC}{\text{{\normalsize${\color{hblue}c}$}}}
\newcommand{\edgeD}{\text{{\normalsize${\color{hblue}d}$}}}
\newcommand{\edgeE}{\text{{\normalsize${\color{hblue}e}$}}}
\newcommand{\edgeF}{\text{{\normalsize${\color{hblue}f}$}}}
\newcommand{\edgeG}{\text{{\normalsize${\color{hblue}g}$}}}
\newcommand{\edgeH}{\text{{\normalsize${\color{hblue}h}$}}}

\newcommand{\brFour}[4]{\mathbin{\hspace{-1.5pt}\big[\hspace{-3.5pt}\big[#1,\!#2,\!#3,\!#4\big]\hspace{-3.5pt}\big]\hspace{-1.75pt}}}

\newcommand{\brEight}[8]{\mathbin{\hspace{-1.5pt}\big[\hspace{-3.5pt}\big[#1,\!#2,\!#3,\!#4,\!#5,\!#6,\!#7,\!#8\big]\hspace{-3.5pt}\big]\hspace{-1.5pt}}}
\newcommand{\br}[1]{\mathbin{\hspace{-1.5pt}\big[\hspace{-3.5pt}\big[#1\big]\hspace{-3.5pt}\big]\hspace{-1.5pt}}}
\newcommand{\ab}[1]{\langle #1 \rangle}
\newcommand{\sqb}[1]{[ #1 ]}
\newcommand{\aMs}[3]{\langle #1 | #2| #3 ]}
\newcommand{\lam}[1]{\lambda_{#1}}
\newcommand{\lamt}[1]{\widetilde{\lambda}_{#1}}

\newcommand{\bfq}{\mathbf{q}}

\hypersetup{citecolor=blue,linkcolor=blue,urlcolor=blue}

\newcommand{\calA}{\mathcal{A}}
\newcommand{\calF}{\mathcal{F}}
\newcommand{\calI}{\mathcal{I}}
\newcommand{\calJ}{\mathcal{J}}

\newcommand{\calN}{\mathcal{N}}
\newcommand{\calO}{\mathcal{O}}
\newcommand{\calQ}{\mathcal{Q}}

\newcommand{\calT}{\mathcal{T}}

\definecolor{varcolor}{rgb}{0.08,0.44,0.2}
\definecolor{functioncolor}{rgb}{0.08,0.28,0.6}

\definecolor{hblue}{rgb}{0,0,0.575}
\definecolor{hred}{rgb}{0.575,0.0,0.225}
\definecolor{dim}{rgb}{0.55,0.55,0.55}
\definecolor{deemph}{rgb}{0.25,0.25,0.25}

\usetikzlibrary{snakes}
\usetikzlibrary{decorations}
\usetikzlibrary{trees}
\usetikzlibrary{decorations.pathmorphing}
\usetikzlibrary{decorations.markings}
\usetikzlibrary{decorations.shapes}
\usetikzlibrary{external}
\usetikzlibrary{intersections}
\usetikzlibrary{shapes,arrows}
\usetikzlibrary{arrows.meta}
\usetikzlibrary{calc}
\usetikzlibrary{shapes.misc}
\usetikzlibrary{decorations.text}
\usetikzlibrary{backgrounds}
\usetikzlibrary{shapes.geometric}
\usetikzlibrary{tikzmark,calc,arrows,shapes,decorations.pathreplacing}

\definecolor{mhvblue}{rgb}{0.6,0.6,0.7765}
\definecolor{ampgrey}{rgb}{0.9,0.9,0.9}
\definecolor{unord}{rgb}{0,0,0}
\definecolor{ord}{rgb}{0,0,0.575}
\definecolor{anchorLeg}{rgb}{0.575,0.0,0.225}

\tikzset{
	graviton/.style={decorate,line width=0.1mm, decoration={snake,amplitude=.5mm, segment length=2mm}},
	photon/.style={decorate, decoration={snake}, draw=red},
	scalar/.style={postaction={decorate},
	},
	massive/.style={postaction={decorate},
		line width=0.75mm,
	},
	massless/.style={postaction={decorate},
    line width=0.5mm
	},
    masslessWithArrow/.style={postaction={decorate},
        line width=0.5mm,
		decoration={
			markings,
			mark=at position .75 with {\arrow{latex}}}
	},
  	masslessOverlap/.style={
		decoration={show path construction, lineto code={
			\draw[white,line width=1.5mm,line cap=round] ($(\tikzinputsegmentfirst)!0.2!(\tikzinputsegmentlast)$) to ($(\tikzinputsegmentfirst)!0.8!(\tikzinputsegmentlast)$);
			\draw (\tikzinputsegmentfirst) to (\tikzinputsegmentlast);
		}},
	decorate
	},
	masslessWithDot/.style={postaction={decorate},
		decoration={
			markings,
			mark=at position 0.5 with {\fill circle (2pt);}}
	},
	massiveWithDot/.style={postaction={decorate},
		line width=0.5mm,
		decoration={
			markings,
			mark=at position 0.5 with {\fill circle (2pt);}}
	},
	massiveLinWithDot/.style={postaction={decorate},
		double,
		thick,
		fill=white,
		decoration={
			markings,
			mark=at position 0.5 with {\fill[black] circle (2pt);}}
	},	
	massiveWithArrow/.style={postaction={decorate},
		line width=0.75mm,
		decoration={
			markings,
			mark=at position .75 with {\arrow{latex}}}
	},
	massiveWithArrow2/.style={postaction={decorate},
		line width=0.75mm,
		decoration={
			markings,
			mark=at position 0.9 with {\arrow{latex}}}
	},
	massiveLin/.style={postaction={decorate},
        line width=0.5mm,
		double,
		fill=white
	},
	massivePhi/.style={postaction={decorate},
		line width=0.75mm,
		dashed
	},
	masslessPhi/.style={postaction={decorate},
		dashed
	},
	unitaryCut/.style={postaction={draw,densely dashed,blue,thin},
		line width = 0.2cm,white
	},
	gluon/.style={decorate, draw=magenta,
		decoration={coil,amplitude=4pt, segment length=5pt}},
	partial ellipse/.style args={#1:#2:#3}{
		insert path={+ (#1:#3) arc (#1:#2:#3)}
	},
	cross/.style={cross out, draw=black, minimum size=2*(#1-\pgflinewidth), inner sep=0pt, outer sep=0pt},
	branchCut/.style={postaction={decorate},
		snake=zigzag,
		decoration = {snake=zigzag,segment length = 2mm, amplitude = 2mm}	
	},
	cross/.default={1pt},
	massiveWithResidue/.style={postaction={decorate},
		decoration={
			markings,
			mark=at position 0.5 with {\draw node[cross,red] {};}}
	},
	massiveLinWithResidue/.style={postaction={decorate},
	double,
	thick,
	fill=white,
	decoration={
	markings,
	mark=at position 0.5 with {\draw node[cross=4pt,red] {};}}	
	}
}
\tikzset{decorate sep/.style 2 args=
{decorate,decoration={shape backgrounds,shape=circle,shape size=#1,shape sep=#2}}}

\colorlet{mred}{black!30!red}
\colorlet{mgreen}{black!30!green}
\colorlet{mblue}{black!30!blue}
\colorlet{morange}{blue!70!red}

\newcommand{\diagEgNotNeeded}{
\begin{tikzpicture}[scale=1]
        \coordinate (e1) at (-.65,0);
		\coordinate (e2) at (-2, .375);
		\coordinate (e3) at (-2, -.375);
        \coordinate (e4) at (2, -.375);
        \coordinate (e5) at (2, .375);

		\coordinate (v1) at (0,0);
        \coordinate (v2) at (-1.5, 0.75);
        \coordinate (v2p) at (-1.5, 0.0);
        \coordinate (v3) at (-1.5, -0.75);
        \coordinate (v4) at (1.5,-.75);
        \coordinate (v4p) at (1.5,0);
        \coordinate (v5) at (1.5,.75);
        \coordinate (v6) at (0,.9);
        \coordinate (v7) at (0,-.9);

        \coordinate (l11) at (-.78,-.75);
        \coordinate (l10) at (-.78,.75);
        \coordinate (l12) at (.33,-.34);
        \coordinate (l9) at (.8,-.75);
        \coordinate (l8) at (.8,.75);
        \coordinate (l13) at (.4,.4);
        
		\draw[massiveLin,red] (e1) -- (v1);
		\draw[massless] (e2) -- (v2p);
        \draw[massless] (e3) -- (v2p);
        \draw[massless] (e4) -- (v4p);
        \draw[massless] (e5) -- (v4p);
        \draw[masslessWithArrow] (v2p) -- (v7);
        \draw[masslessWithArrow] (v4p) -- (v7);
        \draw[masslessWithArrow] (v6) -- (v2p);
        \draw[masslessWithArrow] (v6) -- (v4p);
        \draw[masslessWithArrow] (v1) -- (v6);
        \draw[masslessWithArrow] (v7) -- (v1);
        \path[draw=black,line width=1.5pt, fill=ampgrey] (v2p) circle[radius=0.25];
        \path[draw=black,line width=1.5pt, fill=ampgrey] (v4p) circle[radius=0.25];
        \path[draw=black,line width=1.5pt, fill=white] (v6) circle[radius=0.2];
        \path[draw=black,line width=1.5pt, fill=white] (v7) circle[radius=0.2];
         \path[draw=black,line width=1.5pt, fill=white] (v1) circle[radius=0.2];
         \path[draw=black, line width=1.5pt] (v1) -- +(0.12,0.12);
         \path[draw=black, line width=1.5pt] (v1) -- +(-0.12,-0.12);
         \path[draw=black, line width=1.5pt] (v1) -- +(0.12,-0.12);
         \path[draw=black, line width=1.5pt] (v1) -- +(-0.12,0.12);

        \node[left]  at (e1) {$1$};
        \node[above left] at (e2) {$2$};
        \node[below left] at (e3) {$3$};
        \node[below right] at (e4) {$4$};
        \node[above right] at (e5) {$5$};

        \node at (l13) {$\ell_{13}$};
        \node at (l12) {$\ell_{12}$};
        \node at (l11) {$\ell_{11}$};
        \node at (l10) {$\ell_{10}$};
        \node at (l9) {$\ell_{9}$};
        \node at (l8) {$\ell_{8}$};
\end{tikzpicture}}

\newcommand{\diagOne}{
\begin{tikzpicture}[scale=1]
        \coordinate (e1) at (-.65,0);
		\coordinate (e2) at (-1.875, 1.125);
		\coordinate (e3) at (-1.875, -1.125);
        \coordinate (e4) at (1.875, -1.125);
        \coordinate (e5) at (1.875, 1.125);

		\coordinate (v1) at (0,0);
        \coordinate (v2) at (-1.5, 0.75);
        \coordinate (v3) at (-1.5, -0.75);
        \coordinate (v4) at (1.5,-.75);
        \coordinate (v5) at (1.5,.75);
        \coordinate (v6) at (0,.75);
        \coordinate (v7) at (0,-.75);

        \coordinate (l11) at (-.75,-1.2);
        \coordinate (l10) at (-.75,1.2);
        \coordinate (l12) at (.4,-.4);
        \coordinate (l6) at (-2,0);
        \coordinate (l9) at (.75,-1.2);
        \coordinate (l8) at (.75,1.2);
        \coordinate (l7) at (1.9,0);
        \coordinate (l13) at (.4,.4);
        
		\draw[massiveLin] (e1) -- (v1);
		\draw[massless] (e2) -- (v2);
        \draw[massless] (e3) -- (v3);
        \draw[massless] (e4) -- (v4);
        \draw[massless] (e5) -- (v5);
        \draw[masslessWithArrow] (v2) -- (v3);
        \draw[masslessWithArrow] (v3) -- (v7);
        \draw[masslessWithArrow] (v4) -- (v7);
        \draw[masslessWithArrow] (v5) -- (v4);
        \draw[masslessWithArrow] (v6) -- (v2);
        \draw[masslessWithArrow] (v6) -- (v5);
        \draw[masslessWithArrow] (v1) -- (v6);
        \draw[masslessWithArrow] (v7) -- (v1);
        \path[draw=black,line width=1.5pt, fill=black] (v1) circle[radius=0.1];

        \node[left]  at (e1) {$1$};
        \node[above left] at (e2) {$2$};
        \node[below left] at (e3) {$3$};
        \node[below right] at (e4) {$4$};
        \node[above right] at (e5) {$5$};

        \node at (l13) {$\ell_{13}$};
        \node at (l12) {$\ell_{12}$};
        \node at (l11) {$\ell_{11}$};
        \node at (l10) {$\ell_{10}$};
        \node at (l9) {$\ell_{9}$};
        \node at (l8) {$\ell_{8}$};
        \node at (l7) {$\ell_{7}$};
        \node at (l6) {$\ell_{6}$};          
\end{tikzpicture}}

\newcommand{\diagTwo}{
\begin{tikzpicture}[scale=1]
        \coordinate (e1) at (-.2,0);
		\coordinate (e2) at (-.5,1.5);
		\coordinate (e3) at (-2,0);
        \coordinate (e4) at (-.5,-1.5);
        \coordinate (e5) at (2,0);

        \coordinate (v3) at (-.5,1);
        \coordinate (v4) at (-1.5,0);
        \coordinate (v5) at (-.5,-1);
        \coordinate (v1) at (.4,0);
        \coordinate (v2) at (1,.6);
        \coordinate (v6) at (1,-.6);
        \coordinate (v7) at (1.6,0);

        \coordinate (l11) at (1.6,.5);
        \coordinate (l12) at (.4,.5);
        \coordinate (l6) at (1.6,-.5);
        \coordinate (l10) at (.25,1.1);
        \coordinate (l9) at (-1.25,.8);
        \coordinate (l8) at (-1.25,-.8);
        \coordinate (l7) at (.25,-1.1);
        \coordinate (l13) at (.4,-.45);
        
	\draw[massiveLin] (e1) -- (v1);
	\draw[massless] (e2) -- (v3);
        \draw[massless] (e3) -- (v4);
        \draw[massless] (e4) -- (v5);
        \draw[massless] (e5) -- (v7);
        \draw[masslessWithArrow] (v2) -- (v1);
        \draw[masslessWithArrow] (v3) -- (v2);
        \draw[masslessWithArrow] (v4) -- (v3);
        \draw[masslessWithArrow] (v5) -- (v4);
        \draw[masslessWithArrow] (v6) -- (v5);
        \draw[masslessWithArrow] (v6) -- (v7);
        \draw[masslessWithArrow] (v7) -- (v2);
        \draw[masslessWithArrow] (v1) -- (v6);
        \path[draw=black,line width=1.5pt, fill=black] (v1) circle[radius=0.1];

        \node[left]  at (e1) {$1$};
        \node[above] at (e2) {$2$};
        \node[left] at (e3) {$3$};
        \node[below] at (e4) {$4$};
        \node[right] at (e5) {$5$};

        \node at (l13) {$\ell_{13}$};
        \node at (l12) {$\ell_{12}$};
        \node at (l11) {$\ell_{11}$};
        \node at (l10) {$\ell_{10}$};
        \node at (l9) {$\ell_{9}$};
        \node at (l8) {$\ell_{8}$};
        \node at (l7) {$\ell_{7}$};
        \node at (l6) {$\ell_{6}$};          
\end{tikzpicture}}

\newcommand{\diagThree}{
\begin{tikzpicture}[scale=1]
        \coordinate (e1) at (2,1);
		\coordinate (e2) at (-.5,1.5);
		\coordinate (e3) at (-2,0);
        \coordinate (e4) at (-.5,-1.5);
        \coordinate (e5) at (2,-1);

		\coordinate (v1) at (1.5,.5);
        \coordinate (v2) at (.5,.5);
        \coordinate (v3) at (-.5,1);
        \coordinate (v4) at (-1.5,0);
        \coordinate (v5) at (-.5,-1);
        \coordinate (v6) at (.5,-.5);
        \coordinate (v7) at (1.5,-.5);

        \coordinate (l11) at (2,0);
        \coordinate (l12) at (1,.9);
        \coordinate (l6) at (1,-.9);
        \coordinate (l10) at (.25,1.1);
        \coordinate (l9) at (-1.25,.8);
        \coordinate (l8) at (-1.25,-.8);
        \coordinate (l7) at (.25,-1.1);
        \coordinate (l13) at (0,0);
        
	\draw[massiveLin] (e1) -- (v1);
	\draw[massless] (e2) -- (v3);
        \draw[massless] (e3) -- (v4);
        \draw[massless] (e4) -- (v5);
        \draw[massless] (e5) -- (v7);
        \draw[masslessWithArrow] (v1) -- (v2);
        \draw[masslessWithArrow] (v3) -- (v2);
        \draw[masslessWithArrow] (v4) -- (v3);
        \draw[masslessWithArrow] (v5) -- (v4);
        \draw[masslessWithArrow] (v6) -- (v5);
        \draw[masslessWithArrow] (v6) -- (v7);
        \draw[masslessWithArrow] (v7) -- (v1);
        \draw[masslessWithArrow] (v2) -- (v6);
         \path[draw=black,line width=1.5pt, fill=black] (v1) circle[radius=0.1];
         
        \node[above right]  at (e1) {$1$};
        \node[above] at (e2) {$2$};
        \node[left] at (e3) {$3$};
        \node[below] at (e4) {$4$};
        \node[below right] at (e5) {$5$};

        \node at (l13) {$\ell_{13}$};
        \node at (l12) {$\ell_{12}$};
        \node at (l11) {$\ell_{11}$};
        \node at (l10) {$\ell_{10}$};
        \node at (l9) {$\ell_{9}$};
        \node at (l8) {$\ell_{8}$};
        \node at (l7) {$\ell_{7}$};
        \node at (l6) {$\ell_{6}$};          
\end{tikzpicture}}

\newcommand{\diagFour}{
\begin{tikzpicture}[scale=1]
        \coordinate (e1) at (2,1);
		\coordinate (e2) at (-.5,1.5);
		\coordinate (e3) at (-2,0);
        \coordinate (e4) at (-.5,-1.5);
        \coordinate (e5) at (2,-1);

		\coordinate (v1) at (1.5,.5);
        \coordinate (v2) at (.5,.5);
        \coordinate (v3) at (-.5,1);
        \coordinate (v4) at (-1.5,0);
        \coordinate (v5) at (-.5,-1);
        \coordinate (v6) at (.5,-.5);
        \coordinate (v7) at (1.5,-.5);

        \coordinate (l11) at (2,0);
        \coordinate (l12) at (1,.9);
        \coordinate (l6) at (1,-.9);
        \coordinate (l10) at (.25,1.1);
        \coordinate (l9) at (-1.25,.8);
        \coordinate (l8) at (-1.25,-.8);
        \coordinate (l7) at (.25,-1.1);
        \coordinate (l13) at (0,0);
        
	\draw[massless] (e1) -- (v1);
	\draw[massless] (e2) -- (v3);
        \draw[massiveLin] (e3) -- (v4);
        \draw[massless] (e4) -- (v5);
        \draw[massless] (e5) -- (v7);
        \draw[masslessWithArrow] (v1) -- (v2);
        \draw[masslessWithArrow] (v3) -- (v2);
        \draw[masslessWithArrow] (v4) -- (v3);
        \draw[masslessWithArrow] (v5) -- (v4);
        \draw[masslessWithArrow] (v6) -- (v5);
        \draw[masslessWithArrow] (v6) -- (v7);
        \draw[masslessWithArrow] (v7) -- (v1);
        \draw[masslessWithArrow] (v2) -- (v6);
         \path[draw=black,line width=1.5pt, fill=black] (v4) circle[radius=0.1];

        \node[above right]  at (e1) {$4$};
        \node[above] at (e2) {$5$};
        \node[left] at (e3) {$1$};
        \node[below] at (e4) {$2$};
        \node[below right] at (e5) {$3$};

        \node at (l13) {$\ell_{13}$};
        \node at (l12) {$\ell_{12}$};
        \node at (l11) {$\ell_{11}$};
        \node at (l10) {$\ell_{10}$};
        \node at (l9) {$\ell_{9}$};
        \node at (l8) {$\ell_{8}$};
        \node at (l7) {$\ell_{7}$};
        \node at (l6) {$\ell_{6}$};          
\end{tikzpicture}}

\newcommand{\diagFive}{
\begin{tikzpicture}[scale=1]
        \coordinate (e1) at (2,1);
        \coordinate (e5) at (2,-1);
        \coordinate (e2) at (-1.5,1.4);
		\coordinate (e3) at (-2.2,.65);
        \coordinate (e4) at (-2,-1);

		\coordinate (v1) at (1.5,.65);
        \coordinate (v2) at (0,.65);
        \coordinate (v4) at (-1.5,.65);
        \coordinate (v5) at (-1.5,-.65);
        \coordinate (v6) at (0,-.65);
        \coordinate (v7) at (1.5,-.65);

        \coordinate (l11) at (2,0);
        \coordinate (l12) at (.75,1.1);
        \coordinate (l6) at (.75,-1.1);
        \coordinate (l10) at (-.75,1.1);
        \coordinate (l8) at (-2,0);
        \coordinate (l7) at (-.75,-1.1);
        \coordinate (l13) at (-.5,0);
        
	\draw[massless] (e1) -- (v1);
	\draw[massless] (e2) -- (v4);
        \draw[massiveLin] (e3) -- (v4);
        \draw[massless] (e4) -- (v5);
        \draw[massless] (e5) -- (v7);
        \draw[masslessWithArrow] (v1) -- (v2);
        \draw[masslessWithArrow] (v4) -- (v2);
        \draw[masslessWithArrow] (v5) -- (v4);
        \draw[masslessWithArrow] (v6) -- (v5);
        \draw[masslessWithArrow] (v6) -- (v7);
        \draw[masslessWithArrow] (v7) -- (v1);
        \draw[masslessWithArrow] (v2) -- (v6);
         \path[draw=black,line width=1.5pt, fill=black] (v4) circle[radius=0.1];

        \node[above right]  at (e1) {$4$};
        \node[above] at (e2) {$5$};
        \node[left] at (e3) {$1$};
        \node[below left] at (e4) {$2$};
        \node[below right] at (e5) {$3$};

        \node at (l13) {$\ell_{13}$};
        \node at (l12) {$\ell_{12}$};
        \node at (l11) {$\ell_{11}$};
        \node at (l10) {$\ell_{10}$};
        \node at (l8) {$\ell_{8}$};
        \node at (l7) {$\ell_{7}$};
        \node at (l6) {$\ell_{6}$};          
\end{tikzpicture}}

\newcommand{\diagSix}{
\begin{tikzpicture}[scale=1]
        \coordinate (e1) at (2,1);
        \coordinate (e5) at (2,-1);
        \coordinate (e2) at (-2,1);
	\coordinate (e3) at (-2.2,-.65);
        \coordinate (e4) at (-1.5,-1.4);

	\coordinate (v1) at (1.5,.65);
        \coordinate (v2) at (0,.65);
        \coordinate (v4) at (-1.5,.65);
        \coordinate (v5) at (-1.5,-.65);
        \coordinate (v6) at (0,-.65);
        \coordinate (v7) at (1.5,-.65);

        \coordinate (l11) at (2,0);
        \coordinate (l12) at (.75,1.1);
        \coordinate (l6) at (.75,-1.1);
        \coordinate (l10) at (-.75,1.1);
        \coordinate (l8) at (-2,0);
        \coordinate (l7) at (-.75,-1.1);
        \coordinate (l13) at (-.5,0);
        
	\draw[massiveLin] (e1) -- (v1);
	\draw[massless] (e2) -- (v4);
        \draw[massless] (e3) -- (v5);
        \draw[massless] (e4) -- (v5);
        \draw[massless] (e5) -- (v7);
        \draw[masslessWithArrow] (v1) -- (v2);
        \draw[masslessWithArrow] (v4) -- (v2);
        \draw[masslessWithArrow] (v5) -- (v4);
        \draw[masslessWithArrow] (v6) -- (v5);
        \draw[masslessWithArrow] (v6) -- (v7);
        \draw[masslessWithArrow] (v7) -- (v1);
        \draw[masslessWithArrow] (v2) -- (v6);
         \path[draw=black,line width=1.5pt, fill=black] (v1) circle[radius=0.1];
         \path[draw=black,line width=1.5pt, fill=black] (v5) circle[radius=0.1];
         
        \node[above right]  at (e1) {$1$};
        \node[above left] at (e2) {$2$};
        \node[left] at (e3) {$3$};
        \node[below] at (e4) {$4$};
        \node[below right] at (e5) {$5$};

        \node at (l13) {$\ell_{13}$};
        \node at (l12) {$\ell_{12}$};
        \node at (l11) {$\ell_{11}$};
        \node at (l10) {$\ell_{10}$};
        \node at (l8) {$\ell_{9}$};
        \node at (l7) {$\ell_{7}$};
        \node at (l6) {$\ell_{6}$};          
\end{tikzpicture}}

\newcommand{\diagSeven}{
\begin{tikzpicture}[scale=1]
        \coordinate (e1) at (-.2,0);
	\coordinate (e2) at (-1,1.2);
	\coordinate (e3) at (-1,-1.2);
        \coordinate (e5) at (2.1,0.5);
        \coordinate (e4) at (2.1,-0.5);

        \coordinate (v3) at (-.75,0.8);
        \coordinate (v5) at (-.75,-0.8);
        \coordinate (v1) at (.4,0);
        \coordinate (v2) at (1.1,.8);
        \coordinate (v6) at (1.1,-.8);
        \coordinate (v7) at (1.7,0);

        \coordinate (l8) at (1.65,.5);
        \coordinate (l9) at (1.65,-.5);
        \coordinate (l10) at (.25,1.2);
        \coordinate (l11) at (.25,-1.2);
        \coordinate (l12) at (.4,-0.5);
        \coordinate (l13) at (.4,+.5);
        \coordinate (l6) at (-1.1,0);
		
	\draw[massiveLin] (e1) -- (v1);
	\draw[massless] (e2) -- (v3);
        \draw[massless](e4)--(v7);
        \draw[massless](e3)--(v5);
        \draw[massless] (e5) -- (v7);
        \draw[masslessWithArrow] (v3) -- (v5);
        \draw[masslessWithArrow] (v1) -- (v2);
        \draw[masslessWithArrow] (v2) -- (v3);
        \draw[masslessWithArrow] (v5) -- (v6);
        \draw[masslessWithArrow] (v7) -- (v6);
        \draw[masslessWithArrow] (v2) -- (v7);
        \draw[masslessWithArrow] (v6) -- (v1);
         \path[draw=black,line width=1.5pt, fill=black] (v1) circle[radius=0.1];
         \path[draw=black,line width=1.5pt, fill=black] (v7) circle[radius=0.1];

        \node[left]  at (e1) {$1$};
        \node[above left] at (e2) {$2$};
        \node[below left] at (e3) {$3$};
        \node[below right] at (e4) {$4$};
        \node[above right] at (e5) {$5$};
        \node at (l13) {$\ell_{13}$};
        \node at (l12) {$\ell_{12}$};
        \node at (l11) {$\ell_{11}$};
        \node at (l10) {$\ell_{10}$};
        \node at (l9) {$\ell_{9}$};
        \node at (l8) {$\ell_{8}$};
        \node at (l6) {$\ell_{6}$};   
\end{tikzpicture}}

\newcommand{\diagEight}{
\begin{tikzpicture}[scale=1]
        \coordinate (e1) at (2,0.5);
	\coordinate (e2) at (-.5,1.5);
	\coordinate (e3) at (-2,0);
        \coordinate (e4) at (-.5,-1.5);
        \coordinate (e5) at (2,-0.5);

	    \coordinate (v1) at (1.5,0);
        \coordinate (v2) at (.5,.5);
        \coordinate (v3) at (-.5,1);
        \coordinate (v4) at (-1.5,0);
        \coordinate (v5) at (-.5,-1);
        \coordinate (v6) at (.5,-.5);

        \coordinate (l12) at (1,.6);
        \coordinate (l6) at (1,-.6);
        \coordinate (l10) at (.25,1.1);
        \coordinate (l9) at (-1.25,.8);
        \coordinate (l8) at (-1.25,-.8);
        \coordinate (l7) at (.25,-1.1);
        \coordinate (l13) at (0,0);
        
	\draw[massiveLin] (e1) -- (v1);
	\draw[massless] (e2) -- (v3);
        \draw[massless] (e3) -- (v4);
        \draw[massless] (e4) -- (v5);
        \draw[massless] (e5) -- (v1);
        \draw[masslessWithArrow] (v1) -- (v2);
        \draw[masslessWithArrow] (v3) -- (v2);
        \draw[masslessWithArrow] (v4) -- (v3);
        \draw[masslessWithArrow] (v5) -- (v4);
        \draw[masslessWithArrow] (v6) -- (v5);
         \draw[masslessWithArrow] (v2) -- (v6);
         \draw[masslessWithArrow] (v6) -- (v1);
          \path[draw=black,line width=1.5pt, fill=black] (v1) circle[radius=0.1];

        \node[above right]  at (e1) {$1$};
        \node[above] at (e2) {$2$};
        \node[left] at (e3) {$3$};
        \node[below] at (e4) {$4$};
        \node[below right] at (e5) {$5$};

        \node at (l13) {$\ell_{13}$};
        \node at (l12) {$\ell_{12}$};
        \node at (l10) {$\ell_{10}$};
        \node at (l9) {$\ell_{9}$};
        \node at (l8) {$\ell_{8}$};
        \node at (l7) {$\ell_{7}$};
        \node at (l6) {$\ell_{6}$};          
\end{tikzpicture}}

\newcommand{\diagNine}{
\begin{tikzpicture}[scale=1]
        \coordinate (e1) at (2,0.5);
	\coordinate (e2) at (-.5,1.5);
	\coordinate (e3) at (-2,0);
        \coordinate (e4) at (-.5,-1.5);
        \coordinate (e5) at (2,-0.5);

	\coordinate (v1) at (1.5,0);
        \coordinate (v2) at (.5,.5);
        \coordinate (v3) at (-.5,1);
        \coordinate (v4) at (-1.5,0);
        \coordinate (v5) at (-.5,-1);
        \coordinate (v6) at (.5,-.5);

        \coordinate (l11) at (2,0);
        \coordinate (l12) at (1,.6);
        \coordinate (l6) at (1,-.6);
        \coordinate (l10) at (.25,1.1);
        \coordinate (l9) at (-1.25,.8);
        \coordinate (l8) at (-1.25,-.8);
        \coordinate (l7) at (.25,-1.1);
        \coordinate (l13) at (0,0);
        
	\draw[massless] (e1) -- (v1);
	\draw[massless] (e2) -- (v3);
        \draw[massiveLin] (e3) -- (v4);
        \draw[massless] (e4) -- (v5);
        \draw[massless](e5)--(v1);
        \draw[masslessWithArrow] (v1) -- (v2);
        \draw[masslessWithArrow] (v3) -- (v2);
        \draw[masslessWithArrow] (v4) -- (v3);
        \draw[masslessWithArrow] (v5) -- (v4);
        \draw[masslessWithArrow] (v6) -- (v5);
       \draw[masslessWithArrow] (v2) -- (v6);
       \draw[masslessWithArrow] (v6) -- (v1);
        \path[draw=black,line width=1.5pt, fill=black] (v4) circle[radius=0.1];
        \path[draw=black,line width=1.5pt, fill=black] (v1) circle[radius=0.1];

        \node[above right]  at (e1) {$4$};
        \node[above] at (e2) {$5$};
        \node[left] at (e3) {$1$};
        \node[below] at (e4) {$2$};
        \node[below right] at (e5) {$3$};

        \node at (l13) {$\ell_{13}$};
        \node at (l12) {$\ell_{12}$};
        \node at (l10) {$\ell_{10}$};
        \node at (l9) {$\ell_{9}$};
        \node at (l8) {$\ell_{8}$};
        \node at (l7) {$\ell_{7}$};
        \node at (l6) {$\ell_{6}$};          
\end{tikzpicture}}

\newcommand{\diagTen}{
\begin{tikzpicture}[scale=1]
        \coordinate (e1) at (2,1);
	\coordinate (e2) at (-1,1);
	\coordinate (e3) at (-1,-1);
        \coordinate (e4) at (.5,-1.2);
        \coordinate (e5) at (2,-1);

	\coordinate (v1) at (1.5,.5);
        \coordinate (v2) at (.5,.5);
        \coordinate (v3) at (-.5,.5);
        \coordinate (v5) at (-.5,-.5);
        \coordinate (v6) at (.5,-.5);
        \coordinate (v7) at (1.5,-.5);

        \coordinate (l11) at (2,0);
        \coordinate (l12) at (1,.8);
        \coordinate (l6) at (1,-.9);
        \coordinate (l10) at (0,0.8);
        \coordinate (l9) at (-0.8,.1);
        \coordinate (l8) at (-0.2,-.8);
        \coordinate (l7) at (.25,-1.1);
        \coordinate (l13) at (0.9,0);
        
	\draw[massiveLin] (e1) -- (v1);
	\draw[massless] (e2) -- (v3);
         \draw[massless] (e4) -- (v6);
        \draw[massless] (e3) -- (v5);
        \draw[massless] (e5) -- (v7);
        \draw[masslessWithArrow] (v1) -- (v2);
        \draw[masslessWithArrow] (v3) -- (v2);
         \draw[masslessWithArrow] (v3) -- (v5);
        \draw[masslessWithArrow] (v6) -- (v5);
        \draw[masslessWithArrow] (v6) -- (v7);
        \draw[masslessWithArrow] (v7) -- (v1);
        \draw[masslessWithArrow] (v2) -- (v6);
         \path[draw=black,line width=1.5pt, fill=black] (v1) circle[radius=0.1];
         \path[draw=black,line width=1.5pt, fill=black] (v6) circle[radius=0.1];

        \node[above right]  at (e1) {$1$};
        \node[above left] at (e2) {$2$};
        \node[below left] at (e3) {$3$};
        \node[below] at (e4) {$4$};
        \node[below right] at (e5) {$5$};

        \node at (l13) {$\ell_{13}$};
        \node at (l12) {$\ell_{12}$};
        \node at (l11) {$\ell_{11}$};
        \node at (l10) {$\ell_{10}$};
        \node at (l9) {$\ell_{9}$};
        \node at (l8) {$\ell_{8}$};
        \node at (l6) {$\ell_{6}$};          
\end{tikzpicture}}

\newcommand{\diagEleven}{
\begin{tikzpicture}[scale=1]
        \coordinate (e1) at (0,-1.6);
	\coordinate (e2) at (-1.875, 1.125);
	\coordinate (e3) at (-1.875, -1.125);
        \coordinate (e4) at (1.875, -1.125);
        \coordinate (e5) at (1.875, 1.125);

        \coordinate (v2) at (-1.5, 0.75);
        \coordinate (v3) at (-1.5, -0.75);
        \coordinate (v4) at (1.5,-.75);
        \coordinate (v5) at (1.5,.75);
        \coordinate (v6) at (0,.75);
        \coordinate (v7) at (0,-.75);

        \coordinate (l11) at (-.75,-1.2);
        \coordinate (l10) at (-.75,1.2);
        \coordinate (l12) at (.4,-.4);
        \coordinate (l6) at (-1.8,0);
        \coordinate (l9) at (.75,-1.2);
        \coordinate (l8) at (.75,1.2);
        \coordinate (l7) at (1.8,0);
        \coordinate (l13) at (.4,0);
        
	\draw[massiveLin] (e1) -- (v7);
        \draw[massless] (e2) -- (v2);
        \draw[massless] (e3) -- (v3);
        \draw[massless] (e4) -- (v4);
        \draw[massless] (e5) -- (v5);
        \draw[masslessWithArrow] (v2) -- (v3);
        \draw[masslessWithArrow] (v7) -- (v6);
        \draw[masslessWithArrow] (v3) -- (v7);
        \draw[masslessWithArrow] (v4) -- (v7);
        \draw[masslessWithArrow] (v5) -- (v4);
        \draw[masslessWithArrow] (v6) -- (v2);
        \draw[masslessWithArrow] (v6) -- (v5);
         \path[draw=black,line width=1.5pt, fill=black] (v7) circle[radius=0.1];

        \node[below]  at (e1) {$1$};
        \node[above left] at (e2) {$2$};
        \node[below left] at (e3) {$3$};
        \node[below right] at (e4) {$4$};
        \node[above right] at (e5) {$5$};

        \node at (l13) {$\ell_{13}$};
        \node at (l11) {$\ell_{11}$};
        \node at (l10) {$\ell_{10}$};
        \node at (l9) {$\ell_{9}$};
        \node at (l8) {$\ell_{8}$};
        \node at (l7) {$\ell_{7}$};
        \node at (l6) {$\ell_{6}$};          
\end{tikzpicture}}

\newcommand{\diagTwelve}{
\begin{tikzpicture}[scale=1]
        \coordinate (e1) at (1,1.4);
	\coordinate (e2) at (-.5,1.5);
	\coordinate (e3) at (-2,0);
        \coordinate (e4) at (-.5,-1.5);
        \coordinate (e5) at (2,0);

        \coordinate (v3) at (-.5,1);
        \coordinate (v4) at (-1.5,0);
        \coordinate (v5) at (-.5,-1);
        \coordinate (v2) at (1,.6);
        \coordinate (v6) at (1,-.6);
        \coordinate (v7) at (1.6,0);

        \coordinate (l11) at (1.6,.5);
        \coordinate (l12) at (.4,.5);
        \coordinate (l6) at (1.6,-.5);
        \coordinate (l10) at (.25,1.1);
        \coordinate (l9) at (-1.25,.8);
        \coordinate (l8) at (-1.25,-.8);
        \coordinate (l7) at (.25,-1.1);
        \coordinate (l13) at (.4,0);
        
	\draw[massiveLin] (e1) -- (v2);	
	\draw[massless] (e2) -- (v3);
        \draw[massless] (e3) -- (v4);
        \draw[massless] (e4) -- (v5);
        \draw[massless] (e5) -- (v7);
        \draw[masslessWithArrow] (v2) -- (v6);
        \draw[masslessWithArrow] (v3) -- (v2);
        \draw[masslessWithArrow] (v4) -- (v3);
        \draw[masslessWithArrow] (v5) -- (v4);
        \draw[masslessWithArrow] (v6) -- (v5);
        \draw[masslessWithArrow] (v6) -- (v7);
        \draw[masslessWithArrow] (v7) -- (v2);
         \path[draw=black,line width=1.5pt, fill=black] (v2) circle[radius=0.1];

        \node[above]  at (e1) {$1$};
        \node[above] at (e2) {$2$};
        \node[left] at (e3) {$3$};
        \node[below] at (e4) {$4$};
        \node[right] at (e5) {$5$};

        \node at (l13) {$\ell_{13}$};
        \node at (l11) {$\ell_{11}$};
        \node at (l10) {$\ell_{10}$};
        \node at (l9) {$\ell_{9}$};
        \node at (l8) {$\ell_{8}$};
        \node at (l7) {$\ell_{7}$};
        \node at (l6) {$\ell_{6}$};          
\end{tikzpicture}}

\newcommand{\diagThirteen}{
\begin{tikzpicture}[scale=1]
        \coordinate (e1) at (2,0);
	\coordinate (e2) at (-.5,1.5);
	\coordinate (e3) at (-2,0);
        \coordinate (e4) at (-.5,-1.5);
        \coordinate (e5) at (1,-1);

        \coordinate (v2) at (.5,.5);
        \coordinate (v3) at (-.5,1);
        \coordinate (v4) at (-1.5,0);
        \coordinate (v5) at (-.5,-1);
        \coordinate (v6) at (.5,-.5);
        \coordinate (v7) at (1.5,0);

        \coordinate (l11) at (1.3,-0.5);
        \coordinate (l12) at (1.3,.5);
        \coordinate (l6) at (1,-.9);
        \coordinate (l10) at (.25,1.1);
        \coordinate (l9) at (-1.25,.8);
        \coordinate (l8) at (-1.25,-.8);
        \coordinate (l7) at (.25,-1.1);
        \coordinate (l13) at (0,0);
        
	\draw[massless] (e1) -- (v7);	
	\draw[massless] (e2) -- (v3);
        \draw[massiveLin] (e3) -- (v4);
        \draw[massless] (e4) -- (v5);
        \draw[massless] (e5) -- (v6);
       \draw[masslessWithArrow] (v7) -- (v2);
        \draw[masslessWithArrow] (v3) -- (v2);
        \draw[masslessWithArrow] (v4) -- (v3);
        \draw[masslessWithArrow] (v5) -- (v4);
        \draw[masslessWithArrow] (v6) -- (v5);
        \draw[masslessWithArrow] (v6) -- (v7);
       \draw[masslessWithArrow] (v2) -- (v6);
        \path[draw=black,line width=1.5pt, fill=black] (v4) circle[radius=0.1];
        \path[draw=black,line width=1.5pt, fill=black] (v6) circle[radius=0.1];

        \node[right]  at (e1) {$4$};
        \node[above] at (e2) {$5$};
        \node[left] at (e3) {$1$};
        \node[below] at (e4) {$2$};
        \node[below right] at (e5) {$3$};

        \node at (l13) {$\ell_{13}$};
        \node at (l12) {$\ell_{12}$};
        \node at (l11) {$\ell_{11}$};
        \node at (l10) {$\ell_{10}$};
        \node at (l9) {$\ell_{9}$};
        \node at (l8) {$\ell_{8}$};
        \node at (l7) {$\ell_{7}$};
\end{tikzpicture}}

\newcommand{\diagFourteen}{
\begin{tikzpicture}[scale=1]
        \coordinate (e1) at (2,1);
	\coordinate (e2) at (-.5,1.5);
	\coordinate (e3) at (-2,0);
        \coordinate (e4) at (-.5,-1.5);
        \coordinate (e5) at (2,-1);

	\coordinate (v1) at (1.5,.5);
        \coordinate (v2) at (.5,0);
        \coordinate (v3) at (-.5,1);
        \coordinate (v4) at (-1.5,0);
        \coordinate (v5) at (-.5,-1);
        \coordinate (v7) at (1.5,-.5);

        \coordinate (l11) at (2,0);
        \coordinate (l12) at (1,.7);
        \coordinate (l6) at (1,-.7);
        \coordinate (l10) at (.25,0.7);
        \coordinate (l9) at (-1.25,.8);
        \coordinate (l8) at (-1.25,-.8);
        \coordinate (l7) at (.25,-0.7);
        \coordinate (l13) at (0,0);
        
		\draw[massless] (e1) -- (v1);
		\draw[massless] (e2) -- (v3);
        \draw[massiveLin] (e3) -- (v4);
        \draw[massless] (e4) -- (v5);
        \draw[massless] (e5) -- (v7);
        \draw[masslessWithArrow] (v1) -- (v2);
        \draw[masslessWithArrow] (v3) -- (v2);
        \draw[masslessWithArrow] (v4) -- (v3);
        \draw[masslessWithArrow] (v5) -- (v4);
        \draw[masslessWithArrow] (v2) -- (v5);
        \draw[masslessWithArrow] (v2) -- (v7);
        \draw[masslessWithArrow] (v7) -- (v1);
         \path[draw=black,line width=1.5pt, fill=black] (v4) circle[radius=0.1];
         \path[draw=black,line width=1.5pt, fill=black] (v2) circle[radius=0.1];

        \node[above right]  at (e1) {$4$};
        \node[above] at (e2) {$5$};
        \node[left] at (e3) {$1$};
        \node[below] at (e4) {$2$};
        \node[below right] at (e5) {$3$};

        \node at (l12) {$\ell_{12}$};
        \node at (l11) {$\ell_{11}$};
        \node at (l10) {$\ell_{10}$};
        \node at (l9) {$\ell_{9}$};
        \node at (l8) {$\ell_{8}$};
        \node at (l7) {$\ell_{7}$};
        \node at (l6) {$\ell_{6}$};          
\end{tikzpicture}}

\newcommand{\diagFifteen}{
\begin{tikzpicture}[scale=1]
        \coordinate (e1) at (2,1);
        \coordinate (e5) at (2,-1);
        \coordinate (e2) at (-1.5,1.4);
	\coordinate (e3) at (-2.2,.65);
        \coordinate (e4) at (-2.0,-1.0);

	\coordinate (v1) at (1.5,.65);
        \coordinate (v2) at (0,.65);
        \coordinate (v4) at (-1.5,.65);
        \coordinate (v5) at (-1.5,-.65);
        \coordinate (v6) at (0,-.65);
        \coordinate (v7) at (1.5,-.65);

        \coordinate (l11) at (2,0);
        \coordinate (l12) at (.75,1.1);
        \coordinate (l6) at (.75,-1.1);
        \coordinate (l10) at (-.75,1.1);
        \coordinate (l8) at (-2,0);
        \coordinate (l7) at (-.75,-1.1);
        \coordinate (l13) at (-.5,0);
        
	\draw[massiveLin] (e1) -- (v1);
	\draw[massless] (e2) -- (v4);
        \draw[massless] (e3) -- (v4);
        \draw[massless] (e4) -- (v5);
        \draw[massless] (e5) -- (v7);
        \draw[masslessWithArrow] (v1) -- (v2);
        \draw[masslessWithArrow] (v4) -- (v2);
        \draw[masslessWithArrow] (v5) -- (v4);
        \draw[masslessWithArrow] (v6) -- (v5);
        \draw[masslessWithArrow] (v6) -- (v7);
        \draw[masslessWithArrow] (v7) -- (v1);
        \draw[masslessWithArrow] (v2) -- (v6);
         \path[draw=black,line width=1.5pt, fill=black] (v1) circle[radius=0.1];
         \path[draw=black,line width=1.5pt, fill=black] (v4) circle[radius=0.1];

        \node[above right]  at (e1) {$1$};
        \node[above] at (e2) {$2$};
        \node[left] at (e3) {$3$};
        \node[below left] at (e4) {$4$};
        \node[below right] at (e5) {$5$};

        \node at (l13) {$\ell_{13}$};
        \node at (l12) {$\ell_{12}$};
        \node at (l11) {$\ell_{11}$};
        \node at (l10) {$\ell_{10}$};
        \node at (l8) {$\ell_{8}$};
        \node at (l7) {$\ell_{7}$};
        \node at (l6) {$\ell_{6}$};          
\end{tikzpicture}}

\newcommand{\diagSixteen}{
\begin{tikzpicture}[scale=1]
        \coordinate (e1) at (-.2,0);
	    \coordinate (e2) at (-1,1.2);
	    \coordinate (e3) at (-1.15,-0.8);
        \coordinate (e5) at (2,0);
        \coordinate (e4) at (-0.75,-1.2);

        \coordinate (v3) at (-.75,0.8);
        \coordinate (v5) at (-.75,-0.8);
        \coordinate (v1) at (.4,0);
        \coordinate (v2) at (1.1,.8);
        \coordinate (v6) at (1.1,-.8);
        \coordinate (v7) at (1.7,0);

        \coordinate (l11) at (1.65,.5);
        \coordinate (l6) at (1.65,-.5);
        \coordinate (l10) at (.25,1.2);
        \coordinate (l7) at (.25,-1.2);
        \coordinate (l13) at (.4,-0.5);
        \coordinate (l12) at (.4,+.5);
        \coordinate (l9) at (-1.1,0);
		
	\draw[massiveLin] (e1) -- (v1);
	\draw[massless] (e2) -- (v3);
        \draw[massless](e4)--(v5);
        \draw[massless](e3)--(v5);
        \draw[massless] (e5) -- (v7);
        \draw[masslessWithArrow](v5)--(v3);
        \draw[masslessWithArrow] (v2) -- (v1);
        \draw[masslessWithArrow] (v3) -- (v2);
        \draw[masslessWithArrow] (v6) -- (v5);
        \draw[masslessWithArrow] (v6) -- (v7);
        \draw[masslessWithArrow] (v7) -- (v2);
        \draw[masslessWithArrow] (v1) -- (v6);
         \path[draw=black,line width=1.5pt, fill=black] (v1) circle[radius=0.1];
         \path[draw=black,line width=1.5pt, fill=black] (v5) circle[radius=0.1];

        \node[left]  at (e1) {$1$};
        \node[above] at (e2) {$2$};
        \node[left] at (e3) {$3$};
        \node[below] at (e4) {$4$};
        \node[right] at (e5) {$5$};

        \node at (l13) {$\ell_{13}$};
        \node at (l12) {$\ell_{12}$};
        \node at (l11) {$\ell_{11}$};
        \node at (l10) {$\ell_{10}$};
        \node at (l9) {$\ell_{9}$};
        \node at (l7) {$\ell_{7}$};
        \node at (l6) {$\ell_{6}$};   
\end{tikzpicture}}

\newcommand{\diagSeventeen}{
\begin{tikzpicture}[scale=1]
        \coordinate (e1) at (2,1);
	\coordinate (e2) at (-1.0,0.5);
	\coordinate (e3) at (-1.2,0);
        \coordinate (e4) at (-1.0,-0.5);
        \coordinate (e5) at (2,-1);

	\coordinate (v1) at (1.5,.5);
        \coordinate (v2) at (.5,.5);
        \coordinate (v3) at (-0.6,0);
        \coordinate (v6) at (.5,-.5);
        \coordinate (v7) at (1.5,-.5);

        \coordinate (l11) at (2,0);
        \coordinate (l12) at (1,.9);
        \coordinate (l6) at (1,-.9);
        \coordinate (l10) at (.05,.65);
        \coordinate (l7) at (.05,-0.65);
        \coordinate (l13) at (.9,0);
        
	\draw[massless] (e1) -- (v1);
	\draw[massless] (e2) -- (v3);
        \draw[massiveLin] (e3) -- (v3);
        \draw[massless] (e4) -- (v3);
        \draw[massless] (e5) -- (v7);
        \draw[masslessWithArrow] (v1) -- (v2);
        \draw[masslessWithArrow] (v3) -- (v2);
        \draw[masslessWithArrow] (v6) -- (v3);
        \draw[masslessWithArrow] (v6) -- (v7);
        \draw[masslessWithArrow] (v7) -- (v1);
        \draw[masslessWithArrow] (v2) -- (v6);
         \path[draw=black,line width=1.5pt, fill=black] (v3) circle[radius=0.1];

        \node[above right]  at (e1) {$4$};
        \node[above] at (e2) {$5$};
        \node[left] at (e3) {$1$};
        \node[below] at (e4) {$2$};
        \node[below right] at (e5) {$3$};

        \node at (l13) {$\ell_{13}$};
        \node at (l12) {$\ell_{12}$};
        \node at (l11) {$\ell_{11}$};
        \node at (l10) {$\ell_{10}$};
        \node at (l7) {$\ell_{7}$};
        \node at (l6) {$\ell_{6}$};          
\end{tikzpicture}}

\newcommand{\diagEighteen}{
\begin{tikzpicture}[scale=1]
        \coordinate (e1) at (2,1);
        \coordinate (e5) at (2,-1);
        \coordinate (e2) at (0.4,1.2);
	\coordinate (e3) at (-0.4,1.2);
        \coordinate (e4) at (-1.6,0);

	\coordinate (v1) at (1.5,.65);
        \coordinate (v2) at (0,.65);
        \coordinate (v5) at (-1.2,0);
        \coordinate (v6) at (0,-.65);
        \coordinate (v7) at (1.5,-.65);

        \coordinate (l11) at (2,0);
        \coordinate (l12) at (.85,1);
        \coordinate (l6) at (.85,-1);
        \coordinate (l8) at (-0.9,0.6);
        \coordinate (l7) at (-.75,-0.7);
        \coordinate (l13) at (+0.4,0);
        
		\draw[massless] (e1) -- (v1);
		\draw[massless] (e2) -- (v2);
        \draw[massiveLin] (e3) -- (v2);
        \draw[massless] (e4) -- (v5);
        \draw[massless] (e5) -- (v7);
        \draw[masslessWithArrow] (v1) -- (v2);
        \draw[masslessWithArrow] (v5) -- (v2);
        \draw[masslessWithArrow] (v6) -- (v5);
        \draw[masslessWithArrow] (v6) -- (v7);
        \draw[masslessWithArrow] (v7) -- (v1);
        \draw[masslessWithArrow] (v2) -- (v6);
         \path[draw=black,line width=1.5pt, fill=black] (v2) circle[radius=0.1];

        \node[above right]  at (e1) {$4$};
        \node[above right] at (e2) {$5$};
        \node[above left] at (e3) {$1$};
        \node[left] at (e4) {$2$};
        \node[below right] at (e5) {$3$};

        \node at (l13) {$\ell_{13}$};
        \node at (l12) {$\ell_{12}$};
        \node at (l11) {$\ell_{11}$};
        \node at (l8) {$\ell_{8}$};
        \node at (l7) {$\ell_{7}$};
        \node at (l6) {$\ell_{6}$};          
\end{tikzpicture}}

\newcommand{\diagNineteen}{
\begin{tikzpicture}[scale=1]
        \coordinate (e1) at (0,-1.0);
	\coordinate (e2) at (-1.875, 1.125);
	\coordinate (e3) at (-1.875, -1.125);
        \coordinate (e4) at (1.875, -1.125);
        \coordinate (e5) at (1.875, 1.125);

        \coordinate (v2) at (-1.5, 0.75);
        \coordinate (v3) at (-1.5, -0.75);
        \coordinate (v4) at (1.5,-.75);
        \coordinate (v5) at (1.5,.75);
        \coordinate (v7) at (0,0);

        \coordinate (l11) at (-.75,-0.7);
        \coordinate (l10) at (-.75,0.7);
        \coordinate (l12) at (.4,-.4);
        \coordinate (l6) at (-1.8,0);
        \coordinate (l9) at (.75,-0.7);
        \coordinate (l8) at (.75,0.7);
        \coordinate (l7) at (1.8,0);
        \coordinate (l13) at (.4,0);
        
	\draw[massiveLin] (e1) -- (v7);
        \draw[massless] (e2) -- (v2);
        \draw[massless] (e3) -- (v3);
        \draw[massless] (e4) -- (v4);
        \draw[massless] (e5) -- (v5);
        \draw[masslessWithArrow] (v2) -- (v3);
        \draw[masslessWithArrow] (v7) -- (v2);
        \draw[masslessWithArrow] (v7) -- (v5);
        \draw[masslessWithArrow] (v3) -- (v7);
        \draw[masslessWithArrow] (v4) -- (v7);
        \draw[masslessWithArrow] (v5) -- (v4);
         \path[draw=black,line width=1.5pt, fill=black] (v7) circle[radius=0.1];

        \node[below]  at (e1) {$1$};
        \node[above left] at (e2) {$2$};
        \node[below left] at (e3) {$3$};
        \node[below right] at (e4) {$4$};
        \node[above right] at (e5) {$5$};

        \node at (l11) {$\ell_{11}$};
        \node at (l10) {$\ell_{10}$};
        \node at (l9) {$\ell_{9}$};
        \node at (l8) {$\ell_{8}$};
        \node at (l7) {$\ell_{7}$};
        \node at (l6) {$\ell_{6}$};          
\end{tikzpicture}}

\newcommand{\diagff}{
\begin{tikzpicture}[scale=1]
        \coordinate (e1) at (0,1);
        \coordinate (e2) at (-0.75,-0.75);
	    \coordinate (e4) at (+0.75,-0.75);

	    \coordinate (v1) at (0,0);

		\draw[massless] (e2) -- (v1);
        \draw[massiveLin,red] (e1) -- (v1);
        \draw[massless] (e4) -- (v1);

        \node[above] at (e1) {$q$}; 
        \node[below left] at (e2) {$1$}; 
        \node[below right] at (e4) {$n$}; 

        \node at ($(e2)!.5!(e4)$) {\bf{\ldots}};
        \path[draw=black,line width=1.5pt, fill=white] (v1) circle[radius=0.2];
        \path[draw=black, line width=1.5pt] (v1) -- +(0.12,0.12);
        \path[draw=black, line width=1.5pt] (v1) -- +(-0.12,-0.12);
        \path[draw=black, line width=1.5pt] (v1) -- +(0.12,-0.12);
        \path[draw=black, line width=1.5pt] (v1) -- +(-0.12,0.12);
  \end{tikzpicture}}

\newcommand{\mhvAmp}{
\begin{tikzpicture}[scale=1]
        \coordinate (e1) at (+0.75,+0.75);
        \coordinate (e2) at (-0.75,+0.75);
        \coordinate (e3) at (-0.75,-0.75);
	    \coordinate (e4) at (+0.75,-0.75);

	   \coordinate (v1) at (0,0);

		\draw[massless] (e2) -- (v1);
        \draw[massless] (e1) -- (v1);
        \draw[massless] (e3) -- (v1);
        \draw[massless] (e4) -- (v1);

        \node[above right] at (e1) {$1$}; 
        \node[above left] at (e2) {$2$}; 
        \node[below left] at (e3) {$3$};
        \node[below right] at (e4) {$n$};

        \node at ($(e3)!.5!(e4)$) {\bf{\ldots}};
        \path[draw=black,line width=1.5pt, fill=ampgrey] (v1) circle[radius=0.35];
  \end{tikzpicture}}

\newcommand{\diagmhv}{
\begin{tikzpicture}[scale=1]
        \coordinate (e1) at (0,1);
        \coordinate (e2) at (-0.75,-0.75);
	    \coordinate (e4) at (+0.75,-0.75);

	\coordinate (v1) at (0,0);

		\draw[massless] (e2) -- (v1);
        \draw[massless] (e1) -- (v1);
        \draw[massless] (e4) -- (v1);

        \node[above] at (e1) {$1$}; 
        \node[below left] at (e2) {$2$}; 
        \node[below right] at (e4) {$3$}; 

        \path[draw=black,line width=1.5pt, fill=mhvblue] (v1) circle[radius=0.25];
  \end{tikzpicture}}

  \newcommand{\diagantimhv}{
\begin{tikzpicture}[scale=1]
        \coordinate (e1) at (0,1);
        \coordinate (e2) at (-0.75,-0.75);
	    \coordinate (e4) at (+0.75,-0.75);

	    \coordinate (v1) at (0,0);

		\draw[massless] (e2) -- (v1);
        \draw[massless] (e1) -- (v1);
        \draw[massless] (e4) -- (v1);

        \node[above] at (e1) {$1$}; 
        \node[below left] at (e2) {$2$}; 
        \node[below right] at (e4) {$3$}; 
        \path[draw=black,line width=1.5pt, fill=white] (v1) circle[radius=0.25];
  \end{tikzpicture}}

\newcommand{\ssonea}{
\begin{tikzpicture}[scale=1]
        \coordinate (e1) at (-.65,0);
		\coordinate (e2) at (-1.875, 1.125);
		\coordinate (e3) at (-1.875, -1.125);
        \coordinate (e4) at (1.875, -1.125);
        \coordinate (e5) at (1.875, 1.125);

		\coordinate (v1) at (0,0);
        \coordinate (v2) at (-1.5, 0.75);
        \coordinate (v3) at (-1.5, -0.75);
        \coordinate (v4) at (1.5,-.75);
        \coordinate (v5) at (1.5,.75);
        \coordinate (v6) at (0,.75);
        \coordinate (v7) at (0,-.75);

        \coordinate (l11) at (-.75,-1.2);
        \coordinate (l10) at (-.75,1.2);
        \coordinate (l12) at (.4,-.4);
        \coordinate (l6) at (-2,0);
        \coordinate (l9) at (.75,-1.2);
        \coordinate (l8) at (.75,1.2);
        \coordinate (l7) at (2,0);
        \coordinate (l13) at (.4,.4);
        
	\draw[massiveLin, red] (e1) -- (v1);
	\draw[massless] (e2) -- (v2);
        \draw[massless] (e3) -- (v3);
        \draw[massless] (e4) -- (v4);
        \draw[massless] (e5) -- (v5);
        \draw[masslessWithArrow] (v2) -- (v3);
        \draw[masslessWithArrow] (v3) -- (v7);
        \draw[masslessWithArrow] (v4) -- (v7);
        \draw[masslessWithArrow] (v5) -- (v4);
        \draw[masslessWithArrow] (v6) -- (v2);
        \draw[masslessWithArrow] (v6) -- (v5);
        \draw[masslessWithArrow] (v1) -- (v6);
        \draw[masslessWithArrow] (v7) -- (v1);

        \path[draw=black,line width=1.5pt, fill=white] (v1) circle[radius=0.2];
        \path[draw=black, line width=1.5pt] (v1) -- +(0.12,0.12);
        \path[draw=black, line width=1.5pt] (v1) -- +(-0.12,-0.12);
        \path[draw=black, line width=1.5pt] (v1) -- +(0.12,-0.12);
        \path[draw=black, line width=1.5pt] (v1) -- +(-0.12,0.12);

         \path[draw=black,line width=1.5pt, fill=white] (v2) circle[radius=0.2];
         \path[draw=black,line width=1.5pt, fill=mhvblue] (v3) circle[radius=0.2];
         \path[draw=black,line width=1.5pt, fill=white] (v4) circle[radius=0.2];
         \path[draw=black,line width=1.5pt, fill=mhvblue] (v5) circle[radius=0.2];
         \path[draw=black,line width=1.5pt, fill=white] (v6) circle[radius=0.2];
         \path[draw=black,line width=1.5pt, fill=white] (v7) circle[radius=0.2];

        \node[left]  at (e1) {$1$};
        \node[above left] at (e2) {$2$};
        \node[below left] at (e3) {$3$};
        \node[below right] at (e4) {$4$};
        \node[above right] at (e5) {$5$};

        \node at (l13) {$\ell_{13}$};
        \node at (l12) {$\ell_{12}$};
        \node at (l11) {$\ell_{11}$};
        \node at (l10) {$\ell_{10}$};
        \node at (l9) {$\ell_{9}$};
        \node at (l8) {$\ell_{8}$};
        \node at (l7) {$\ell_{7}$};
        \node at (l6) {$\ell_{6}$};          
\end{tikzpicture}}

\newcommand{\ssoneb}{
\begin{tikzpicture}[scale=1]
        \coordinate (e1) at (-.65,0);
		\coordinate (e2) at (-1.875, 1.125);
		\coordinate (e3) at (-1.875, -1.125);
        \coordinate (e4) at (1.875, -1.125);
        \coordinate (e5) at (1.875, 1.125);

		\coordinate (v1) at (0,0);
        \coordinate (v2) at (-1.5, 0.75);
        \coordinate (v3) at (-1.5, -0.75);
        \coordinate (v4) at (1.5,-.75);
        \coordinate (v5) at (1.5,.75);
        \coordinate (v6) at (0,.75);
        \coordinate (v7) at (0,-.75);

        \coordinate (l11) at (-.75,-1.2);
        \coordinate (l10) at (-.75,1.2);
        \coordinate (l12) at (.4,-.4);
        \coordinate (l6) at (-2,0);
        \coordinate (l9) at (.75,-1.2);
        \coordinate (l8) at (.75,1.2);
        \coordinate (l7) at (2,0);
        \coordinate (l13) at (.4,.4);
        
	\draw[massiveLin, red] (e1) -- (v1);
	\draw[massless] (e2) -- (v2);
        \draw[massless] (e3) -- (v3);
        \draw[massless] (e4) -- (v4);
        \draw[massless] (e5) -- (v5);
        \draw[masslessWithArrow] (v2) -- (v3);
        \draw[masslessWithArrow] (v3) -- (v7);
        \draw[masslessWithArrow] (v4) -- (v7);
        \draw[masslessWithArrow] (v5) -- (v4);
        \draw[masslessWithArrow] (v6) -- (v2);
        \draw[masslessWithArrow] (v6) -- (v5);
        \draw[masslessWithArrow] (v1) -- (v6);
        \draw[masslessWithArrow] (v7) -- (v1);

        \path[draw=black,line width=1.5pt, fill=white] (v1) circle[radius=0.2];
        \path[draw=black, line width=1.5pt] (v1) -- +(0.12,0.12);
        \path[draw=black, line width=1.5pt] (v1) -- +(-0.12,-0.12);
        \path[draw=black, line width=1.5pt] (v1) -- +(0.12,-0.12);
        \path[draw=black, line width=1.5pt] (v1) -- +(-0.12,0.12);

         \path[draw=black,line width=1.5pt, fill=mhvblue] (v2) circle[radius=0.2];
         \path[draw=black,line width=1.5pt, fill=white] (v3) circle[radius=0.2];
         \path[draw=black,line width=1.5pt, fill=mhvblue] (v4) circle[radius=0.2];
         \path[draw=black,line width=1.5pt, fill=white] (v5) circle[radius=0.2];
         \path[draw=black,line width=1.5pt, fill=white] (v6) circle[radius=0.2];
         \path[draw=black,line width=1.5pt, fill=white] (v7) circle[radius=0.2];

        \node[left]  at (e1) {$1$};
        \node[above left] at (e2) {$2$};
        \node[below left] at (e3) {$3$};
        \node[below right] at (e4) {$4$};
        \node[above right] at (e5) {$5$};

        \node at (l13) {$\ell_{13}$};
        \node at (l12) {$\ell_{12}$};
        \node at (l11) {$\ell_{11}$};
        \node at (l10) {$\ell_{10}$};
        \node at (l9) {$\ell_{9}$};
        \node at (l8) {$\ell_{8}$};
        \node at (l7) {$\ell_{7}$};
        \node at (l6) {$\ell_{6}$};          
\end{tikzpicture}}

\newcommand{\oneloopboxnumGen}{
\begin{tikzpicture}[scale=1]
        \coordinate (e1) at (-2,1);
        \coordinate (e2) at (-2,-1);
        \coordinate (e3) at (0.5, -1);
        \coordinate (e4) at (0.5, 1);

        \coordinate (v1) at (-1.5,.65);
        \coordinate (v2) at (-1.5,-.65);
        \coordinate (v3) at (0,-.65);
        \coordinate (v4) at (0,.65);

        \coordinate (l10) at (-.75,1);
        \coordinate (l8) at (-1.8,0);
        \coordinate (l7) at (-.75,-1.1);
        \coordinate (l13) at (.45,0);
        
	\draw[massless] (e2) -- (v2);
       \draw[massless] (e4) -- (v4);
        \draw[masslessWithArrow] (v2) -- (v1);
        \draw[masslessWithArrow] (v1) -- (v4);
        \draw[masslessWithArrow] (v4) -- (v3);
        \draw[masslessWithArrow] (v3) -- (v2);
        \fill (v1) -- (-2.2,1) -- (-2,1.25);
        \fill (v3) -- (0.7,-0.95) -- (0.5,-1.2);

        \path[draw=black,line width=1.5pt, fill=black] (v1) circle[radius=0.1];
        \path[draw=black,line width=1.5pt, fill=black] (v3) circle[radius=0.1];

        \node[below left] at (e2) {$2$};
        \node[above right]  at (e4) {$4$};
        \node[above left] at (e1) {$1$};
        \node[below right] at (e3) {$3$};

        \node at (l13) {$\ell_{c}$};
        \node at (l10) {$\ell_{b}$};
        \node at (l8) {$\ell_{a}$};
        \node at (l7) {$\ell_{d}$};      
\end{tikzpicture}}

\newcommand{\oneloopboxMHVGen}{
\begin{tikzpicture}[scale=1]
        \coordinate (e1) at (-2,1);
        \coordinate (e2) at (-2,-1);
        \coordinate (e3) at (0.5, -1);
        \coordinate (e4) at (0.5, 1);

        \coordinate (v1) at (-1.5,.65);
        \coordinate (v2) at (-1.5,-.65);
        \coordinate (v3) at (0,-.65);
        \coordinate (v4) at (0,.65);

        \coordinate (l10) at (-.75,1);
        \coordinate (l8) at (-1.8,0);
        \coordinate (l7) at (-.75,-1.1);
        \coordinate (l13) at (.45,0);
        
	\draw[massless] (e2) -- (v2);
        \draw[massless] (e4) -- (v4);
        \draw[masslessWithArrow] (v2) -- (v1);
        \draw[masslessWithArrow] (v1) -- (v4);
        \draw[masslessWithArrow] (v4) -- (v3);
        \draw[masslessWithArrow] (v3) -- (v2);
       \fill (v1) -- (-2.25,1.0) -- (-2.05,1.3);
        \fill (v3) -- (0.7,-1.0) -- (0.5,-1.25);

        \path[draw=black,line width=1.5pt, fill=white] (v2) circle[radius=0.2];
        \path[draw=black,line width=1.5pt, fill=white] (v4) circle[radius=0.2];
         \path[draw=black,line width=1.5pt, fill=black] (v1) circle[radius=0.1];
        \path[draw=black,line width=1.5pt, fill=black] (v3) circle[radius=0.1];

        \node[below left] at (e2) {$2$};
        \node[above right]  at (e4) {$4$};
        \node[above left] at (e1) {$1$};
        \node[below right] at (e3) {$3$};

        \node at (l13) {$\ell_{c}$};
        \node at (l10) {$\ell_{b}$};
        \node at (l8) {$\ell_{a}$};
        \node at (l7) {$\ell_{d}$};      
\end{tikzpicture}}

\newcommand{\oneloopboxMHVbarGen}{
\begin{tikzpicture}[scale=1]
        \coordinate (e1) at (-2,1);
        \coordinate (e2) at (-2,-1);
        \coordinate (e3) at (0.5, -1);
        \coordinate (e4) at (0.5, 1);

        \coordinate (v1) at (-1.5,.65);
        \coordinate (v2) at (-1.5,-.65);
        \coordinate (v3) at (0,-.65);
        \coordinate (v4) at (0,.65);

        \coordinate (l10) at (-.75,1);
        \coordinate (l8) at (-1.8,0);
        \coordinate (l7) at (-.75,-1.1);
        \coordinate (l13) at (.45,0);
        
	\draw[massless] (e2) -- (v2);
        \draw[massless] (e4) -- (v4);
        \draw[masslessWithArrow] (v2) -- (v1);
        \draw[masslessWithArrow] (v1) -- (v4);
        \draw[masslessWithArrow] (v4) -- (v3);
        \draw[masslessWithArrow] (v3) -- (v2);
        \fill (v1) -- (-2.25,1.0) -- (-2.05,1.3);
        \fill (v3) -- (0.7,-1.0) -- (0.5,-1.25);

        \path[draw=black,line width=1.5pt, fill=mhvblue] (v2) circle[radius=0.2];
        \path[draw=black,line width=1.5pt, fill=mhvblue] (v4) circle[radius=0.2];
         \path[draw=black,line width=1.5pt, fill=black] (v1) circle[radius=0.1];
        \path[draw=black,line width=1.5pt, fill=black] (v3) circle[radius=0.1];

        \node[below left] at (e2) {$2$};
        \node[above right]  at (e4) {$4$};
        \node[above left] at (e1) {$1$};
        \node[below right] at (e3) {$3$};

        \node at (l13) {$\ell_{c}$};
        \node at (l10) {$\ell_{b}$};
        \node at (l8) {$\ell_{a}$};
        \node at (l7) {$\ell_{d}$};      
\end{tikzpicture}}

\newcommand{\oneloopbox}{
\begin{tikzpicture}[scale=1]
        \coordinate (e1) at (0.5, 1);
        \coordinate (e2) at (-1.5,1.4);
	\coordinate (e3) at (-2.4,.65);
        \coordinate (e4) at (-2,-1);
        \coordinate (e5) at (0, -1.35);
        \coordinate (e6) at (0.7, -0.65);
        \coordinate (e7) at (-2.2,1.2);

        \coordinate (v2) at (0,.65);
        \coordinate (v4) at (-1.5,.65);
        \coordinate (v5) at (-1.5,-.65);
        \coordinate (v6) at (0,-.65);

        \coordinate (l10) at (-.75,1.1);
        \coordinate (l8) at (-2,0);
        \coordinate (l7) at (-.75,-1.1);
        \coordinate (l13) at (.5,0);
        
	\draw[massless] (e1) -- (v2);
	\draw[massless] (e3) -- (v4);
        \draw[massiveLin,red] (e7) -- (v4);
        \draw[massless] (e2) -- (v4);
        \draw[massless] (e4) -- (v5);
        \draw[massless] (e5) -- (v6);
        \draw[massless] (e6) -- (v6);
        \draw[masslessWithArrow] (v4) -- (v2);
        \draw[masslessWithArrow] (v5) -- (v4);
        \draw[masslessWithArrow] (v6) -- (v5);
        \draw[masslessWithArrow] (v2) -- (v6);

        \path[draw=black,line width=1.5pt, fill=white] (v4) circle[radius=0.2];
        \path[draw=black, line width=1.5pt] (v4) -- +(0.12,0.12);
        \path[draw=black, line width=1.5pt] (v4) -- +(-0.12,-0.12);
        \path[draw=black, line width=1.5pt] (v4) -- +(0.12,-0.12);
        \path[draw=black, line width=1.5pt] (v4) -- +(-0.12,0.12);

        \path[draw=black,line width=1.5pt, fill=ampgrey] (v6) circle[radius=0.25];
        \path[draw=black,line width=1.5pt, fill=white] (v2) circle[radius=0.2];
         \path[draw=black,line width=1.5pt, fill=white] (v5) circle[radius=0.2];

        \node[above right]  at (e1) {$j$};
        \node[above left] at (e7) {$1$};
        \node[left] at (e3) {$i-1$};
        \node[below left] at (e4) {$i$};
        \node[below] at (e5) {$i+1$};
        \node[right] at (e6) {$j-1$};
        \node[above] at (e2) {$j+1$};
        \node at (0.45, -0.9) [circle,fill,inner sep=1 pt]{};
        \node at (0.25, -1.15) [circle,fill,inner sep=1 pt]{};
        \node at (-1.75, 1.15) [circle,fill,inner sep=1 pt]{};
        \node at (-2, 0.8) [circle,fill,inner sep=1 pt]{};

        \node at (l13) {$\ell_{c}$};
        \node at (l10) {$\ell_{b}$};
        \node at (l8) {$\ell_{a}$};
        \node at (l7) {$\ell_{d}$};
\end{tikzpicture}}

\newcommand{\oneloopboxMHVBar}{
\begin{tikzpicture}[scale=1]
        \coordinate (e1) at (0.5, 1);
        \coordinate (e2) at (-1.5,1.4);
	\coordinate (e3) at (-2.4,.65);
        \coordinate (e4) at (-2,-1);
        \coordinate (e5) at (0, -1.35);
        \coordinate (e6) at (0.7, -0.65);
        \coordinate (e7) at (-2.2,1.2);

        \coordinate (v2) at (0,.65);
        \coordinate (v4) at (-1.5,.65);
        \coordinate (v5) at (-1.5,-.65);
        \coordinate (v6) at (0,-.65);

        \coordinate (l10) at (-.75,1.1);
        \coordinate (l8) at (-2,0);
        \coordinate (l7) at (-.75,-1.1);
        \coordinate (l13) at (.5,0);
        
	\draw[massless] (e1) -- (v2);
	\draw[massless] (e3) -- (v4);
        \draw[massiveLin,red] (e7) -- (v4);
        \draw[massless] (e2) -- (v4);
        \draw[massless] (e4) -- (v5);
        \draw[massless] (e5) -- (v6);
        \draw[massless] (e6) -- (v6);
        \draw[masslessWithArrow] (v4) -- (v2);
        \draw[masslessWithArrow] (v5) -- (v4);
        \draw[masslessWithArrow] (v6) -- (v5);
        \draw[masslessWithArrow] (v2) -- (v6);

        \path[draw=black,line width=1.5pt, fill=white] (v4) circle[radius=0.2];
        \path[draw=black, line width=1.5pt] (v4) -- +(0.12,0.12);
        \path[draw=black, line width=1.5pt] (v4) -- +(-0.12,-0.12);
        \path[draw=black, line width=1.5pt] (v4) -- +(0.12,-0.12);
        \path[draw=black, line width=1.5pt] (v4) -- +(-0.12,0.12);

        \path[draw=black,line width=1.5pt, fill=ampgrey] (v6) circle[radius=0.25];
        \path[draw=black,line width=1.5pt, fill=mhvblue] (v2) circle[radius=0.2];
        \path[draw=black,line width=1.5pt, fill=mhvblue] (v5) circle[radius=0.2];

        \node[above right]  at (e1) {$j$};
        \node[above left] at (e7) {$1$};
        \node[left] at (e3) {$i-1$};
        \node[below left] at (e4) {$i$};
        \node[below] at (e5) {$i+1$};
        \node[right] at (e6) {$j-1$};
        \node[above] at (e2) {$j+1$};
        \node at (0.45, -0.9) [circle,fill,inner sep=1 pt]{};
        \node at (0.25, -1.15) [circle,fill,inner sep=1 pt]{};
        \node at (-1.75, 1.15) [circle,fill,inner sep=1 pt]{};
        \node at (-2, 0.8) [circle,fill,inner sep=1 pt]{};

        \node at (l13) {$\ell_{c}$};
        \node at (l10) {$\ell_{b}$};
        \node at (l8) {$\ell_{a}$};
        \node at (l7) {$\ell_{d}$};
\end{tikzpicture}}

\newcommand{\oneloopboxnum}{
\begin{tikzpicture}[scale=1]
        \coordinate (e1) at (0.5, 1);
        \coordinate (e2) at (-1.5,1.4);
	\coordinate (e3) at (-2.4,.65);
        \coordinate (e4) at (-2,-1);
        \coordinate (e5) at (0, -1.35);
        \coordinate (e6) at (0.7, -0.65);
        \coordinate (e7) at (-2.2,1.2);

        \coordinate (v2) at (0,.65);
        \coordinate (v4) at (-1.5,.65);
        \coordinate (v5) at (-1.5,-.65);
        \coordinate (v6) at (0,-.65);

        \coordinate (l10) at (-.75,1.1);
        \coordinate (l8) at (-2,0);
        \coordinate (l7) at (-.75,-1.1);
        \coordinate (l13) at (.5,0);
        
	\draw[massless] (e1) -- (v2);
	\draw[massless] (e3) -- (v4);
        \draw[massiveLin] (e7) -- (v4);
        \draw[massless] (e2) -- (v4);
        \draw[massless] (e4) -- (v5);
        \draw[massless] (e5) -- (v6);
        \draw[massless] (e6) -- (v6);
        \draw[masslessWithArrow] (v4) -- (v2);
        \draw[masslessWithArrow] (v5) -- (v4);
        \draw[masslessWithArrow] (v6) -- (v5);
        \draw[masslessWithArrow] (v2) -- (v6);

        \path[draw=black,line width=1.5pt, fill=black] (v4) circle[radius=0.1];
        \path[draw=black,line width=1.5pt, fill=black] (v6) circle[radius=0.1];

        \node[above right]  at (e1) {$j$};
        \node[above left] at (e7) {$1$};
        \node[left] at (e3) {$i-1$};
        \node[below left] at (e4) {$i$};
        \node[below] at (e5) {$i+1$};
        \node[right] at (e6) {$j-1$};
        \node[above] at (e2) {$j+1$};
        \node at (0.45, -0.9) [circle,fill,inner sep=1 pt]{};
        \node at (0.25, -1.15) [circle,fill,inner sep=1 pt]{};
        \node at (-1.75, 1.15) [circle,fill,inner sep=1 pt]{};
        \node at (-2, 0.8) [circle,fill,inner sep=1 pt]{};

        \node at (l13) {$\ell_{c}$};
        \node at (l10) {$\ell_{b}$};
        \node at (l8) {$\ell_{a}$};
        \node at (l7) {$\ell_{d}$};
\end{tikzpicture}}

\newcommand{\onelooptrinumTwoMass}{
\begin{tikzpicture}[scale=1]
        \coordinate (e1) at (-2,0);
	\coordinate (e2) at (.5,-1);
        \coordinate (e3) at (.5,1);

        \coordinate (v1) at (-1.5,0);
        \coordinate (v2) at (0,-.65);
        \coordinate (v3) at (0,.65);

        \coordinate (l10) at (-0.75,.68);
        \coordinate (l7) at (-0.75,-0.65);
        \coordinate (l13) at (.4,0);
        
	    \draw[massiveLin] (e1) -- (v1);
        \draw[massiveLin] (e2) -- (v2);
        \draw[massless] (e3) -- (v3);
        \draw[masslessWithArrow] (v2) -- (v1);
        \draw[masslessWithArrow] (v3) -- (v2);
        \draw[masslessWithArrow] (v1) -- (v3);
        \path[draw=black,line width=1.5pt, fill=black] (v1) circle[radius=0.1];
        \path[draw=black,line width=1.5pt, fill=black] (v2) circle[radius=0.1];
        \fill (v1)--(-2.1,-0.2)--(-2.1,0.2);
        \fill (v2)--(0.45,-1.2)--(0.65,-0.9);

        \node[left]  at (e1) {$1+2$};
        \node[below right] at (e2) {$3$};
        \node[above right] at (e3) {$4$};
        \node at (l13) {$\ell_{c}$};
        \node at (l10) {$\ell_{b}$};
        \node at (l7) {$\ell_{d}$};   
\end{tikzpicture}}

\newcommand{\onelooptrinumOneMass}{
\begin{tikzpicture}[scale=1]
        \coordinate (e1) at (-2,0);
	\coordinate (e2) at (.5,-1);
        \coordinate (e3) at (.5,1);

        \coordinate (v1) at (-1.5,0);
        \coordinate (v2) at (0,-.65);
        \coordinate (v3) at (0,.65);

        \coordinate (l10) at (-0.75,.68);
        \coordinate (l7) at (-0.75,-0.65);
        \coordinate (l13) at (.4,0);
        
	    \draw[massiveLin] (e1) -- (v1);
        \draw[massless] (e2) -- (v2);
        \draw[massless] (e3) -- (v3);
        \draw[masslessWithArrow] (v2) -- (v1);
        \draw[masslessWithArrow] (v3) -- (v2);
        \draw[masslessWithArrow] (v1) -- (v3);
        \path[draw=black,line width=1.5pt, fill=black] (v1) circle[radius=0.1];
        \path[draw=black,line width=1.5pt, fill=black] (v2) circle[radius=0.1];
        \fill (v1)--(-2.1,-0.2)--(-2.1,0.2);

        \node[left]  at (e1) {$Q$};
        \node[below right] at (e2) {$i$};
        \node[above right] at (e3) {$j$};
        \node at (l13) {$\ell_{c}$};
        \node at (l10) {$\ell_{b}$};
        \node at (l7) {$\ell_{d}$};   
\end{tikzpicture}}

\newcommand{\onelooptri}{
\begin{tikzpicture}[scale=1]
        \coordinate (e1) at (1,1);
	\coordinate (e2) at (-1.1,0.7);
        \coordinate (e3) at (1, -1);
        \coordinate (e4) at (-1.1,-0.7);
        \coordinate (e5) at (-1.4,0);

        \coordinate (v2) at (.5,.5);
        \coordinate (v3) at (-0.6,0);
        \coordinate (v6) at (.5,-.5);

        \coordinate (l10) at (-0.1,.60);
        \coordinate (l7) at (-0.1,-0.60);
        \coordinate (l13) at (.9,0);
        
	\draw[massless] (e4) -- (v3);
        \draw[massless] (e1) -- (v2);
        \draw[massiveLin, red] (e5) -- (v3);
        \draw[massless] (e2)--(v3);
        \draw[massless] (e3) -- (v6);
        \draw[masslessWithArrow] (v3) -- (v2);
        \draw[masslessWithArrow] (v6) -- (v3);
        \draw[masslessWithArrow] (v2) -- (v6);

        \path[draw=black,line width=1.5pt, fill=white] (v3) circle[radius=0.2];
        \path[draw=black, line width=1.5pt] (v3) -- +(0.12,0.12);
        \path[draw=black, line width=1.5pt] (v3) -- +(-0.12,-0.12);
        \path[draw=black, line width=1.5pt] (v3) -- +(0.12,-0.12);
        \path[draw=black, line width=1.5pt] (v3) -- +(-0.12,0.12);

        \path[draw=black,line width=1.5pt, fill=white] (v2) circle[radius=0.2];
        \path[draw=black,line width=1.5pt, fill=mhvblue] (v6) circle[radius=0.2];

        \node[above right]  at (e1) {$i+1$};
        \node[above left] at (e2) {$i+2$};
        \node[left] at (e5) {$1$};
        \node[below left] at (e4) {$i-1$};
        \node[below right] at (e3) {$i$};
        \node at (-1.1, -0.3) [circle,fill,inner sep=1 pt]{};
        \node at (-1.1, +0.3) [circle,fill,inner sep=1 pt]{};
        \node at (l13) {$\ell_{c}$};
        \node at (l10) {$\ell_{b}$};
        \node at (l7) {$\ell_{d}$};  
\end{tikzpicture}}

\newcommand{\onelooptriint}{
\begin{tikzpicture}[scale=1]
        \coordinate (e1) at (1,1);
	\coordinate (e2) at (-1.1,0.7);
        \coordinate (e3) at (1, -1);
        \coordinate (e4) at (-1.1,-0.7);
        \coordinate (e5) at (-1.4,0);

        \coordinate (v2) at (.5,.5);
        \coordinate (v3) at (-0.6,0);
        \coordinate (v6) at (.5,-.5);

        \coordinate (l10) at (-0.1,.60);
        \coordinate (l7) at (-0.1,-0.60);
        \coordinate (l13) at (.9,0);
        
	\draw[massless] (e4) -- (v3);
        \draw[massless] (e1) -- (v2);
        \draw[massiveLin] (e5) -- (v3);
        \draw[massless] (e2)--(v3);
        \draw[massless] (e3) -- (v6);
        \draw[masslessWithArrow] (v3) -- (v2);
        \draw[masslessWithArrow] (v6) -- (v3);
        \draw[masslessWithArrow] (v2) -- (v6);

        \path[draw=black,line width=1.5pt, fill=black] (v3) circle[radius=0.1];

        \node[above right]  at (e1) {$i+1$};
        \node[above left] at (e2) {$i+2$};
        \node[left] at (e5) {$1$};
        \node[below left] at (e4) {$i-1$};
        \node[below right] at (e3) {$i$};
        \node at (-1.1, -0.3) [circle,fill,inner sep=1 pt]{};
        \node at (-1.1, +0.3) [circle,fill,inner sep=1 pt]{};
        \node at (l13) {$\ell_{c}$};
        \node at (l10) {$\ell_{b}$};
        \node at (l7) {$\ell_{d}$};  
\end{tikzpicture}}

\newcommand{\oneloopblob}{
\begin{tikzpicture}[scale=1]
        \coordinate (e1) at (1.5,1.5);
	\coordinate (e2) at (-1.5,1.5);
        \coordinate (e3) at (1.5, -1.5);
        \coordinate (e4) at (-1.5,-1.5);

	\coordinate (v1) at (0,0);
        \coordinate (v2) at (0,0);

	\draw[massless] (e1) -- (v1);
        \draw[massless] (e2) -- (v1);
        \draw[massless] (e3) -- (v1);
        \draw[massless] (e4) -- (v1);
        \path[draw=black,line width=1.5pt, fill=ampgrey] (v2) circle[radius=0.75];
        \path[draw=black,line width=1.5pt, fill=white] (v1) circle[radius=0.3];    

\end{tikzpicture}}

\newcommand{\oneloopcut}{
\begin{tikzpicture}[scale=1]
        \coordinate (e1) at (-2.0,1.5);
	\coordinate (e2) at (-2.0,-1.5);
        \coordinate (e3) at (2.0, 1.5);
        \coordinate (e4) at (2.0,-1.5);
        \coordinate (e5) at (-1,0.5);
	\coordinate (e6) at (-1.0,-0.5);
        \coordinate (e7) at (1, 0.5);
        \coordinate (e8) at (1,-0.5);

	\coordinate (v1) at (0,0);
        \coordinate (v2) at (0,0);

	\draw[massless] (e1) -- (e5);
        \draw[massless] (e2) -- (e6);
        \draw[massless] (e3) -- (e7);
        \draw[massless] (e4) -- (e8);
        \draw[massless] (e5) -- (e7);
        \draw[massless] (e6) -- (e8);
        \node[ellipse,draw = black,fill = ampgrey, minimum width = 1cm, minimum height = 1.8cm] (e) at (-1,0) {};
         \node[ellipse,draw = black,fill = ampgrey, minimum width = 1cm, minimum height = 1.8cm] (f) at (1,0) {};
         \draw[dashed, -] (0,1.1) to (0,-1.1);
\end{tikzpicture}}

\newcommand{\onelooptrisoft}{
\begin{tikzpicture}[scale=1]
        \coordinate (e1) at (1,1);
	    \coordinate (e2) at (-1.1,0.7);
        \coordinate (e3) at (1, -1);
        \coordinate (e4) at (-1.1,-0.7);
        \coordinate (e5) at (-1.4,0);

        \coordinate (v2) at (.5,.5);
        \coordinate (v3) at (-0.6,0);
        \coordinate (v6) at (.5,-.5);

        \coordinate (l10) at (-0.1,.60);
        \coordinate (l7) at (-0.1,-0.60);
        \coordinate (l13) at (1.2,0);
        
	    \draw[massless] (e4) -- (v3);
        \draw[massless] (e1) -- (v2);
        \draw[massless](e2)--(v3);
        \draw[massiveLin,red] (e5) -- (v3);
        \draw[massless] (e3) -- (v6);
        \draw[masslessWithArrow] (v3) -- (v2);
        \draw[masslessWithArrow] (v6) -- (v3);

         \path[draw=black,line width=1.5pt, fill=white] (v3) circle[radius=0.2];
         \path[draw=black, line width=1.5pt] (v3) -- +(0.12,0.12);
         \path[draw=black, line width=1.5pt] (v3) -- +(-0.12,-0.12);
         \path[draw=black, line width=1.5pt] (v3) -- +(0.12,-0.12);
         \path[draw=black, line width=1.5pt] (v3) -- +(-0.12,0.12);

        \path[draw=black,line width=1.5pt, fill=black] (v2) circle[radius=0.1];
        \path[draw=black,line width=1.5pt, fill=black] (v6) circle[radius=0.1];
        \draw[decorate sep={0.5 mm}{1.2 mm},fill]  (v2) -- (v6);

        \node[above right]  at (e1) {$i+1$};
        \node[above left] at (e2) {$i+2$};
        \node[below left] at (e4) {$i-1$};
        \node[below right] at (e3) {$i$};
        \node[left] at (e5) {$1$};
        \node at (-1.1, -0.3) [circle,fill,inner sep=1 pt]{};
        \node at (-1.1, +0.3) [circle,fill,inner sep=1 pt]{};
        \node at (l13) {$\ell_{c}=0$};
        \node at (l10) {$\ell_{b}$};
        \node at (l7) {$\ell_{d}$};   
\end{tikzpicture}}

\begin{document}

\unitlength = .8mm
\renewcommand{\thesubfigure}{(\arabic{subfigure})}

\begin{titlepage}

\begin{center}

\hfill \\
\hfill \\
\vskip 1cm

\title{The planar two-loop four-point form factor in maximally supersymmetric Yang-Mills theory}

\author{Tushar Gopalka$^1$, Enrico Herrmann$^1$}

\address{
${}^1$Mani L. Bhaumik Institute for Theoretical Physics,
University of California at Los Angeles,
Los Angeles, CA 90095, USA}
\end{center}

\vskip 1 cm 

\abstract{We derive the four-dimensional integrand of the maximal-helicity-violating four-particle form factor for the chiral part of the stress-tensor supermultiplet in planar $\mathcal{N}=4$ super-Yang-Mills theory at two loops. In our integrand construction, we adopt a special integrand basis with triangle powercounting where each basis element has unit leading singularities on all co-dimension eight residues. This basis was first constructed in the context of two-loop $n$-point scattering amplitudes in $\mathcal{N}=4$ beyond the planar limit and we describe here, how to directly utilize it for the form factor computation. Our result sets the stage for an independent confirmation of the symbol level bootstrap results of Dixon et al.~via direct evaluation. The imminent availability of all relevant two-loop five-point, one-mass master integrals will permit further investigations of a set of marvelous (self-)dualities between form factors and scattering amplitudes in planar $\mathcal{N}=4$ in the near future. 
}

\vfill

\end{titlepage}

\eject

\begingroup
    \setcounter{tocdepth}{2}
    \hypersetup{linkcolor=black}
    \tableofcontents
\endgroup

\vspace{1cm}

\section{Introduction}
\label{sec:intro}

The discovery of the incredible simplicity of the tree-level multi-gluon Parke-Taylor scattering amplitude \cite{parkeTaylor1985,Parke:1986gb,Mangano:1987xk,Mangano:1990by} in the mid 1980s ignited tremendous progress in our understanding of perturbative \emph{scattering amplitudes} in quantum field theory. On the one hand, these insights exposed, for example, connections to modern developments in mathematics~\cite{Lusztig,Postnikov:2006kva,Postnikov2009matching,Williams:2003a,Goncharov:2011hp,KLS}, the mathematical structures behind functions and numbers appearing in perturbation theory~\cite{Broadhurst:1996kc,Kreimer:1997dp,Bloch:2005bh,Aluffi:2008sy,Brown:2009ta,Marcolli:2009zy,Brown:2009rc,Brown:2010bw,Bogner:2014mha,Brown:2015fyf,Panzer:2015ida,Caron-Huot:2019bsq,Gurdogan:2020ppd}, and shed light on the ultraviolet properties of quantum gravity~\cite{Bern:1998ug,Bern:2007xj,Bern:2012uf,Bern:2015xsa}. On the other hand, advanced computational tools have been applied to phenomenologically relevant questions ranging from particle physics to classical gravitational wave science. The majority of these discoveries originated from concrete computations, often in simplified toy models, that pushed the boundaries of our theoretical reach. Several simplifying structures in turn led to new, more powerful computational tools. Exemplary milestones of this virtuous circle of progress include the Britto Cachazo Feng Witten (BCFW) on-shell recursion relations at tree-~\cite{BCF, BCFW}, and loop-level~\cite{ArkaniHamed:2010kv}, as well as various new features of scattering amplitudes in planar maximally supersymmetric gauge theory ($\calN=4$ sYM): The twistor string \cite{Witten:2003nn,Roiban:2004vt}, the discovery of dual conformal invariance \cite{Drummond:2006rz,Alday:2007hr,Drummond:2008vq}, the duality of amplitudes to Wilson loops and correlation functions \cite{Drummond:2007aua,Brandhuber:2007yx,Drummond:2007au,Mason:2010yk,CaronHuot:2010ek,Alday:2010zy,Eden:2010zz,Eden:2010ce}, the connection to Grassmannian geometry \cite{ArkaniHamed:2012nw,ArkaniHamed:2009dn,ArkaniHamed:2009vw,Mason:2009qx,ArkaniHamed:2009dg,ArkaniHamed:2009sx,Huang:2013owa}, and the amplituhedron \cite{ArkaniHamed:2010gg,Arkani-Hamed:2013jha}, together with various bootstrap methods \cite{Dixon:2011pw,Dixon:2011nj,Dixon:2013eka,Duhr:2011zq}. Beyond the planar limit, the double-copy between gauge theory and gravity amplitudes \cite{Bern:2008qj,Bern:2010ue,Bern:2019prr} opened a new path to perform high-loop gravity calculations both for quantum gravity \cite{Bern:2010tq,Carrasco:2011mn,Bern:2011rj,Boucher-Veronneau:2011rlc, Bern:2018jmv} and for classical gravitational wave applications \cite{Cheung:2018wkq, Kosower:2018adc, Bern:2019nnu}, see also \cite{Bern:2022wqg, Bjerrum-Bohr:2022blt, Kosower:2022yvp} for reviews. For a broad overview of these developments, see e.g.~\cite{Bern:1996je,Cachazo:2005ga,Beisert:2010jr,Elvang:2013cua,Dixon:2013uaa,Henn:2014yza,Duhr:2014woa,Weinzierl:2016bus,Travaglini:2022uwo, Brandhuber:2022qbk}.

\newpage

Besides on-shell scattering amplitudes, form factors are another interesting class of objects in quantum field theory. In many aspects, form factors represent a bridge between fully on-shell amplitudes, where some of the recent developments have been recounted above, and off-shell correlation functions of local operators which, especially for conformal field theories, attracted significant interest in recent years, see e.g.~\cite{Simmons-Duffin:2016gjk} and references therein. For a given gauge-invariant local composite operator $\calO(x)$, the form factor $\calF_{\calO}$ is defined as the overlap of the off-shell state created by $\calO$ (inserted at spacetime point $x$) from the vacuum $\ket{0}$ with an on-shell $n$-particle state $\bra{1,\ldots,n}$. Fourier transforming to momentum space, the operator $\calO$ inserts an off-shell momentum\footnote{We highlight off-shell momenta in boldface typeset.} $\bfq$, s.t.~$\bfq^2 \neq 0$, and the form factor is defined as
\begin{align}
\calF_{n,\calO}(\bfq; 1,\ldots,n)\equivR \int d^4x\, e^{-i\, \bfq x}\, \langle 1,\ldots,n| \calO(x)|0\rangle\,.
\end{align}
Form factors play a crucial role in a variety of physical situations. Some of the earliest applications of form factors appeared in the context of deep inelastic scattering. Additionally, the $e^+e^- \to X$ annihilation process into a hadronic final state $X$ is governed by the form factor of the electromagnetic current. 
The Sudakov form factor \cite{Mueller:1979ih,Sen:1981sd,Collins:1980ih} governs the universal structure of infrared divergences of scattering amplitudes, see e.g.~\cite{Becher:2014oda}, and also appears in factorized cross sections where it allows the resummation of large logarithms to all orders in perturbation theory. For precision Higgs physics at the LHC, an accurate understanding of higher order corrections to Higgs boson cross sections for $gg \to  H + X$ processes is crucial. One ingredient is the virtual correction to the effective Higgs-gluon-gluon vertex which is presently known to four-loop order in Quantum Chromodynamics (QCD)~\cite{Lee:2022nhh}. Besides the few concrete examples highlighted here, there are numerous other phenomenological applications of form factors which we are not going to summarize here. 

In our work, we focus on the simpler maximally supersymmetric cousin of QCD: $\calN=4$ sYM. In this theory, form factors were first studied more than thirty years ago \cite{vanNeerven:1985ja} and have received increased attention both at weak coupling \cite{Brandhuber:2010ad,Bork:2010wf,Brandhuber:2011tv,Bork:2011cj,Henn:2011by,Gehrmann:2011xn,Brandhuber:2012vm,Bork:2012tt,Boels:2012ew,Penante:2014sza,Wilhelm:2014qua,Nandan:2014oga,Loebbert:2015ova,Frassek:2015rka,Brandhuber:2017bkg,Bianchi:2018rrj,Brandhuber:2018xzk,Yang:2016ear,Sever:2020jjx,Sever:2021nsq} and at strong coupling \cite{Alday:2007he,Maldacena:2010kp,Gao:2013dza,Sever:2021xga}. In perturbation theory, they can be calculated by efficient on-shell techniques that were originally developed in the context of scattering amplitudes. In particular, BCFW \cite{BCF, BCFW} recursion relations can be applied to construct higher-point form factors at tree level \cite{Brandhuber:2010ad,Brandhuber:2011tv} that impact loop-level calculations via generalized unitarity \cite{Bern:1994zx,Bern:1994cg,Britto:2004nc}. All these investigations have shown that a surprising simplicity persists for form factors when viewed in the light of modern on-shell techniques. For a review of form factors and their properties, see e.g.~\cite{Yang:2019vag,Brandhuber:2022qbk}.

In an unexpected recent development, novel bootstrap results have uncovered a mysterious `antipodal duality' between the six-particle maximally-helicity violating (MHV) on-shell scattering amplitude and the MHV three-particle form factor of a single chiral stress tensor supermultiplet operator in planar maximally supersymmetric Yang-Mills theory~\cite{Dixon:2021tdw}. (See also \cite{Liu:2022vck} for related work.) The antipode refers to a special operation at the level of symbols or co-products defined on polylogarithmic functions \cite{goncharov2001multiple, goncharov2004galois, Goncharov:2009kx, Goncharov:2010jf, Brown:2011ik, Duhr:2011zq, Duhr:2012fh,Duhr:2014woa} and roughly corresponds to exchanging derivatives with discontinuities. Subsequently, an antipodal self-duality of the two-loop four-particle MHV form factor of the chiral stress tensor supermultiplet (also obtained by a bootstrap approach) has been observed at the symbol level in Ref.~\cite{Dixon:2022xqh}. In the light-like limit where the momentum of the operator becomes null ($q^2=0$), the result had been known from Ref.~\cite{Guo:2022qgv} partially based on the availability of the results for the massless two-loop five-particle master integrals, see e.g.~\cite{Abreu:2018aqd,Chicherin:2018old,Chicherin:2020oor}.

In this work, we construct the four-dimensional loop integrand of the two-loop four-particle MHV form factor of the chiral stress tensor multiplet using a tailored variant of generalized unitarity \cite{Bern:1994zx,Bern:1994cg,Britto:2004nc} that leads to extremely compact and clean expressions with many desirable properties~\cite{Bourjaily:2019iqr,Bourjaily:2019gqu}. The basic idea behind generalized unitarity is quite simple: Loop integrands are rational functions that can be determined by their residues. Furthermore, we can expand any $n$-particle $L$-loop amplitude or form factor integrand in a complete basis of rational functions\footnote{In this general discussion, we schematically write $\calA_n^{(L)}$ which denotes both $L$-loop amplitude and form factor integrands alike. Similarly, when we write `amplitude' in the context of the unitarity construction, the statements hold for form factors as well.}:
\begin{align}
\label{eq:gen_unitarity_expansion}
    \calA_n^{(L)}=\sum_{k} c_k\,\calI^{(L)}_k\,.
\end{align}
The size of the basis depends on the spacetime dimension and the power counting of the quantum field theory in question, see e.g.~the detailed discussions in \cite{Bourjaily:2020qca}. Given a complete (but a priori arbitrary) basis $\{\mathcal{I}^{(L)}_k\}$, the coefficients $c_k$ of the loop amplitude/form factor $\calA^{(L)}_n$ are determined by a linear algebra problem where we equate the residues of the basis expansion on the r.h.s. of (\ref{eq:gen_unitarity_expansion}) with those of field theory defined as the product of lower loop (and ultimately tree-level) on-shell amplitudes summed over all states that can cross the cuts. We will be more precise about this matching procedure in sections \ref{sec:unitarity} and \ref{sec:results}. As with any such linear algebra problem, the form of the resulting solution strongly depends on the choice of basis. Some bases are better than others. In Refs.~\cite{Bourjaily:2019iqr,Bourjaily:2019gqu}, one of the authors constructed a clean basis of integrands for scattering amplitudes in $\calN=4$ sYM \emph{beyond the planar limit}, which can be directly recycled in the \emph{planar} form factor computation. Because we consider the form factor of a color-singlet operator, even the leading color contribution requires non-planar kinematic integrands such as
\begin{align}
\label{eg:fig_np_penta_box}
    \vcenter{\hbox{\scalebox{.8}{\diagOne}}}\,.
\end{align}
The key idea of Refs.~\cite{Bourjaily:2019iqr,Bourjaily:2019gqu} was to construct a \emph{prescriptive} \cite{Bourjaily:2017wjl} basis of (four-dimensional) two-loop integrands with triangle powercounting (the multi-loop generalization of the scaling of one-loop scalar triangle integrals), see e.g.~\cite{Bourjaily:2020qca}, where each integrand basis element matches \emph{exactly} one field-theory cut out of a spanning set. Since we consider the form factor of a supersymmetry protected operator that does not get renormalized, we do not need any bubble topologies that would introduce UV divergences. Owing to this fact, the integrand bases constructed in Refs.~\cite{Bourjaily:2019iqr,Bourjaily:2019gqu} are perfectly suited for the problem at hand. The ``good'' integrand building blocks are chiral in the sense of Ref.~\cite{ArkaniHamed:2010gh} and have unit leading singularities, i.e. all co-dimension eight residues are either $\pm 1$ or 0. Empirically, several of these integrands have desirable properties upon integration, see e.g.~\cite{Arkani-Hamed:2014via,Abreu:2020jxa,Henn:2014qga,Badger:2023eqz}.

Integrating the two-loop four-particle MHV form factor integrand constructed here is beyond the scope of this work and would require the evaluation of state-of-the art two-loop five-point one-mass Feynman integrals. (One would also have to consider the compatibility of our four-dimensional integrand with $d$-dimensional loop integration techniques that are used in the majority of modern approaches to evaluate Feynman integrals, see e.g.~\cite{Smirnov:2004ym,Smirnov:2006ry,Smirnov:2012gma}. This is not a priori straightforward, but at the level of pure bases of master integrals, there has been considerable progress \cite{Abreu:2020jxa}.) Interestingly, all but the single two-loop five particle integral family depicted in (\ref{eg:fig_np_penta_box}) are already known in the literature, see \cite{Abreu:2020jxa,Badger:2021nhg} and references therein.\footnote{We learned from private communication with Ben Page that this family will be available in the near future.} With the imminent availability of analytic results for all relevant Feynman integrals, we would be in a position to amend the symbol level bootstrap result for the two-loop form factor of Ref.~\cite{Dixon:2022xqh} by transcendental constants and check the antipodal duality at the level of the co-action. A direct integration of the planar two-loop four-point form factor in $\calN=4$ sYM is not only the simplest physical setting in which to test the new analytic results of the master integrals, but also constitutes an independent verification and check of the bootstrap assumptions.

The remainder of this work is structured as follows: Section \ref{sec:unitarity} summarizes the main tools used in our computation. We first provide an overview of generalized unitarity in subsection \ref{subsec:unitarity_intro} and comment, in subsection \ref{subsec:integrand_bases}, on a refinement of unitarity in terms of a clean basis that was originally constructed in the context of scattering amplitudes in $\calN=4$ sYM beyond the planar limit \cite{Bourjaily:2019iqr,Bourjaily:2019gqu}. We introduce all relevant building blocks and explicitly evaluate some examples of the field-theory cuts as products of tree-level amplitudes and tree-level form factors summed over on-shell (super-)states in subsection \ref{subsec:supersum_gluing}. In section \ref{sec:results}, we perform the unitarity matching equation and present the results for the two-loop four-particle MHV form factor integrand of the chiral stress tensor supermultiplet in planar $\calN=4$ sYM. In section \ref{sec:conclusions} we offer our conclusions and an outlook of further directions. A technical appendix \ref{sec:Appendix} summarizes the relevant two-loop integral topologies and their (loop-)momentum dependent kinematic numerators. We provide our results in computer readable form as an ancillary file.

\section{Integrand Construction via Generalized Unitarity}
\label{sec:unitarity}

\subsection{Generalized Unitarity Overview}
\label{subsec:unitarity_intro}

Many computations of scattering amplitudes or form factors in perturbative quantum field theories can roughly be split into two distinct parts. First, one derives the \emph{loop integrand}, before performing integral reduction to a set of `master integrals' and their subsequent evaluation. Here, we focus on the first step: The integrand construction. Traditionally, the QFT textbook procedure involves the evaluation of all Feynman loop diagrams. However, as alluded to in the introduction, modern on-shell methods have highlighted even at tree-level that such a procedure misses important simplifications that should be exploited. Instead, one ought to focus on physical, gauge invariant on-shell quantities as much as possible.

The same philosophy has been reinforced in the generalized unitarity method  \cite{Bern:1994zx,Bern:1994cg,Britto:2004nc} originally developed by Bern, Dixon, Dunbar, and Kosower in the mid 1990s. This method ushered in decades of theoretical progress and provided phenomenologically relevant computational tools \cite{Ita:2011wn} for precision collider physics predictions by recycling simple on-shell tree-level results in the construction of loop integrands. In hindsight, the basic idea behind generalized unitarity is very simple. Feynman diagrams prior to loop integration are rational functions of external and loop momenta. Therefore, the sum of Feynman diagrams, i.e.~the loop integrand of a scattering amplitude/form factor is also a rational function. Viewed as a rational function, we are allowed to think about the loop integrand not only as a sum of Feynman diagrams, but more generally as an expansion into a sufficiently large basis of rational functions (see Eq.~(\ref{eq:gen_unitarity_expansion})) with coefficients that are determined by residues.  More concretely, we view bases of integrands $\{\mathcal{I}_k\}$ as functions of the loop momenta $\ell_i$, keeping the external momenta $p_j$ fixed $\{\mathcal{I}_k = \mathcal{I}_k(\ell_i,p_j)\}$. The coefficients, $c_k$, in Eq.~(\ref{eq:gen_unitarity_expansion}) are determined by the criterion that the right hand side matches field theory on all residues (of arbitrary co-dimension) in the loop variables as function of the external kinematics, (color-)charges, and polarizations of the states participating in the process under consideration\footnote{In principle, we can also consider residues in external kinematics. In fact, this is exactly the point of view that gives the BCFW recursion relations at tree level, but we will keep the two distinct for the purpose of loop level unitarity discussed here.}. Depending on the choice of basis $\{\mathcal{I}_k\}$, the coefficients $c_k$ in (\ref{eq:gen_unitarity_expansion}) may be individual residues or arbitrarily complicated linear combinations thereof.

A key aspect is based on the unitarity of the $S$-matrix, in a form that is familiar from the optical theorem in quantum mechanics, or the Cutkowsky rule studied in the 1960s. Consider $2\to 2$ scattering: The discontinuity, or `cut', of the one-loop amplitude is the product of two trees summed over the on-shell Lorentz invariant phase space (LIPS) and the exchanged states
\begin{align}
    \textbf{Cut}  
    \left[\vcenter{\hbox{\scalebox{.8}{\oneloopblob}}}\right] 
    = 
    \sum_{{\rm states}}\int d\text{LIPS} \vcenter{\hbox{\scalebox{.8}{\oneloopcut}}} \,.
\end{align}
In its most powerful form, we can repeatedly use this `factorization' at the integrand level, i.e. without performing the $d$LIPS integration. At the integrand level, taking the discontinuity is replaced by taking residues of the corresponding propagators that go on-shell. The resulting cuts of loop amplitudes are gauge invariant \emph{on-shell functions} and will be discussed in more detail in subsection (\ref{subsec:supersum_gluing}) as one of the two main building blocks in the unitarity method. This approach is quite universal and can in principle be used to find a representation of any perturbative amplitude or form factor in any quantum field theory. For recent applications of the unitarity method as well as integrand based reduction algorithms, see e.g.~\cite{Passarino:1978jh,Ossola:2006us,Mastrolia:2010nb,Badger:2012dv,Mastrolia:2012an,Badger:2013sta,Ita:2015tya,Mastrolia:2016dhn}. In certain cases, the unitarity approach allows to construct physical quantities that would traditionally involve infinitely many Feynman diagrams (e.g. scattering amplitudes with arbitrarily many external legs) thereby highlighting the immense efficiency of modern on-shell methods. 

Even though the majority of applications of generalized unitarity are for fully on-shell scattering amplitudes, there are various examples in the literature that have used the same technology for form factor calculations, see e.g.~\cite{Brandhuber:2010ad, Brandhuber:2011tv, Bork:2011cj, Bork:2012tt, Brandhuber:2012vm, Boels:2012ew, Yang:2016ear}. As alluded to in the introduction, despite the on-shell efficiency, there is an inherent arbitrariness in the choice of integrand bases, where ``bad'' bases can lead to increased complexity of the linear algebra problem and sub-optimal expressions for the coefficients. In our approach, we follow the refinements in the basis construction first discussed for scattering amplitudes \cite{Bourjaily:2017wjl,Bourjaily:2019iqr,Bourjaily:2019gqu} and apply them to the case of form factors. We give a more detailed overview of the good basis for the planar two-loop four-point form factor of the chiral stress tensor supermultiplet in subsection \ref{subsec:integrand_bases}.

\subsection{Clean Basis of Loop Integrands}
\label{subsec:integrand_bases}

In order to specify \emph{a} basis of integrands, we need two pieces of information, see e.g.~\cite{Bourjaily:2020qca}. First, we have to specify the spacetime dimension of interest (here, we work in strictly $d=4$) and second, we have to either know, or guess the powercounting of the theory under consideration. In general, powercounting is somewhat hard to define in situations that involve kinematic integrands with non-planar diagram topologies due to the lack of canonical loop momentum labels. However, \cite{Bourjaily:2020qca} proposed an operational, graph theoretic definition that we follow here. Note that if one assumes a too optimistic powercounting (schematically, one allows too few powers of loop momentum in the diagram numerators), one would not be able to consistently match a spanning set of field theory cuts in the generalized unitarity framework. Since the chiral stress tensor supermultiplet is protected from renormalization by supersymmetry, we do not expect any UV divergences that arise diagrammatically from bubble topologies. Therefore, it is suggestive, that it suffices to consider a basis with triangle powercounting where every integrand basis element scales at worst like a scalar triangle as loop momenta $\ell_i$ are sent to infinity, see \cite{Bourjaily:2020qca} for details. Fortunately, a basis of triangle powercounting that covers all integrand topologies we need for our two-loop form factor computation, has already been constructed in Refs.~\cite{Bourjaily:2019iqr,Bourjaily:2019gqu} in the context of two-loop scattering amplitudes in $\calN=4$ sYM beyond the planar limit. We recall some of the features of that basis and how it has to be adapted to our form factor computation. 

As alluded to above, generally, there is a substantial difference between constructing \emph{some} integrand basis, and a \emph{good} integrand basis. One measure of ``goodness'' would be the maximization of the number of vanishing coefficients (for some amplitudes of interest) in Eq.~(\ref{eq:gen_unitarity_expansion}). Even though this is a reasonable strategy, it would lead to a very different representation from the one that has been obtained for non-planar amplitudes in sYM \cite{Bourjaily:2019iqr,Bourjaily:2019gqu}, which we follow closely here.  Instead, as in \cite{Bourjaily:2019iqr,Bourjaily:2019gqu}, we choose integrands that match as many physical singularities as possible. In this basis, the composite leading singularities \cite{ArkaniHamed:2012nw}, such as the one in (\ref{eq:1loop_1m_tri_soft_composit}), directly match the soft-collinear divergences of loop amplitudes or form factors by infrared-divergent integrals. Diagonalizing such an integrand basis tends to render a maximal subset of integrands infrared finite which is a quite attractive feature. Since soft-collinear divergences are associated with parity-even residues in loop-momentum space that localize loop momenta to become either soft $\ell_i \to 0$ or to become collinear to massless external lines $\ell_j \propto p_i$, \emph{chiral} integrands often vanish on such cuts. 

Before moving on, let us clarify some terminology that is repeatedly used in unitarity computations. Residues with maximal co-dimension (which either involve $d\!\times\!L$ internal propagators, or any $d\!\times\!L$ cut conditions among fewer propagators) have no internal degrees of freedom and are also called {\it leading singularities} \cite{Cachazo:2008vp}. Residues for which the number of cut conditions exceeds the number of internal propagators are called {\it composite}. Residues which depend on $r$ internal degrees of freedom are referred to as {\it sub-leading}, or (next-to)$^r$-maximal singularities. Closely related to (sub-)leading singularities are `maximal cuts' \cite{Bern:2007ct} (see also \cite{Bosma:2017ens}). Maximal cuts are those residues which cut the maximum number of internal propagators of an amplitude or form factor. This number depends on multiplicity, but can be substantially less than \mbox{$d\!\times\!L$}.

Before we describe the details of our integrand basis, we remind the reader of some useful notation discussed in \cite{Bourjaily:2019iqr,Bourjaily:2019gqu}. We have a `bracket' built out of traces of (pairs of) four-momenta expressed in terms of $2\times2$ matrices (by dotting any $p^\mu$ into the Pauli matrices $\sigma_\mu^{\alpha\,\dot\alpha}$):
\begin{align}
\brEight{a_1}{a_2}{b_1}{b_2}{\cdot}{\cdot}{c_1}{c_2}\equivR\Big[(a_1\!\cdot\!a_2)^{\alpha}_{\phantom{\alpha}\beta}(b_1\!\cdot\!b_2)^{\beta}_{\phantom{\beta}\gamma}\cdots(c_1\!\cdot\!c_2)^{\delta}_{\phantom{\gamma}\alpha}\Big] \, ,
\label{definition_of_br}
\end{align}
where $(a_1\!\cdot\!a_2)^{\alpha}_{\phantom{\alpha}\beta}\equivR a_1^{\alpha\,\dot{\alpha}}\epsilon_{\dot{\alpha}\dot{\gamma}}a_2^{\dot{\gamma}\gamma}\epsilon_{\gamma\beta}$. This bracket has also been called a chiral `Dirac trace', `$\text{tr}_{+}[\cdots]$'$\equivR\br{\cdots}$ in e.g.~\cite{Badger:2013gxa,Badger:2015lda,Badger:2016ozq}.

\subsubsection{Chiral Numerators for Box Integrands and One-Loop Basis}

In order to discuss the aforementioned features of `chirality' and triangle powercounting more explicitly, we go back to the simpler one-loop case.  One simple example relevant to our discussion of the MHV form factor is the `two-mass-easy' box integrand. Prior to introducing any nontrivial loop-momentum dependent numerators, this integrand is just determined by its propagator structure, pictorially represented as
\begin{align}
     \vcenter{\hbox{\scalebox{.8}{\oneloopboxnumGen}}} = \frac{1}{\ell^2_a\, \ell^2_b\,\ell^2_c\,\ell^2_d\,}\,,
     \label{eq:scalar_2me_box}
\end{align}
where $\ell_a,\ell_b,\ell_c,\ell_d$ represent the momenta flowing through the propagators of the graph. Of course, momentum conservation requires (in all-outgoing conventions) that 
\begin{align}
\ell_a+\mathbf{p}_{1}=\ell_b,\quad \ell_b+p_{4}=\ell_c,\quad \ell_c+\mathbf{p}_{3}=\ell_d,\quad \ell_{d}+p_{2}=\ell_a\,.
\end{align}
Note that in this general setup, the massive momenta $\mathbf{p}_{1,3}$ ($\mathbf{p}^2_{1,3}\neq 0$), highlighted by `wedges' in (\ref{eq:scalar_2me_box}), can represent arbitrary sums of individual massless and massive momenta. In degenerate cases, they are also allowed to become a single massless leg. This variability is encoded in our `wedge' notation in diagram (\ref{eq:scalar_2me_box}). The precise leg arrangement in these `generic' corners will not matter for the discussion of the kinematic numerators below. Momentum conservation can always be solved for example by declaring that $\ell = \ell_a$, but it can be useful to \emph{not} require any particular solution to momentum conservation. 

If we want to define a one-loop integrand like that of (\ref{eq:scalar_2me_box}) with `triangle' power counting, we follow the definition of power-counting of \cite{Bourjaily:2017wjl}, which leads to a numerator of the form 
\begin{align}
\frac{\mathfrak{n}^{(1)}(\ell)}{\ell_a^2\,\ell_b^2\,\ell_c^2\,\ell_d^2}\quad\text{with}\quad \mathfrak{n}^{(1)}(\ell)\in\text{span}_{q}\!\big\{\!(\ell_a+q)^2\!\big\}\,,
\label{space_of_triangle_power_counting_box_numerators}
\end{align}
where $q$ can be any spacetime vector. One can show that this numerator space is spanned by a Lorentz-invariant degree-two polynomial in $\ell$, 
\eq{\text{span}_{q}\!\big\{\!(\ell_a+q)^2\!\big\}\simeq\text{span}\!\big\{\ell^2,\ell^\mu,1\big\}\,,}
where we have ignored overall factors of mass-dimension, and used a spanning set of coordinate vectors $e_\mu$ to express each component of $\ell^\mu\equivR(\ell\!\cdot\!e^\mu)$. From expansion, one can easily read off that the rank of such numerators is $(d+2)$ for (integer) $d$ spacetime dimensions. 

A very natural subspace of numerators for the integral (\ref{eq:scalar_2me_box}) corresponds to taking $q\in\{0,\mathbf{p}_1,\mathbf{p}_{\mathbf{1}4},\mathbf{p}_{\mathbf{13}4}\}$, where $p_{i\ldots j}:= p_i+{\cdots}+ p_j$. These elements are \emph{contact terms} where the numerator cancels a propagator, leading to e.g. the diagram with the $1/\ell^2_a$ propagator collapsed:
\begin{align}
     \vcenter{\hbox{\scalebox{.8}{\onelooptrinumTwoMass}}}\, .
     \label{eq:scalar_2m_tri}
\end{align}
Looking at the mass dimension of the integrand in $d=4$, in order to produce a basis element that is normalized to have unit modulus on co-dimension four residues (leading singularity), the numerator has to include some \emph{external} kinematic dependent factor that scales like $p_i\cdot p_j$. We are not allowed to use any loop momentum dependence, as this would go beyond triangle powercounting and is exclude by construction. It turns out that we never have to explicitly consider the two-mass triangle integrand (\ref{eq:scalar_2m_tri}). The reason is as follows: The triangle integrand in (\ref{eq:scalar_2m_tri}) has one massless external leg. For this configuration, we can take a (composite) residue where the loop momentum becomes collinear with the massless external line $\ell_b =\alpha\, p_4$ by cutting the two propagators $\ell^2_b=\ell^2_c=0$ and further cutting the resulting Jacobian. If one takes an additional residue in $\alpha$ such that $\ell^2_d$ is on shell, one can show that scattering amplitudes or form factors in field theory never have support on such singularities \cite{ArkaniHamed:book}. The basis of integrands we are going to construct here allows us to directly set the coefficient of any one-loop two-mass triangle integrand to zero, and the only case we ever need are one-mass triangles
\begin{align}
     \vcenter{\hbox{\scalebox{.8}{\onelooptrinumOneMass}}}  
     \quad \bigger{\Leftrightarrow}\quad 
     \frac{\mathfrak{n}^{(1)}_3 = s_{ij}}{\ell^2_b \ell^2_c\ell^2_d}\, ,
     \label{eq:scalar_1m_tri}
\end{align}
where $s_{ij}=(p_i+p_j)^2$. This integrand basis element is also tied to infrared singularities and has support on a residue that localizes the loop momentum $\ell_c$ in the soft region, where $\ell_c=0$. The role of this integrand will be to match the infrared structure of the one-loop form factor. 

After taking into account contact terms within the space of numerators, there are $(d-2)$ degrees of freedom remaining. Let us now describe a very natural choice for the two remaining numerators in four spacetime dimensions. To do so, notice that all contact terms vanish on solutions to the `quadruple cut' $\ell_a^2=\ell_b^2=\ell_c^2=\ell_d^2=0$. In four dimensions, there are precisely two solutions to the quadruple cut equations,  
\begin{align}
\vcenter{\hbox{\scalebox{.8}{\oneloopboxMHVGen}}}
\bigger{\Leftrightarrow}\;{\color{hred}\ell_{a,1}^*} = \frac{\lambda_2\, \bfq_{1}|4\rangle}{\ab{24}}\,, 
\quad
\vcenter{\hbox{\scalebox{.8}{\oneloopboxMHVbarGen}}}
\bigger{\Leftrightarrow}\;{\color{hblue}\ell_{a,2}^*} = \frac{[4|\bfq_{1}\,\widetilde{\lambda}_2}{\sqb{24}}\, .
\label{quad_cuts_of_2me}
\end{align}
The cuts above have been decorated in order to emphasize that ${\color{hred}\ell_{a,1}^*}$ is the solution for which the $\lambda$'s of the massless on-shell legs $\{\ell_{a},\ell_{d},p_2\}$ and $\{\ell_b,\ell_c,p_4\}$ are mutually parallel  (and therefore support non-vanishing $\overline{\text{MHV}}$ amplitudes at those vertices). Similarly, ${\color{hblue}\ell_{a,2}^*}$ is the solution for which the $\widetilde{\lambda}$'s of $\{\ell_{a},\ell_{d},p_2\}$ and $\{\ell_b,\ell_c,p_4\}$  are mutually parallel (and therefore support non-vanishing MHV amplitudes at those vertices). For the unitarity cut construction of the form factor integrands, it will be useful that MHV form factors have support exclusively on the cut ${\color{hred}\ell_{a,1}^*}$ and vanish on the cut ${\color{hblue}\ell_{a,2}^*}$.  We have the freedom to choose the remaining two numerator degrees of freedom of the box integrand to be \emph{chiral}. By `chiral', we mean that the integrands have support on one solution or the other, but not both. This is easy to engineer by e.g. writing 
\begin{align}
\label{eq:box_chiral_nums_v1}
\mathfrak{n}^{(1)}_1 = s_{24}(\ell_a-{\color{hblue}\ell_{a,2}^*})^2\,,
\quad\text{and}\quad
\mathfrak{n}^{(1)}_2 = s_{24}(\ell_a-{\color{hred}\ell_{a,1}^*})^2\,,
\end{align}
where we have included the normalization $s_{24}$ so that the residues on the maximal cut (\ref{quad_cuts_of_2me}) have unit magnitude. In terms of the `brackets' defined in (\ref{definition_of_br}), the numerators in (\ref{eq:box_chiral_nums_v1}) can be recast to
\begin{align}
\label{eq:chrial_box_nums}
\mathfrak{n}_1=\br{2,\ell_a,\ell_c,4}\,,
\quad\text{and}\quad
\mathfrak{n}_2=\br{\ell_a,\ell_c,4,2}\, .
\end{align}
In earlier literature (see e.g.\ \cite{ArkaniHamed:2010kv, Drummond:2010mb}), these numerators were referred to as (`magic') `wavy-line' and `dashed-line' numerators, respectively. To keep new notation to a minimum, here, we choose \emph{not} to include special graphic notation to highlight which corners of an integrand are `chiralized' by the respective numerators. Instead, we take a numerator to be present for all integrands implicitly. For the relevant two-loop topologies required to match the leading color form factor, the kinematic numerators are conveniently summarized in appendix \ref{sec:Appendix} and also included in computer readable format as ancillary file. The bracket notation is easily seen to encode the chiral constraints, e.g.~, 
\begin{itemize}
\item $\brFour{2}{\ell_a}{\cdot}{\cdot}$ enforces that the numerator vanishes when $\widetilde{\lambda}_{\ell_a}\propto\widetilde{\lambda}_{2}$,
\item $\brFour{\cdot}{\cdot}{\ell_c}{4}$ enforces that the numerator vanishes when $\widetilde{\lambda}_{\ell_c}\propto \widetilde{\lambda}_{4}$,
%
\end{itemize}
and so on. These examples clearly demonstrate the versatility of the bracket notation. Crucially, due to their `chiral' nature, the numerators are such that they vanish in the collinear regions where e.g. $\ell_a \propto p_2$ or $\ell_b \propto p_4$ and are therefore infrared finite (as well as UV finite from simple power counting) upon loop integration.

\subsubsection{Chiral Numerators with Triangle Power counting at Two-Loops}

Building upon the philosophy introduced for the chiral one-loop triangle powercounting integrands, let us outline some of the features of the two-loop integrands relevant to match the MHV four-point form factor of the chiral stress tensor supermultiplet in planar $\calN=4$ sYM. As mentioned in the introduction, since we are calculating the form factor of a color singlet operator, even for the leading color contribution in the large $N$ limit, we are required to consider non-planar integrand topologies. At the eight propagator level, there are four basic topologies
\begin{align}
\label{eq:2loop_mhv_8prop_topos}
\vcenter{\hbox{\scalebox{.7}{\diagOne}}}
\vcenter{\hbox{\scalebox{.7}{\diagTwo}}}
\vcenter{\hbox{\scalebox{.7}{\diagThree}}}
\vcenter{\hbox{\scalebox{.7}{\diagFour}}} \, .
\end{align}
Here we label the momentum inserted by the off-shell operator with a double-line notation for which the associated momentum $\mathbf{p}^2_1\neq0$, whereas the thin lines represent on-shell momenta for which $p^2_i=0$. Furthermore, we indicate the labels of the internal edges without solving momentum conservation. In principle, one could have envisioned one further planar penta-box with a different attachment of the off-shell leg. However, this integrand does not contribute to the MHV form factor. 

It is important to realize that the diagram topologies in (\ref{eq:2loop_mhv_8prop_topos}) already appeared in the context of \emph{scattering amplitudes} beyond the planar limit (leading color) of $\calN=4$ sYM~\cite{Bourjaily:2019iqr,Bourjaily:2019gqu}. There, one of the authors was involved in the construction of a chiral basis of diagram numerators with triangle powercounting, i.e. numerators involving loop momenta such that each integrand scales (in each sub loop) at worst like a scalar triangle. Even though the notion of powercounting is a little less clear for nonplanar integrand topologies, \cite{Bourjaily:2020qca} gave a graph theoretic prescription on how to define general $n$-gon powercounting at higher loops. Furthermore, the basis constructed in \cite{Bourjaily:2019iqr,Bourjaily:2019gqu} was \emph{prescriptive}, meaning that each integrand basis element matched a single unitarity cut chosen from a special spanning set of cuts.  

Note the form factor of the chiral stress-tensor supermultiplet is protected by supersymmetry from UV renormalization, it is natural to expect that the triangle powercounting basis of \cite{Bourjaily:2019iqr,Bourjaily:2019gqu} is a good starting point for a unitarity based cut matching computation. As we will see shortly in section \ref{sec:results}, this is indeed the case. With this in mind, let us briefly describe how to convert the amplitudes basis elements to the ones for the form factor. The complete list of diagram topologies needed to match the MHV four-point form factor is given in Fig.~\ref{fig:diags_all} and the associated diagram numerators are listed in appendix \ref{sec:Appendix}. Here, we only discuss a few representative examples, the rest can be obtained analogously. 

As an example, consider the first topology in (\ref{eq:2loop_mhv_8prop_topos}). The numerator can be obtained from the representations of \cite{Bourjaily:2019iqr,Bourjaily:2019gqu} with the appropriate identification of external momenta and relabeling of internal lines and is denoted by `double-pentagon A' in Table III of \cite{Bourjaily:2019gqu}
\begin{align}
\label{eq:doublePentA_amp_eg}
\hspace{-1cm}
\vcenter{\includegraphics[trim={0cm 6cm 0cm 0cm},clip,scale=.35]{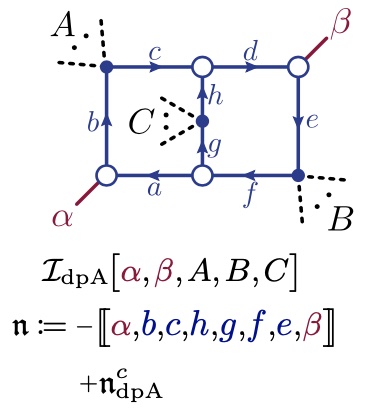}}
&
\hspace{-11cm}\bigger{\Leftrightarrow}\;
\calI_{\text{dpA}} [{\color{hred}\alpha},{\color{hred}\beta},A,B,C] = \frac{{-}\,\br{{\color{hred}\alpha},\!\edgeB,\!\edgeC,\!\edgeH,\!\edgeG,\!\edgeF,\!\edgeE,\!{\color{hred}\beta}} + \mathfrak{n}_{\text{dpA}}^c}{
\edgeA^2\, \edgeB^2\,\edgeC^2\,\edgeD^2\, \edgeE^2 \edgeF^2 \edgeG^2 \edgeH^2}\,,
\hspace{+2cm}\\[-20pt]
&\hspace{-11cm}
\mathfrak{n}_{\text{dpA}}^c = \frac{1}{2} 
\begin{cases}
    0^{\phantom{2}}\phantom{\edgeF^2} & ,\text{if }A\!\neq\!\emptyset, B\!\neq\!\emptyset
    \\[2pt]
    \edgeB^2(\edgeG^2{+}\edgeH^2)\hspace{0.3pt}\br{\hspace{-0.3pt}{\color{hred}\beta},\!\edgeE,\!\hspace{-0.5pt}\edgeF\!,\!{\color{hred}\alpha}}&,\text{if }A\!=\!\emptyset, B\!\neq\!\emptyset
    \\[2pt]
    \edgeE^2(\edgeG^2{+}\edgeH^2)\br{{\color{hred}\alpha},\!\edgeB,\!\edgeC,\!{\color{hred}\beta}}&,\text{if }B\!=\!\emptyset\\
\end{cases}\,. \nonumber
\hspace{-2cm}
\end{align}
In order to get to the desired leg arrangement for the four-point form factor, we take the generic leg ranges in $A$ and $B$ to be single massless edges. There, the numerator instruction tells us that we do not have to add a contact term to the bracket. However, since the first topology in (\ref{eq:2loop_mhv_8prop_topos}) has an additional symmetry, we add the symmetric image of the numerator in case $A$ and $B$ are massless legs in (\ref{eq:doublePentA_amp_eg}). Explicitly, we have
\begin{align}
\label{eq:diag_1_num_integrand_section}
\hspace{-.5cm}
\vcenter{\hbox{\scalebox{.7}{\diagOne}}}
\quad
\mathfrak{n}^{(2)}_{1}= 
    -[\![p_3,\ell_6,\ell_{10},\ell _{13},\ell_{12},\ell_{9},\ell_7,p_5]\!]
    -[\![p_4,\ell_7,\ell_8,\ell_{13},\ell _{12},\ell _{11},\ell _6,p_2]\!] \,.
\hspace{-.4cm}
\end{align}
Conveniently, the numerators in \cite{Bourjaily:2019gqu} were chosen such that contact topologies arise from the eight-propagator integrand basis elements by simple `degenerations' where some of the multi-leg ranges are allowed to become empty (i.e.~\emph{no} momentum flowing into that vertex). Since this property has been described in detail in Ref.~\cite{Bourjaily:2019gqu}, we refer the interested reader directly to that article for explicit examples. Here, we used this degeneration property to also construct integrand topologies with fewer than eight internal edges. In some of those cases, one has to sum the numerator degenerations that can come from multiple eight-propagator families. The explicit forms of the kinematic numerators $\mathfrak{n}^{(2)}_i$ for all 19 diagram topologies needed to match the MHV form factor are conveniently summarized in appendix \ref{sec:Appendix}.

As stated several times, one of the key features of the numerators constructed in \cite{Bourjaily:2019gqu} is that they ensure that the integrands have unit leading singularities, i.e. all maximal co-dimension residues are $\pm 1$ or 0. Furthermore, as explained in a bit more detail in the on-loop chiral box example, the bracket structure directly highlights which vertices of the diagram are protected in regions where the adjacent momenta become collinear and which cut solution is supported. 
 
Having discussed the basis of kinematic integrands, we now turn our attention to the second important ingredient in the generalized unitarity approach: The on-shell functions.  


\subsection{Field Theory Cuts and Supersymmetric State Sums}
\label{subsec:supersum_gluing}

Given a basis of loop integrands, such as the one discussed in the previous subsection, the criterion used to fix the coefficients $c_k$ in the generalized unitarity based expansion (\ref{eq:gen_unitarity_expansion}) is that the residues of the right hand side match those of field theory. Let us review the residues of field theory amplitude or form factor integrands (for brevity let us denote them jointly as field theory integrands) for the sake of clarity. 

The residues of field theory integrands are those rational functions obtained from an off-shell ({\it e.g.}\ Feynman) loop integrand by setting a subset of internal particles on-shell.  These residues correspond to graphs with lower-loop (often tree-level) amplitudes\footnote{In the case of form factors of a single off-shell operator, one of the amplitudes is replaced by a lower-loop (often tree level) form factor. In this general discussion we avoid this specialization, but come back to it when we discuss the explicit construction of the one- and two-loop form factor integrands.} at each vertex separated by on-shell, internal states. Functions corresponding to such graphs are called {\it on-shell functions}, and they have played a key role in many recent developments on scattering amplitudes. 

On-shell functions can be defined (and computed) in many ways and even though they have most prominently featured in the realm of supersymmetric theories \cite{ArkaniHamed:2012nw,Arkani-Hamed:2014bca,Franco:2015rma,Frassek:2016wlg,Heslop:2016plj,Herrmann:2016qea} (see {\it e.g.} \cite{Benincasa:2015zna,Benincasa:2016awv} for some exceptions), they can generally be defined from first principles without reference to (off-shell) loop integrands in any quantum field theory. When represented as a graph $\Gamma$ of amplitudes at vertices indexed by $v$, connected by edges indexed by $i$ (representing on-shell, but internal physical states), the corresponding on-shell function $f_\Gamma$ can be defined as:
\begin{align}
f_\Gamma\equiv\prod_{i}\Big(\!\!\sum_{\text{states}}\int\!\!d^{d-1}\text{LIPS}_i\!\Big)\!\!\prod_{v}\mathcal{A}_v\, .
\label{general_on_shell_function_definition}
\end{align}

This definition follows immediately from locality and unitarity. In (\ref{general_on_shell_function_definition}), `$d\text{LIPS}$' denotes the measure over the `Lorentz invariant phase space' of each on-shell, \emph{internal} particle, and the summation over `states' means all non-kinematical quantum labels distinguishing particles in the theory---helicity, colour, etc. The definition in (\ref{general_on_shell_function_definition}) has been written to make it explicit that these objects can be defined in an arbitrary numbers of dimensions.

For many of the on-shell functions important to this work, the phase space integrations in (\ref{general_on_shell_function_definition}) are not entirely localized by the momentum conservation at the vertices; when this happens, $f_\Gamma$ becomes an (unspecified) integral over the on-shell degrees of freedom.

As discussed above, on-shell functions may be equivalently defined as the iterated residues of off-shell loop amplitudes obtained by putting each edge in the diagram on-shell. On-shell functions defined in this way (as residues of loop amplitudes) appeared first in the context of generalized unitarity. While we prefer the first-principles definition (\ref{general_on_shell_function_definition}), this historical view is also useful. For example, considering on-shell functions as iterative residues makes it easy to count how many `internal' degrees of freedom exist for a given diagram: an $L$-loop diagram with $n_I$ internal edges corresponds to a co-dimension $n_I$ residue of a $(d\!\times\!L)$-dimensional form---resulting in a function of $(d\!\times\!L- n_I)$ remaining degrees of freedom. For a more concrete understanding of the on-shell functions for planar sYM, we refer the reader to more thorough discussions in the literature (see {\it e.g.}\, \cite{ArkaniHamed:2012nw,ArkaniHamed:book} or the appendices of \cite{Bourjaily:2015jna}), and to the computer packages described in \mbox{refs.\ \cite{Bourjaily:2010wh,Bourjaily:2012gy,Dixon:2010ik}}. In this work we will mostly be concerned with $d\!=\!4$ quantum field theories and will also provide explicit examples of the on-shell functions relevant to our case in subsections (\ref{subsubsec:ssg_one_loop_eg}) and (\ref{subsubsec:ssg_two_loop_eg}).

In this paper, we look at $\calN = 4 $ sYM and hence we consider the scattering of supermultiplets, instead of individual particles of specific helicity. The super-amplitude (or form factor) computed in this way efficiently packages all the information about scattering of component fields. The on-shell superfield is given as\cite{Nair:1988bq}:
\begin{align}
  \Phi(p,\eta) = g^+(p) + \eta_A \lambda^A(p) + \frac{\eta_A \eta_B}{2!} \phi^{AB}(p) 
                + \frac{\varepsilon^{ABCD}\eta_A\eta_B\eta_C}{3!} \bar{\lambda}_D(p)+ \eta_1\eta_2\eta_3\eta_4 g^-(p) \, ,
\end{align}
where $(g^+, \lambda^{A}, \phi^{AB}, \bar{\lambda}_D, g^{-})$ are the component fields. Similarly to the on-shell superfields, we also have super form-factors which package all states of gauge invariant superoperators. Here, we look at the super form factor of the chiral part of the stress-tensor supermultiplet, which preserves half of the supersymmetries off-shell \cite{Eden:2011yp,Eden:2011ku}. The supersymmetric stress-tensor is naturally written in harmonic  superspace (see e.g.~\cite{Galperin:2001seg}) in terms of the additional bosonic coordinates $(u^{+a}_{A}, \tred{u^{-\dot{a}}_{A}})$, where $A=1,\ldots4$ is an $SU(4)_R$ $R$-symmetry index, and $a,\dot{a}=1,2$ are the indices of two different $SU(2)$ groups. The additional $\pm$ characterizes a U(1) charge (see Appendix A of \cite{Bork:2012tt}, and Section 3 of \cite{Eden:2011yp} for more details). These variables allow us to  define the harmonic \emph{projections} of e.g.~the fermionic superspace coordinates 
\begin{align}
\begin{split}
    \theta^A_\alpha & 
    \quad \longrightarrow\quad
      \theta^{+\, a}_\alpha  = u^{+\, a}_A \, \theta^A_\alpha\,,
      \qquad
      \tred{\theta^{-\, \dot{a}}_\alpha
      = u^{-\, 
      \dot{a}}_A\, \theta^A_\alpha }\,,
    \\
    \bar{\theta}^{\dot{\alpha}}_A & 
    \quad\longrightarrow\quad
      \tred{\bar{\theta}^{\dot{\alpha}}_{+\,a}  = \bar{u}^A_{+\, a} \,\bar{\theta}^{\dot{\alpha}}_A }\,,
      \qquad
      \bar{\theta}^{\dot{\alpha}}_{-\,\dot{a}} 
      = \bar{u}^A_{-\, \dot{a}} \bar{\theta}^{\dot{\alpha}}_A\,.
\end{split}      
\end{align}
Similar projections can also be defined for the Grassmann-odd on-shell momentum superspace coordinates $\eta$. In harmonic superspace, the \emph{non-chiral} stress tensor $\calT=\calT(x,\theta^+,\bar{\theta}_{-};u)$ does not depend on the conjugate variables $\tred{\theta^-, \bar{\theta}_+}$ as a consequence of the half-BPS condition.

The \emph{chiral} stress-tensor operator is obtained from the non-chiral one by setting half of the remaining fermionic coordinates to zero: $\calT (x, \theta^{+})=\calT (x, \theta^{+}, \bar{\theta}_{-}=0;u)$. Explicitly, one finds:
\begin{align}
     \calT (x, \theta^{+}) & =  
        \text{Tr}(\phi^{++}\phi^{++}) 
        \nonumber\\&
        +i\, 2\sqrt{2}\theta_{\alpha}^{+a} \,\text{Tr}(\lambda_{a}^{+\alpha} \phi^{++}) 
        \nonumber\\ &
        +\theta_{\alpha}^{+a} \epsilon_{ab}\theta_{\beta}^{+b} \, \text{Tr}(\lambda^{+c ( \alpha} \lambda_{c}^{+\beta)}-i\sqrt{2}F^{\alpha\beta}\phi^{++})
        -\theta_{\alpha}^{+a}\epsilon^{\alpha\beta}\theta_{\beta}^{b}\,\text{Tr}\,(\lambda_{(a}^{+\gamma} \lambda_{b)\gamma}^{+} -g \sqrt{2}[\phi^{+C}_{(a}, \bar{\phi}_{C +b)}]\phi^{++}) 
        \nonumber\\ & 
       -\frac{4}{3} (\theta^{+})^{3\, a}_{\alpha} \, \text{Tr} (F^{\alpha}_{\beta} \lambda_{a}^{+\beta} + ig[\phi_{a}^{+B},\bar{\phi}_{BC}]\lambda^{C \alpha})
       \nonumber\\& 
       +\frac{1}{3} (\theta^{+})^4 \cal{L} \, .
\end{align}
The $(\theta^{+})^0$ component is the single trace operator $\text{Tr}(\phi^{++}\phi^{++})$, while the $(\theta^+)^4$ component is the chiral on-shell Lagrangian (see e.g. Eq.~(2.38) of \cite{Eden:2011yp}).
The super form factor of $\calT(x,\theta^+)$ is subsequently defined as:
\begin{align}
   \calF_{\calT}(\bfq,\gamma_+;1 \ldots n) := \int d^4x \, d^4 \theta^+ 
   e^{-i(\bfq {\cdot} x + \theta^{+a}_{\alpha} \gamma_{+a}^{\alpha})} \bra{1 \ldots n} \calT (x, \theta^+) \ket{0}  \, ,
\end{align}
where $\gamma_{+a}^{\alpha}$ is the Fourier conjugate variable to $\theta^{+a}_{\alpha}$. There is no $\gamma_{-\dot{a}}^{\alpha}$ since $\theta^{-\dot{a}} _{\alpha}$ has been set to zero. Solving the Ward identities obeyed by the form-factor, it can be written as:
\begin{align}
    \calF_{\calT}(\bfq,\gamma_+;1 \ldots n) = \delta^{(4)}\left(\bfq+\sum^n_{i=1} \lambda_{i}\widetilde{\lambda}_{i}\right)\delta^{(4)}\left(\gamma_+ + \sum^n_{i=1}  \eta_{+,i} \, \lambda_i \right)\delta^{(4)}\left(\sum^n_{i=1} \eta_{-,i} \, \lambda_{i} \right) \large{R}\,,
\end{align}
where we have suppressed all little-group indices and most indices of (chiral) harmonic superspace to avoid clutter. For the MHV form factor R is a function of bosonic external kinematic variables only while for N$^k$MHV form factor, R would be a polynomial of degree $4k$ in the Grassmann variables $\eta$.

As explained above, field theory residues of amplitude or form factor integrands ultimately factorize into a product of tree level objects (see Eq.~(\ref{general_on_shell_function_definition})). In the supersymmetric setting relevant for our work, they can be ``glued'' together by a superspace integration over the Grassmann variables associated to the on-shell internal edges, thereby performing the sum over on-shell states. For the MHV form factor integrand construction, we encounter the following basic tree-level building blocks (We suppress the chiral stress tensor $\calT$ index of the form factor.):
\begin{align}
\label{eq:os_building_blocks}
    \vcenter{\hbox{\scalebox{.8}{\diagff}}} 
    &  \quad \bigger{\Leftrightarrow}\quad
\calF^{(0),\text{MHV}}_{n} (\textbf{q},\gamma_+; 1,2,...,n) = \frac{
\delta^{(4)}(P)\, \delta^{(8)}(\calQ)}
{\ab{12}\ab{23}\cdots \ab{n1}}\,,
\\
\vcenter{\hbox{\scalebox{.7}{\mhvAmp}}}
&  \quad \bigger{\Leftrightarrow}\quad
\calA_{n}^{\text{MHV}}(1,\ldots,n) =  \frac{
\delta^{(4)}(\sum_{i} \lambda_{i}\widetilde{\lambda}_{i})\,
\delta^{(8)}(\sum_{i}\lambda_{i}\eta_{i})}
{\vev{12}\vev{23}\cdots \vev{n1}}\,,
\\
\vcenter{\hbox{\scalebox{.7}{\diagmhv}}}
&  \quad \bigger{\Leftrightarrow}\quad 
\calA_{3}^{\text{MHV}}(1,2,3) =  \frac{
\delta^{(4)}(\sum_{i} \lambda_{i}\widetilde{\lambda}_{i})\,
\delta^{(8)}(\sum_{i}\lambda_{i}\eta_{i})}
{\vev{12}\vev{23} \vev{31}}\,,
\\
\vcenter{\hbox{\scalebox{.7}{\diagantimhv}}}
& \quad \bigger{\Leftrightarrow}\quad 
\calA_{3}^{\overline{\text{MHV}}}(1,2,3) = \frac{
\delta^{(4)}(\sum_{i} \lambda_{i}\widetilde{\lambda}_{i})\,
\delta^{(4)}([12]\eta_3 + [23] \eta_1 + [31] \eta_2)}
{\sqb{12}\sqb{23}\sqb{31}}\,.
\end{align}
We work in an all-outgoing momentum convention and define the following shorthand notation
\begin{align}
\delta^{(4)}(P) := \delta^{(4)}\left(\bfq+\sum^n_{i=1} \lambda_{i}\widetilde{\lambda}_{i}\right)\,,
\quad
\delta^{(8)}(\calQ):=\delta^{(4)}\left(\gamma_+ + \sum^n_{i=1} \eta_{+,i} \, \lambda_i \right)\,
\delta^{(4)}\left(\sum^n_{i=1} \eta_{-,i}\, \lambda_{i} \right)\,,
\end{align}
which encode momentum and form factor super-momentum conservation, respectively. Since we are interested in the leading color contribution to the form factor integrands, all tree-level building blocks required are color stripped and the on-shell superfields are cyclically ordered. Again, since the chiral stress tensor is a color-singlet, its momentum position is immaterial.

\subsubsection{One Loop Examples of Form Factor On-Shell Functions}
\label{subsubsec:ssg_one_loop_eg}

We are now in the position to discuss the field theory residues of the form factor integrand of the chiral stress tensor supermultiplet to higher loop order in terms of the tree-level on-shell building blocks listed in Eq.~(\ref{eq:os_building_blocks}).

As a first example, we look at the on-shell functions required to match the $n$-point one-loop MHV integrand (see section \ref{sec:results})  $\calF^{(1),\text{MHV}}_n(\mathbf{1},\gamma_+; 2,\ldots n+1)$. (The appearance of $n+1$ is due to our choice of labeling the off-shell leg associated to the operator insertion as $\mathbf{p}_1$ instead of $\bfq$). For the MHV one-loop $n$-point form factor integrand, only the `two-mass-easy' configuration (and its possible one-mass degeneration) is relevant. 
\begin{align}
    \text{cut}^{(1)}_1 
    = \underset{\ell^2_a=\ell^2_b=\ell^2_c=\ell^2_d=0}{\Res}  \left[
    \calF^{(1),\text{MHV}}_n(\mathbf{1},\gamma_+; 2,\ldots n{+}1)
    \right] 
    =
    \vcenter{\hbox{\scalebox{.8}{\oneloopbox}}} \,,
\end{align}
where we pick the solution to the on-shell conditions where $\lam{\ell_a} \sim \lam{i}$ and $\lam{\ell_b} \sim \lam{j}$ (see  Eq.~(\ref{quad_cuts_of_2me})):
\begin{align}
  \ell_a &= \frac{\lam{i} \, \mathbf{q}_A\ket{j}}{\ab{ij}}, \quad 
  \ell_b = -\frac{\lam{j} \, \mathbf{q}_B\ket{i}}{\ab{ij}}, \quad
  \ell_c = -\frac{\lam{j} \, \mathbf{q}_A\ket{i}}{\ab{ij}}, \quad 
  \ell_d = \frac{\lam{i} \, \mathbf{q}_B\ket{j}}{\ab{ij}}\,.
\end{align}
We define $\mathbf{q}_A=p_{j+1}+\ldots + \mathbf{p}_1 + \ldots + p_{i-1}$ as the total external momentum of the form factor vertex of the off-shell momentum $\mathbf{p}_1$ and $0\leq n_1 \leq n{-}3$ on-shell momenta $\{p_{j+1},\ldots,p_{i-1}\}$. Similarly $\mathbf{q}_B = p_{i{+}1}+\ldots+p_{j{-}1}$ is the total external momentum of the $n{-}2 \geq n_2 \geq 1$ on-shell momenta attached to the MHV vertex. The massless leg partition is always such that $n_1+n_2+2 = n$. As one can see, the loop momentum is completely localized due to momentum conservation at each vertex and the four on-shell constraints $\ell^2_a=\ell^2_b=\ell^2_c=\ell^2_d=0$. Taking the residue, we also get a Jacobian, which is denoted by $\calJ$ and for the cut above is given as: 
\begin{align}
\label{eq:1loop_2me_MHV_jac}
    \calJ  & = - [i|\mathbf{q}_B\ket{j} \, [j|\mathbf{q}_A \ket{i}\,.
\end{align}
The residue of the MHV form factor on the opposite chirality solution, where $\lamt{a} \sim \lamt{i}$ and $\lamt{b}\sim\lamt{j}$ is zero. In the above figure, we have $n$ massless legs, representing the on-shell superfields $\Phi(p_i,\eta_i)$ which are cyclically ordered and a single massive leg corresponding to the stress-tensor (super-)momentum which we always take to be $\mathbf{p}_1$. Note that the massless leg `$2$' is allowed to be attached in any position of the on-shell function. The explicit labels of all other legs are then determined from the cyclic ordering. We can sum over states as an integral over Grassmann variables, yielding the following `supersum gluing' \cite{Brandhuber:2011tv, Bork:2011cj, Bork:2012tt}:
\begin{align}
\label{eq:1loop_2me_mhv_box_cut}
\begin{split}
     \text{cut}^{(1)}_1 &= 
    \frac{1}{\calJ} \int\! d \eta_{\ell_{a}}\,d \eta_{\ell_{b}}\,d \eta_{\ell_{c}}\,d \eta_{\ell_{d}}\,
     \calF^{(0),\text{MHV}}_{n_1+2}({\bf 1},\gamma_+;\ldots,i-1,-\ell_{a},\ell_{b},j+1,\ldots)
     \\
     & \hspace{1.5cm}\times
     \calA^{\overline{\text{MHV}}}_3(i,-\ell_{d},\ell_{a})
     \calA^{\text{MHV}}_{n_2+2}(i+1,\ldots,j-1,-\ell_{c},\ell_{d})
     \calA^{\overline{\text{MHV}}}_3(j,-\ell_{b},-\ell_{c}) 
     \\
     & = \delta^{(4)}(P)\, \delta^{(8)}(\calQ) \times \widetilde{\text{cut}}^{(1)}_1  \,.
\end{split}     
\end{align}  
Carrying out the above integration over Grassmann variables results in:
\begin{align}
    \widetilde{\text{cut}}^{(1)}_1 = \frac{1}{\calJ}\frac{1}{\vev{\ell_{a} \, \ell_{b}}\vev{\ell_{b} \, j+1}\ldots \vev{i-1 \, \ell_{a}}} 
    \frac{\sqb{i \, \ell_{d}}^3}{\sqb{\ell_{d} \, \ell_{a}}\sqb{\ell_{a} \, i}}
    \frac{\vev{\ell_{c} \, \ell_{d}}^3}{\vev{\ell_{d} \, i+1}\ldots \vev{j-1 \, \ell_{c}}} 
    \frac{\sqb{\ell_{c}\,  j}^3}{\sqb{j \, \ell_{b}}\sqb{\ell_{b} \,  \ell_{c}}}\,. 
    \label{eq: ssgoneloopbox}
\end{align}
Substituting the cut solution from (\ref{quad_cuts_of_2me}) and the Jacobian (\ref{eq:1loop_2me_MHV_jac}) in the above equation, we get: 
\begin{align}
\label{eq:1_loop_2me_MHV_ssg_tilde}
      \widetilde{\text{cut}}_1 = \frac{1}{\ab{2\,3}\ldots\ab{i{-}1 \, i}\ab{i \, i{+}1}\ldots\ab{j{-}1 \, j}\ab{j\, j{+}1}\ldots\ab{n{+}1\, 2}}\,.
\end{align}
Next, we look at one of the sub-leading singularities associated to the one-mass triangle 
\begin{align}
\label{eq:1loop_1m_triangle_cut}
       \text{cut}^{(1)}_2 
    = \underset{\ell^2_b=\ell^2_c=\ell^2_d=0}{\Res}  \left[
    \calF^{(1),\text{MHV}}_n(\mathbf{1},\gamma_+; 2,\ldots n{+}1)
    \right] 
    = \vcenter{\hbox{\scalebox{.8}{\onelooptri}}}\,,
\end{align}
where we have chosen the solution to the triple-cut $\ell^2_b=\ell^2_c=\ell^2_c=0$ to be:
\begin{align}
\label{eq:trip_cut_sol}
  \ell_b^{*} = \lam{i+1} (\alpha \, \lamt{i} + \lamt{i+1}), \quad 
  \ell_c^{*} = \alpha \lam{i+1} \, \lamt{i}, \quad
  \ell_d^{*} = (\alpha \lam{i+1} - \lam{i}) \, \lamt{i}  \,.
\end{align}
The above solution leaves one unfixed loop degree of freedom which we parameterize by $\alpha$ and the Jacobian is $\calJ= s_{i\, i{+}1}$. (The second solution to the on-shell conditions $\ell^\ast_{c,2} \sim \lam{i} \lamt{i+1}$  is also relevant for the MHV form factor and can be trivially obtained by appropriate relabeling from the one discussed here.) The corresponding on-shell function is gives as: 
\begin{align}
\begin{split}
     \text{cut}^{(1)}_2  &=  \frac{1}{\calJ}
     \int\! d \eta_{\ell_{b}}\,d \eta_{\ell_{c}}\,d \eta_{\ell_{d}}\,
     \calF^{(0),\text{MHV}}_{n}({\bf 1},\gamma_+;\ldots,i-1,-\ell_{d},\ell_{b},i+2,\ldots)
     \\
     & \hspace{1.5cm}\times
     \calA^{\overline{\text{MHV}}}_3(i+1,-\ell_{b},\ell_{c})
     \calA^{\text{MHV}}_3(i,-\ell_{c},\ell_{d}) 
     \\
     & = \delta^{(4)}(P)\, \delta^{(8)}(\calQ) \times \widetilde{\text{cut}}^{(1)}_2  \,.
\end{split}     
\end{align}  
Evaluating the fermionic integration using the basic building blocks from Eq.~(\ref{eq:os_building_blocks}), we get:
\begin{align}
    \widetilde{\text{cut}}^{(1)}_2  = \frac{1}{\calJ}\frac{1}{\ab{\ell_{b} \, \ell_{d}}\vev{\ell_{b} \, i+2} \ldots \vev{i-1 \, \ell_{d}}} 
    \frac{\sqb{i+1 \, \ell_{c}}^3}{\sqb{\ell_{b} \, \ell_{c}}\sqb{\ell_{b} \, i+1}}
    \frac{\vev{\ell_{c} \, \ell_{d}}^3}{\vev{\ell_{d} \, i}\ldots \vev{i \, \ell_{c}}} \, . 
    \label{eq: ssgonelooptri}
\end{align}
Substituting the cut solution (\ref{eq:trip_cut_sol}) into Eq.~(\ref{eq: ssgonelooptri}), we get:
\begin{align}
  \widetilde{\text{cut}}^{(1)}_2  =  
  \frac{1}{\ab{2 \, 3} \cdots \ab{i{-}1 \, i} \ab{i \, i{+}1} \ab{i{+}1 \, i{+}2} \cdots \ab{n{+}1 \, 2}}
  \frac{1}{\big(1-\alpha \frac{\ab{i{-}1 \, i{+}1}}{\ab{i{-}1 \, i}} \big)\,  \alpha}\, ,
\end{align}
which is nothing but the (super-)BCFW shift of the tree-level $n$-point MHV form factor. Being a one-parametric function of $\alpha$, we still have the freedom to take further `composite' residues. In this case, we can approach the soft limit where the loop momentum $\ell_{c}$ becomes zero, by taking a residue at $\alpha = 0$. We graphically denote this composite residue as
\begin{align}
\label{eq:1loop_1m_tri_soft_composit}
\underset{\alpha=0}{\Res} \left[
    \vcenter{\hbox{\scalebox{.8}{\onelooptri}}}
    \hspace{-.85cm}
    \hbox{\scalebox{.8}{$ =\alpha \,\lam{i+1}\lamt{i}$} }
\right]
 =   \vcenter{\hbox{\scalebox{.8}{\onelooptrisoft}}}
 = \calF^{(0),\text{MHV}}_{n} (\textbf{1},\gamma_+; 2,\ldots,n{+}1)\,,
\end{align}
or analytically 
\begin{align}
  \underset{\alpha=0}{\Res} \left[ \widetilde{\text{cut}}^{(1)}_2 \right]= 
     \frac{1}{\ab{2 \, 3}\ldots\ab{i-1 \, i}\ab{i \, i+1}\ab{i+1 \, i+2}\, \ldots \ab{n{+}1 \, 2}}\,.
\end{align}
Note that all co-dimension four residues are proportional to the tree-level MHV form factor.

\subsubsection{Two Loop Example of Form Factor On-Shell Functions}
\label{subsubsec:ssg_two_loop_eg}

Next, we look at an example for the on-shell function of the two-loop four-point MHV integrand $\calF^{(2),\text{MHV}}_{4}(\mathbf{1},\gamma_+; 2,3,4,5)$. In this section, we explicitly discuss the leading singularity of the double pentagon in Fig.~\ref{fig:diag1}. (Other on-shell functions can be computed in the same way.) The leading singularity corresponds to setting all the internal edges on-shell ($\ell^2_i=0$ for $i \in \{6,7,\ldots,13\}$). There are 8 solutions to these on-shell conditions, but the MHV form factor is nonzero only on two of them. The corresponding kinematic cut solutions can also be labeled by the following on-shell diagrams:
\begin{align}
\label{eq:ssglue_2loop_maxCut_eg}
\text{cut}^{(2)}_{1,a}   =  \vcenter{\hbox{\scalebox{.8}{\ssonea}}}\,,
\qquad
\text{cut}^{(2)}_{1,b} =\vcenter{\hbox{\scalebox{.8}{\ssoneb}}}\,,
\end{align}
where the blue vertices indicate that the $\lamt{}$ of all the adjacent edges are proportional and a white vertex indicates the proportionality of the $\lam{}$'s. Similar to the one-loop case, $\text{cut}^{(2)}_{1,a}$ is simply the residue of $\calF^{(2),\text{MHV}}_{4}(\mathbf{1},\gamma_+; 2,3,4,5)$ when all internal momenta are taken on-shell. Furthermore, the same picture also represent the chirality of the amplitudes entering the unitarity cut. As a concrete example, we consider the supersum gluing of the first diagram, denoted by cut$^{(2)}_{1,a}$. The second cut would follow by an appropriate relabeling. In the case of the maximal cuts, all `internal' momenta $\ell_i= \lambda_{\ell_i} \widetilde{\lambda}_{\ell_i}$ are completely determined  in terms of external kinematics because of the constraints of momentum conservation and the on-shell conditions. As an explicit example, for the $\text{cut}^{(2)}_{1,a}$, we have:
\begin{align}
\begin{split}
\label{eq: ossoln}
 \ell^{1,a}_6 &= \frac{\ab{34}}{\ab{24}} \lam{2}\lamt{3}\,, \ 
 \ell^{1,a}_7 = -\frac{\ab{25}}{\ab{24}} \lam{4}\lamt{5}\,, \
 \ell^{1,a}_8 = -\frac{\ab{45}}{\ab{24}} \lam{2}\lamt{5}\,, \ 
 \ell^{1,a}_9 = \frac{\lam{4}\ p_{45}\ket{2}}{\ab{24}}\,,
 \\
 \ell^{1,a}_{10} &= \frac{\lam{2}\  p_{23} \ket{4}}{\ab{24}}\,,\ 
 \ell^{1,a}_{11} = -\frac{\ab{23}}{\ab{24}} \lam{4}\lamt{3}\,, \
 \ell^{1,a}_{12} = - \frac{\lam{4}\ \mathbf{p_1} \ket{2}}{\ab{24}}\,, \
 \ell^{1,a}_{13} = -\frac{\lam{2}\ \mathbf{p_1} \ket{4}}{\ab{24}}\,.
\end{split}
\end{align}
The Jacobian of the cut is given by:
\begin{align}
    \label{eq: jacobian2loop} 
    \calJ^{(2)}_{1,a} =  s_{23} s_{45} \frac{\ab{34}\ab{52} \aMs{2}{\mathbf{p_1}}{3} \aMs{4}{\mathbf{p_1}}{5}}{\ab{24}^{2}}\,.
\end{align}
The cut is schematically given as follow: 
\begin{align}
\begin{split}
     \text{cut}^{(2)}_{1,a} &=  \frac{1}{\calJ^{(2)}_{1,a}} \,
     \prod_{i=6}^{13} \!\int\! d \eta_{\ell_{i}}\,
     \calF^{\text{MHV}}_2({\bf 1},-\ell_{12},\ell_{13}) \,
     \calA^{\text{MHV}}_3(5,-\ell_8,\ell_7) \,
     \calA^{\text{MHV}}_3(3,\ell_{11},-\ell_6) \,
     \\
     & \hspace{0.5cm}\times
     \calA^{\overline{\text{MHV}}}_3(2,\ell_{6},-\ell_{10}) \,
     \calA^{\overline{\text{MHV}}}_3(4,-\ell_{7},\ell_{9}) \,
     \calA^{\overline{\text{MHV}}}_3(-\ell_{13},\ell_{8},\ell_{10}) \,
     \calA^{\overline{\text{MHV}}}_3(\ell_{12},-\ell_{11},-\ell_9)  \,
     \\
     & = \delta^{(4)}(P)\, \delta^{(8)}(\calQ) \times \widetilde{\text{cut}}^{(2)}_{1,a}  \, . 
\end{split}     
\end{align}  
Carrying out the above integration gives us the following result (up to an overall sign):
\begin{align}
\hspace{-.5cm}
    \widetilde{\text{cut}}^{(2)}_{1,a} = \frac{1}{\calJ^{(2)}_{1,a}}\frac{1}{\vev{\ell_{12} \ell_{13}}^2} 
    \frac{\vev{\ell_{8} \ell_{7}}^3}{\vev{5 \ell_{7}}\vev{\ell_{8} 5}} 
    \frac{\vev{\ell_{6} \ell_{11}}^3}{\vev{\ell_{6} 3}\vev{3 \ell_{11}}} 
    \frac{\sqb{2 \ell_{6}}^3}{\sqb{2 \ell_{10}}\sqb{\ell_{10} \ell_{6}}} 
    \frac{\sqb{\ell_{8} \ell_{10}}^3}{\sqb{\ell_{13} \ell_{10}}\sqb{\ell_{8} \ell_{13}}} 
    \frac{\sqb{\ell_{9} \ell_{11}}^3}{\sqb{\ell_{12} \ell_{9}}\sqb{\ell_{11} \ell_{12}}} 
    \frac{\sqb{4 \ell_{7}}^3}{\sqb{\ell_{7} \ell_{9}}\sqb{\ell_{9} 4}}\, .
\hspace{-.5cm}    
\end{align}
Finally, plugging in the on-shell solution for the internal loop momenta from (\ref{eq: ossoln}) and the Jacobian from (\ref{eq: jacobian2loop}), the $ \widetilde{\text{cut}}^{(2)}_{1,a}$ gets evaluated purely in terms of external momenta to be:
\begin{align}
   \widetilde{\text{cut}}^{(2)}_{1,a} =  \frac{1}{\ab{23}\ab{34}\ab{45}\ab{52}} \, .
\end{align}
Again, the  leading singularities of all planar MHV form factor integrands are proportional to the tree-level Parke-Taylor structure which encodes the only available rational coefficient. This concludes our discussion of both integrand bases and on-shell functions. Next, we combine the two to fix the planar MHV form factor integrands of the chiral stress tensor in $\calN=4$ sYM at one and two loops.

\section{Results}
\label{sec:results}

Given the two ingredients of the unitarity method: (a) a basis of integrands and (b) the residues of field theory, which were introduced in subsections \ref{subsec:integrand_bases} and \ref{subsec:supersum_gluing}, respectively, we are now in the position to build the form-factor integrands at one and two loops. For the sake of clarity,  we discuss the one-loop case in more detail and only sketch the two-loop matching which proceeds in complete analogy.

\subsection{One-Loop $n$-Point MHV Super Form Factor}
\label{subsec:one_loop_results}

Our one-loop triangle powercounting integrand basis for the MHV form factor consists of the chiral box integrands with the two numerators (\ref{eq:chrial_box_nums}) as well as the scalar one-mass triangle integrand (\ref{eq:scalar_1m_tri}) with a single numerator degree of freedom (the overall scale of the triangle integrand). In order to fix the three coefficients of this integrand basis, we could match our ansatz on the three associated field theory cuts discussed in (\ref{eq:1loop_2me_mhv_box_cut}), its conjugate, and (\ref{eq:1loop_1m_tri_soft_composit}). Determining the coefficients of the two box integrands is easy, the linear algebra problem is:
\begin{align}
\label{eq:1loop-box-match-MHV}
 \vcenter{\hbox{\scalebox{.8}{\oneloopbox}}}  & = \underset{\ell_a = \ell^*_{a,1}}{\Res} \left[
        c_1 \frac{\overbrace{\br{i,\ell_a,\ell_c,j}}^{\mathfrak{n}^{(1)}_1}}{\ell^2_a\ell^2_b\ell^2_c\ell^2_d }
       +c_2 \frac{\overbrace{\br{\ell_a,\ell_c,j,i}}^{\mathfrak{n}^{(1)}_2}}{\ell^2_a\ell^2_b\ell^2_c\ell^2_d } \right] = c_1\,,
 \\
\label{eq:1loop-box-match-MHVbar}
0 =  \vcenter{\hbox{\scalebox{.8}{\oneloopboxMHVBar}}} & = 
\underset{\ell_a = \ell^*_{a,2}}{\Res} \left[
        c_1 \frac{\overbrace{\br{i,\ell_a,\ell_c,j}}^{\mathfrak{n}^{(1)}_1}}{\ell^2_a\ell^2_b\ell^2_c\ell^2_d }
       +c_2 \frac{\overbrace{\br{\ell_a,\ell_c,j,i}}^{\mathfrak{n}^{(1)}_2}}{\ell^2_a\ell^2_b\ell^2_c\ell^2_d } \right] = c_2\,.
\end{align}
The locations of the quadruple-cut residues $\ell^*_{a,1}$ and $\ell^*_{a,2}$ are the appropriate adaptations of the kinematic cut solutions in Eq.~(\ref{quad_cuts_of_2me}). As is directly applicable either from the representation of the chiral numerators $\mathfrak{n}^{(1)}_{1,2}$ in (\ref{eq:box_chiral_nums_v1}) or equivalently from our bracket notation, numerator $\mathfrak{n}^{(1)}_2$ vanishes on solution $\ell^*_{a,1}$, whereas numerator $\mathfrak{n}^{(1)}_1$ vanishes on solution $\ell^*_{a,2}$, so that the $2\times 2$ unitarity cut matching system diagonalizes.  From (\ref{eq:1loop-box-match-MHVbar}) we conclude that $c_2=0$. One of the defining features of our chiral integrand basis with unit leading singularities is that all leading singularities are $\pm 1,0$. In the one loop example, this property is easy to check by simpliy plugging in $\ell^*_{a,1}$ into $\mathfrak{n}^{(1)}_1$ and taking the Jacobian $\calJ$ in Eq.~(\ref{eq:1loop_2me_MHV_jac}) into account. We conclude from (\ref{eq:1_loop_2me_MHV_ssg_tilde}) and (\ref{eq:1loop-box-match-MHV}) that 
\begin{align}
 c_1 = \frac{\delta^{(4)}(P)\, \delta^{(8)}(\calQ)}
    {\ab{2\,3}\ldots\ab{i{-}1 \, i}\ab{i \, i{+}1}\ldots
          \ab{j{-}1 \, j}\ab{j\, j{+}1}\ldots\ab{n{+}1\, 2}}
 = \calF^{(0),\text{MHV}}_{n} (\textbf{1},\gamma_+; 2,\ldots,n{+}1)\,.
\end{align}
With the box coefficients fixed, we are left with the single coefficient of the scalar one-mass triangle. Instead of matching the triple-cut (\ref{eq:1loop_1m_triangle_cut}) as a one-parametric function of $\alpha$, it is more convenient to directly compare the composite leading singularity (\ref{eq:1loop_1m_tri_soft_composit}). Crucially, the advantage of this procedure is that the chiral boxes are not contributing to such a cut, because the numerator of the one-mass degeneration of the two-mass-easy box vanishes for $\ell_c=0$ and we directly find:
\begin{align}
 \vcenter{\hbox{\scalebox{.8}{\onelooptrisoft}}} = \underset{\ell^2_b= \ell^2_c=\ell^2_d=0; \,\alpha =0}{\Res} \left[c_3 \frac{\overbrace{s_{i,i{+}1}}^{\mathfrak{n}^{(1)}_3}}{\ell^2_b \ell^2_c  \ell^2_d}\right] = c_3 \, .
\end{align}
This concludes the matching procedure. In principle, one can still proceed and check further unitarity cuts as a consistency check. We have explicitly checked bubble cuts and found that our integrand matches the field theory cuts. This also reaffirms our expectation that a triangle powercounting basis is sufficient to expand the form factor integrand of the chiral stress-tensor supermultiplet which is protected from renormalization. Explicitly, the one-loop $n$-point MHV form factor integrand of the chiral stress-tensor supermultiplet is given by: 
\begin{align}
\label{eq:1loop_MHV_FF_final}
\hspace{-.5cm}
 \frac{\calF^{(1),\text{MHV}}_{n} (\textbf{1},\gamma_+; 2,{\ldots},n{+}1)}        
      {\calF^{(0),\text{MHV}}_{n} (\textbf{1},\gamma_+; 2,{\ldots},n{+}1)} 
 =  \hspace{-.5cm}
     \underset{{\rm cyclic(2,\ldots,n{+}1)}}{\widetilde{\sum}}  \left[
     \hspace{+2cm}\mathfrak{n}^{(1)}_1 \hspace{-2.7cm}
     \vcenter{\hbox{\scalebox{.8}{\oneloopboxnum}}} 
    +
    \mathfrak{n}^{(1)}_3 \hspace{-0.4cm}
    \vcenter{\hbox{\scalebox{.8}{\onelooptriint}}}\!
    \right] \, 
\hspace{-.7cm}    
\end{align}
where we were able to factor out the tree-level (super-)form factor and write the result as a sum of chiral boxes and scalar one-mass triangles. The $\widetilde{\sum}$ instructs to sum over inequivalent diagrams in the cyclic sum, thereby preventing an overcounting of symmetric integrands. 

The one-loop MHV $n$-point super form factor is of course known (See \cite{Brandhuber:2010ad, Bork:2011cj}). Note, however, that our representation nicely separates the infrared divergent part from the finite part of the answer. Since our chiral box integrands vanish in all collinear regions, where the loop momentum becomes collinear to one of the external massless legs $p_i$ or $p_j$, it is clear that these integrands evaluate to IR finite answers. (With triangle powercounting, UV finiteness is also guaranteed.)

\subsection{Two-Loop Four-Point MHV Super Form Factor}
\label{subsec:two_loop_results}

The key strategy to determine the two-loop four-point MHV  super form factor of the chiral stress tensor supermultiplet is analogous to the the one-loop case. Since we already have a clean basis of chiral, unit leading singularity integrands from subsection \ref{subsec:integrand_bases} together with the ability to evaluate (supersymmetric) on-shell functions as described in subsection \ref{subsec:supersum_gluing}, we proceed as before: Determine a `spanning set of cuts', i.e. a list of unitarity cuts that allow to uniquely fix all coefficients of an integrand ansatz, and solve the matching equations. 

In a little more detail, it turns out that our integrand basis adapted from the one used for scattering amplitudes in $\calN=4$ sYM beyond the planar limit has even more desirable features. For example, we only need to explicitly work with the 19 graph topologies summarized in Fig.~\ref{fig:diags_all} where their explicit numerators $\mathfrak{n}^{(2)}_i$ are collected in appendix \ref{sec:Appendix}. Naively, the number of possible topologies is far greater. As an illustration, there are MHV cuts of the form factor, e.g. 
\begin{align}
\label{eq:eg_graph_not_on_list}
\vcenter{\hbox{\scalebox{.8}{\diagEgNotNeeded}}}\, ,
\end{align}
where the associated integrand topology is \emph{not} on the list of graphs in Fig.~\ref{fig:diags_all} even though it would be allowed by triangle powercounting. In all such cases, our representation of the integrand is already `smart' enough such that the cut is matched implicitly by other diagram topologies.

%
\begin{figure}[]
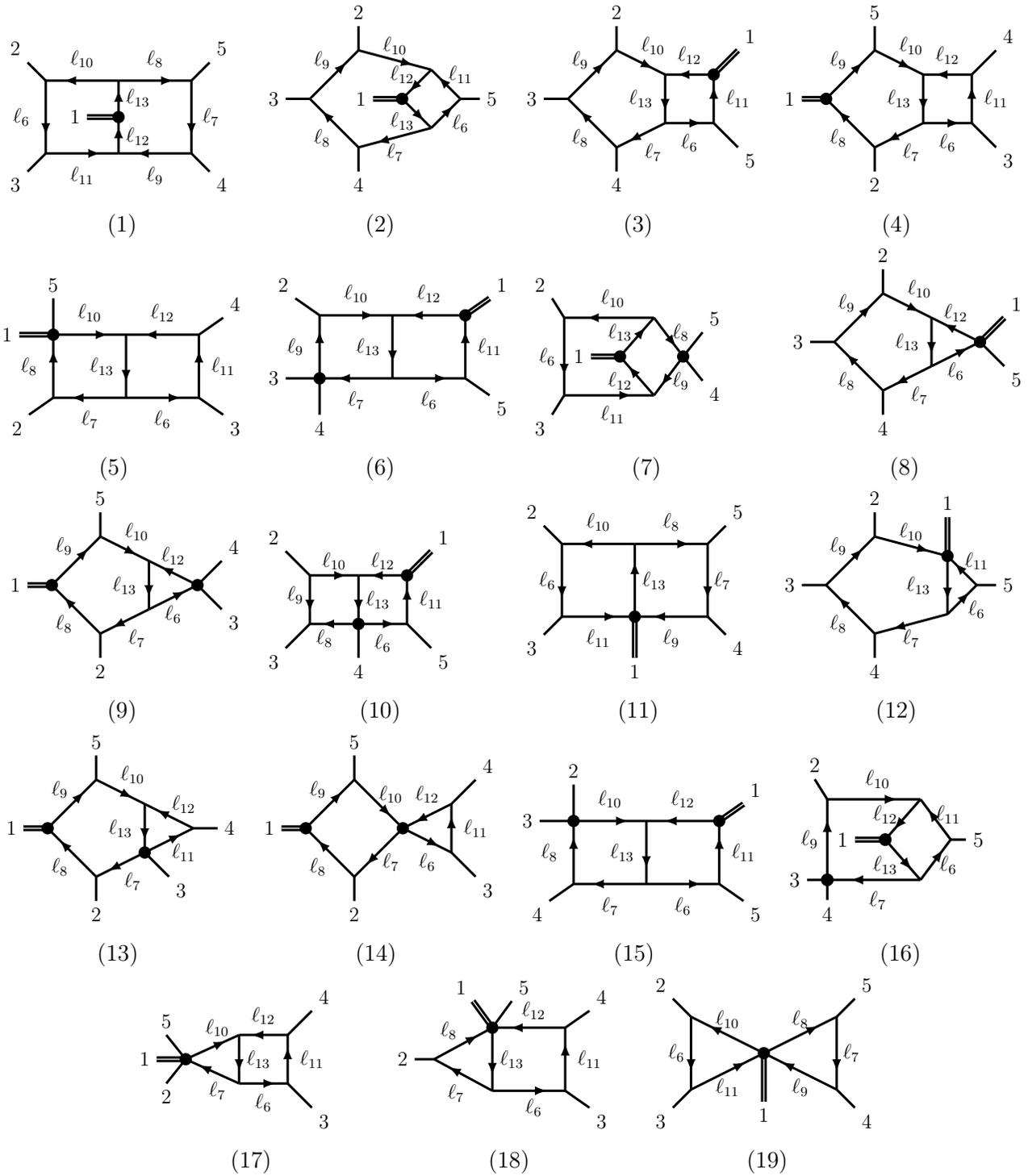

\centering
\begin{subfigure}[b]{.24\linewidth}
\hbox{\scalebox{.8}{\diagOne}}
\caption{}\label{fig:diag1}
\end{subfigure}
\begin{subfigure}[b]{.24\linewidth}
\hbox{\scalebox{.8}{\diagTwo}}
\caption{}\label{fig:diag2}
\end{subfigure}
\begin{subfigure}[b]{.24\linewidth}
\hbox{\scalebox{.8}{\diagThree}}
\caption{}\label{fig:diag3}
\end{subfigure}
\begin{subfigure}[b]{.24\linewidth}
\hbox{\scalebox{.8}{\diagFour}}
\caption{}\label{fig:diag4}
\end{subfigure}
\begin{subfigure}[b]{.24\linewidth}
\hbox{\scalebox{.8}{\diagFive}}
\caption{}\label{fig:diag5}
\end{subfigure}
\begin{subfigure}[b]{.24\linewidth}
\hbox{\scalebox{.8}{\diagSix}}
\caption{}\label{fig:diag6}
\end{subfigure}
\begin{subfigure}[b]{.24\linewidth}
\hbox{\scalebox{.8}{\diagSeven}}
\caption{}\label{fig:diag7}
\end{subfigure}
\begin{subfigure}[b]{.24\linewidth}
\hbox{\scalebox{.8}{\diagEight}}
\caption{}\label{fig:diag8}
\end{subfigure}
\begin{subfigure}[b]{.24\linewidth}
\hbox{\scalebox{.8}{\diagNine}}
\caption{}\label{fig:diag9}
\end{subfigure}
\begin{subfigure}[b]{.24\linewidth}
\hbox{\scalebox{.8}{\diagTen}}
\caption{}\label{fig:diag10}
\end{subfigure}
\begin{subfigure}[b]{.24\linewidth}
\hbox{\scalebox{.8}{\diagEleven}}
\caption{}\label{fig:diag11}
\end{subfigure}
\begin{subfigure}[b]{.24\linewidth}
\hbox{\scalebox{.8}{\diagTwelve}}
\caption{}\label{fig:diag12}
\end{subfigure}
\begin{subfigure}[b]{.24\linewidth}
\hbox{\scalebox{.8}{\diagThirteen}}
\caption{}\label{fig:diag13}
\end{subfigure}
\begin{subfigure}[b]{.24\linewidth}
\hbox{\scalebox{.8}{\diagFourteen}}
\caption{}\label{fig:diag14}
\end{subfigure}
\begin{subfigure}[b]{.24\linewidth}
\hbox{\scalebox{.8}{\diagFifteen}}
\caption{}\label{fig:diag15}
\end{subfigure}
\begin{subfigure}[b]{.24\linewidth}
\hbox{\scalebox{.83}{\diagSixteen}}
\caption{}\label{fig:diag16}
\end{subfigure}
\begin{subfigure}[b]{.24\linewidth}
\hbox{\scalebox{.8}{\diagSeventeen}}
\caption{}\label{fig:diag17}
\end{subfigure}
\begin{subfigure}[b]{.24\linewidth}
\hbox{\scalebox{.8}{\diagEighteen}}
\caption{}\label{fig:diag18}
\end{subfigure}
\begin{subfigure}[b]{.24\linewidth}
\hbox{\scalebox{.8}{\diagNineteen}}
\caption{}\label{fig:diag19}
\end{subfigure}
\caption{Diagrams with nonzero numerators contributing to the MHV four-particle form factor for the chiral stress-tensor supermultiplet in planar $\calN=4$ super-Yang-Mills theory at two loops.}
\label{fig:diags_all}
\end{figure}

Before getting ahead of ourselves, let us demonstrate the cut matching procedure on one very simple two-loop example at the co-dimension eight level related to the on-shell function discussed around Eq.~(\ref{eq:ssglue_2loop_maxCut_eg})
\begin{align}
\text{cut}^{(2)}_{1,a}  =  \vcenter{\hbox{\scalebox{.8}{\ssonea}}}\,,
\qquad
\text{cut}^{(2)}_{1,b} =    \vcenter{\hbox{\scalebox{.8}{\ssoneb}}}\,.
\end{align}
Both leading singularities have to be matched by a single integrand basis element: the double-pentagon introduced in Eq.~(\ref{eq:diag_1_num_integrand_section})
\begin{align}
\hspace{-.5cm}
\vcenter{\hbox{\scalebox{.7}{\diagOne}}}
\quad
\mathfrak{n}^{(2)}_{1}= 
    -[\![p_3,\ell_6,\ell_{10},\ell _{13},\ell_{12},\ell_{9},\ell_7,p_5]\!]
    -[\![p_4,\ell_7,\ell_8,\ell_{13},\ell _{12},\ell _{11},\ell _6,p_2]\!]\,.
\hspace{-.5cm}
\end{align}
Incidentally, the two terms of $\mathfrak{n}^{(2)}_1$ are of such a form that the first one vanishes on $\text{cut}^{(2)}_{1,a}$, whereas the second one vanishes on $\text{cut}^{(2)}_{1,b}$. We now use the field theory residues to fix the coefficients associated with the numerators. 
\begin{align}
\begin{split}
    \label{eq:2loop-match-MHV}
 \vcenter{\hbox{\scalebox{.8}{\ssonea}}}  & = \underset{\ell_6 = \ell^{*}_6,\ell_7 = \ell^{*}_7}{\Res} \left[
        c_1 \, \frac{{\mathfrak{n}^{(2)}_1}}{\ell^2_6\ell^2_7\ell^2_8\ell^2_9 \ell^2_{10}\ell^2_{11}\ell^2_{12}\ell^2_{13}}
 \right] 
        = c_1\,,
\end{split}
\end{align}
where $\ell^*_{6}, \ell^*_7$ refers to the cut solution in Eq.~(\ref{eq:ssglue_2loop_maxCut_eg}). On this eight-propagator cut, only the single integrand basis element contributes. Plugging the cut solution into $\mathfrak{n}^{(2)}_1$ and taking into account the Jacobian factor of (\ref{eq: jacobian2loop}), we get the coefficient $c_1$ to be:
\begin{align}
    c_1 = \frac{\delta^{(4)}(P)\, \delta^{(8)}(\calQ)}{\ab{23}\ab{34}\ab{45}\ab{52}}
 = \calF^{(0),\text{MHV}}_{4} (\textbf{1},\gamma_+; 2,3,4,5)\,.
\end{align}
In analogy to the one-loop case discussed in subsection \ref{subsec:one_loop_results}, we consider a spanning set of unitarity cuts that determine all other coefficients in our integrand basis. Since all ingredients of the cut matching procedure have been described, we spare the mechanical details of the calculation. As expected, we find that all coefficients are proportional to the tree-level super form factor:
\begin{align}
\label{eq:2loop_MHV_FF_final}
\hspace{-.5cm}
 \frac{\calF^{(2),\text{MHV}}_{4} (\textbf{1},\gamma_+; 2,{\ldots},5)}        
      {\calF^{(0),\text{MHV}}_{4} (\textbf{1},\gamma_+; 2,{\ldots},5)} 
 =  \hspace{-.2cm}
      \underset{{\rm cyclic(2,\ldots,5)}}{\widetilde{\sum}} \left[
     \mathfrak{n}^{(2)}_1 \hspace{-.2cm}
     \vcenter{\hbox{\scalebox{.7}{\diagOne}}} 
     \hspace{-.2cm}
    +
    \cdots 
    + 
     \mathfrak{n}^{(2)}_{19} \hspace{-.2cm}
     \vcenter{\hbox{\scalebox{.7}{\diagNineteen}}} 
    \right].
\hspace{-.5cm}    
\end{align}
As further consistency check on our result, we have verified that our two-loop integrand also matches additional unitarity cuts that were not needed to fix the undetermined coefficients. 

The final result for the four-dimensional planar two-loop four-point MHV form factor of the chiral stress tensor in $\calN=4$ sYM is extremely simple and only involves (the cyclic sum of) 19 integrand basis elements with compact loop-momentum dependent numerators $\mathfrak{n}_i$ that are organized according to their properties in infrared singular regions and were obtained directly from similar expressions constructed for on-shell scattering amplitudes in the same theory beyond the leading color approximation. The ease of our integrand construction highlights, this time in the context of form factors, the advantages of good bases of loop integrands. In order to interface with modern integration tools based on dimensional regularization, it would be interesting to find a streamlined way to augment extra-dimensional terms to the current version of the integrand or directly subtract all IR divergences and manifestly evaluate the resulting expression in $d=4$. Such investigations are beyond the scope of this work and are left for the future.

\section{Conclusions and Outlook}
\label{sec:conclusions}

Recently, the form factors of the chiral stress tensor supermultiplets in planar $\calN=4$ super Yang-Mills theory have garnered increased attention due to the discovery of a yet-to-be fully understood set of dualities between three-point form factors and scattering amplitdues \cite{Dixon:2021tdw},  together with a self-duality property \cite{Dixon:2022xqh} of the four-particle form factor. These observations were gleaned from the `bootstrapped' answers in terms of the symbol of generalized polylogarithmic functions where the duality acts as `antipodal map' which corresponds to `reading the symbol backwards', or exchanging discontinuities with derivatives. These fascinating observations and their full implications deserve further study.

In this work we took an initial step in that direction. In particular, we have constructed the four-dimensional integrand for the planar two-loop four-point MHV \emph{form factor} of the chiral stress tensor supermultiplet in maximally supersymmetric Yang-Mills theory. Our construction is based on a tailored version of generalized unitarity \cite{Bern:1994zx,Bern:1994cg,Britto:2004nc} where we recycle the integrand basis with `triangle powercounting' that has been used to construct two-loop MHV scattering amplitudes in $\calN=4$ sYM beyond the planar limit \cite{Bourjaily:2019iqr,Bourjaily:2019gqu}. Since the chiral stress-tensor supermultiplet is a color singlet operator that is protected from UV renormalization, it turned out that this basis of integrands was ideally suited for our form factor calculation. All the desirable features, such as the property of unit leading singularities on all co-dimension eight residues were directly exported which led to extremely simple generalized unitarity matching equations that compare the residues of field theory (see subsection \ref{subsec:supersum_gluing}) to the ones of the integrand basis. We gave explicit examples of the cut matching procedure in subsections \ref{subsec:one_loop_results} and \ref{subsec:two_loop_results} which led to the compact representation of the integrand in Eq.~(\ref{eq:2loop_MHV_FF_final}). 

With our integrand at hand, there is a clear path for future progress. With the imminent availability of all relevant two-loop five-particle one-mass master integrals, the planar two-loop four-point MHV form factor in $\calN=4$ sYM is arguably the simplest physically interesting object to integrate. Note that there remains the related interesting open question of how to systematically ensure the compatibility of the four-dimensional integrand technology used in this work and the standard integration techniques that usually rely on dimensional regularization. Barring the complications of adding extra-dimensional (or `$\mu$') terms (which could be obtained by an honest $d$-dimensional cut matching procedure), it is important to stress that we expect several of our four-dimensional integrands to evaluate to pure, uniform transcendental functions. Integrating the form factors by an explicit evaluation in terms of master integrals will enable us to independently cross-check the bootstrapped symbol-level results \cite{Dixon:2021tdw,Dixon:2022xqh} and upgrade them to function level, including the information about all transcendental constants. Such a result will allow us to check the (self-)duality properties of the form factor at the co-product level. 

On the integrand side, there are two fairly obvious next steps. First, one can extend our result for the planar two-loop four-point MHV (super-)form factor to arbitrary numbers of external states. Furthermore, it should also be possible to take into account subleading color effects and work beyond the planar limit. The amplitude integrand basis \cite{Bourjaily:2019iqr,Bourjaily:2019gqu} originally constructed for scattering amplitudes beyond the planar limit should also streamline this analysis. The only minor complication arises from controlling the color-dressed on-shell functions appearing as coefficients. It would also be interesting to study whether the antipodal self-duality extends beyond the planar limit. Another interesting direction to pursue would be to understand how to apply the double copy procedure to obtain gravitational form factors. It was recently shown that, besides the BCJ relations, one requires additional `operator-induced' relations to get the correct gauge invariant gravitational form factors \cite{Lin:2021pne, Lin:2022jrp, Lin:2023rwe}. Other than increasing the number of external legs, or going beyond the leading color approximation, we can also increase the helicity charge of the form factor and construct N$^k$MHV super form factors using similar ideas to the ones presented in this work, by adapting the amplitude integrand bases such as the one constructed in \cite{Bourjaily:2021iyq}. Beyond the MHV sector, the supersymmetric state sums and the resulting integrand coefficients will be more complicated, but nothing stops us in principle. It would also be interesting to investigate the possible rational structures that can appear in N$^k$MHV super form factors which would be a crucial input in any future bootstrap approach. 

\paragraph{Acknowledgement}
We thank Lance Dixon, Andrew McLeod, and Andy Liu for valuable discussions on the antipodal duality and their form factor computations. E.H. is grateful to Jake Bourjaily and Jaroslav Trnka for collaboration on numerous related integrand ideas and to Ben Page for discussions on the two-loop five-particle one-mass master integrals. T.G. and E.H. are supported by the U.S. Department of Energy (DOE) under award number DE-SC0009937. We are also grateful to the Mani L. Bhaumik Institute of Theoretical Physics for support.

\newpage
\appendix
\section{Appendix}
\label{sec:Appendix}
In this appendix, we summarize the 19 diagram topologies and their kinematic numerators that are required to match the two-loop four-particle integrand of the planar form factor of the chiral stress tensor multiplet in $\calN=4$ sYM. The numerators can be obtained from those of Ref.~\cite{Bourjaily:2021iyq} by appropriate identification of the form factor leg and symmetrizing over diagram automorphims. For diagrams with fewer than eight propagators, one has to carefully take into account the `degeneration' rules described in \cite{Bourjaily:2021iyq}.

\begin{center}
\begin{tabular}{|c|c|} 
 \hline
$\vcenter{\hbox{\scalebox{.7}{\diagOne}}}$
& $ \mathfrak{n}^{(2)}_1= -[\![p_4,\ell _7,\ell _8,\ell _{13},\ell _{12},\ell _{11},\ell _6,p_2]\!]
-[\![p_5,\ell _7,\ell _9,\ell _{12},\ell _{13},\ell _{10},\ell _6,p_3]\!]$ \\
\hline
$\vcenter{\hbox{\scalebox{.7}{\diagTwo}}}$
& $\mathfrak{n}^{(2)}_2=-[\![p_4,\ell _8,\ell _9,p_2]\!] {\Big(}[\![\ell _{10},\ell _{12},\ell _{13},\ell _7]\!]-\ell _7^2 \ell _{12}^2-\ell _{10}^2 \ell _{13}^2{\Big )}$ \\ 
\hline
$\vcenter{\hbox{\scalebox{.7}{\diagThree}}}$
& $\mathfrak{n}^{(2)}_3= \frac{1}{2} {\Big(}[\![\ell _{10},\ell _{12},\ell _{11},p_5,p_4,\ell _8,\ell _9,p_2]\!]-2 [\![p_4,\ell _8,\ell _9,p_2]\!] [\![\ell _{10},\ell _{12},\ell _{11},p_5]\!]{\Big)} $ \\ 
\hline
$\vcenter{\hbox{\scalebox{.7}{\diagFour}}}$
& $\mathfrak{n}^{(2)}_4=
\frac{1}{2} {\Big(}
2 [\![p_5,\ell _9,\ell _8,p_2]\!] 
{\big( }[\![\ell _7,\ell_6,\ell_{11},p_4]\!]
       -[\![\ell _{10},\ell _{12},\ell _{11},p_3]\!]
{\big)}$
\\[-20pt] & 
$\hspace{1cm}-[\![\ell _7,\ell_6,\ell _{11},p_4,p_5,\ell _9,\ell _8,p_2]\!]
+[\![\ell _{10},\ell _{12},\ell _{11},p_3,p_2,\ell_8,\ell _9,p_5]\!]
{\Big) }$\\[20pt]
\hline
$\vcenter{\hbox{\scalebox{.7}{\diagFive}}}$
& $\mathfrak{n}^{(2)}_5=
\frac{1}{2} {\Big(}2 {\big(}-s_{34} [\![\ell _{12},\ell _{10},\ell _8,p_2]\!]+\ell _6^2
   [\![p_4,\ell _{10},\ell _8,p_2]\!]+[\![p_4,\ell _{11},\ell _6,\ell _{10},\ell
   _8,p_2]\!]{\big)} 
   $ \\[-20pt] &  $\hspace{1cm}
   -[\![p_4,p_3,p_2,\ell _8,\ell _{10},\ell _{12}]\!]{\Big)}$ \\[20pt]
\hline
\end{tabular}
\end{center}

\begin{center}
\begin{tabular}{ |c|c| } 
\hline
$\vcenter{\hbox{\scalebox{.7}{\diagSix}}}$
& $\mathfrak{n}^{(2)}_6=
[\![p_5,\ell _{11},\ell _{12},\ell _7,\ell _9,p_2]\!]
$\\
\hline
$\vcenter{\hbox{\scalebox{.7}{\diagSeven}}}$
& $\mathfrak{n}^{(2)}_7=
-s_{23} {\Big(}[\![\ell _{11},\ell _9,\ell _8,\ell _{10}]\!]-\ell _9^2 \ell _{10}^2-\ell _8^2 \ell_{11}^2{\Big)}$\\
\hline
$\vcenter{\hbox{\scalebox{.7}{\diagEight}}}$
& $\mathfrak{n}^{(2)}_8=s_{15} [\![p_4,\ell _8,\ell _9,p_2]\!]$\\
\hline
$\vcenter{\hbox{\scalebox{.7}{\diagNine}}}$
& $\mathfrak{n}^{(2)}_9=s_{34}[\![p_5,\ell _9,\ell _8,p_2]\!]$\\
\hline
$\vcenter{\hbox{\scalebox{.7}{\diagTen}}}$
& $\mathfrak{n}^{(2)}_{10}= -s_{23} [\![\ell _{10},\ell _{12},\ell _{11},p_5]\!]+\ell _8^2 [\![p_5,\ell _{11},\ell
   _{12},p_2]\!]-[\![p_2,p_3,p_5,\ell _{11},\ell _{12},\ell _{10}]\!]$\\
\hline
$\vcenter{\hbox{\scalebox{.7}{\diagEleven}}}$
& $\mathfrak{n}^{(2)}_{11}= \frac{1}{2} {\Big( }2s_{23} [\![\ell _{10},\ell _8,\ell _7,p_4]\!]+2 s_{45} [\![\ell
   _8,\ell _{10},\ell _6,p_3]\!]+\ell _9^2 [\![p_5,\ell _{10},\ell _6,p_3]\!]$
   \\[-20pt]
   &  $ +\ell _{11}^2[\![p_4,\ell _7,\ell _8,p_2]\!]+[\![p_2,p_3,p_4,\ell _7,\ell _8,\ell
   _{10}]\!]+[\![p_5,p_4,p_3,\ell _6,\ell _{10},\ell _8]\!]{\Big)}$
   \\[20pt]
\hline
$\vcenter{\hbox{\scalebox{.7}{\diagTwelve}}}$
& $\mathfrak{n}^{(2)}_{12}= \frac{1}{2} {\Big( }[\![p_4,\ell _8,\ell _9,p_2,p_5,\ell _7]\!]-2 [\![p_5,\ell _7]\!][\![p_4,\ell _8,\ell _9,p_2]\!]{\Big)}$\\
\hline
$\vcenter{\hbox{\scalebox{.7}{\diagThirteen}}}$
& $\mathfrak{n}^{(2)}_{13}= [\![p_4,\ell _{10}]\!] [\![p_5,\ell _9,\ell _8,p_2]\!] -\frac{1}{2} \ell _8^2 [\![p_5,\ell _9,\ell _7,p_4]\!]$\\
\hline
\end{tabular}
\end{center}

\begin{center}
\begin{tabular}{ |c|c| } 
\hline
$\vcenter{\hbox{\scalebox{.7}{\diagFourteen}}}$
& $\mathfrak{n}^{(2)}_{14}= s_{34} [\![p_5,\ell _9,\ell _8,p_2]\!]$\\
\hline
$\vcenter{\hbox{\scalebox{.7}{\diagFifteen}}}$
& $\mathfrak{n}^{(2)}_{15}= -\frac{1}{2} [\![p_5,\ell _{11},\ell _{12},\ell _{10},\ell _8,p_4]\!]$\\
\hline
$\vcenter{\hbox{\scalebox{.7}{\diagSixteen}}}$
& $\mathfrak{n}^{(2)}_{16}=\frac{1}{2} {\Big(} \ell _{12}^2 [\![p_5,\ell _7,\ell _9,p_2]\!]+2 [\![p_5,\ell
   _{12},\ell _{13},\ell _7,\ell _9,p_2]\!]{\Big)} $\\
\hline
$\vcenter{\hbox{\scalebox{.7}{\diagSeventeen}}}$
& $\mathfrak{n}^{(2)}_{17}= s_{125} s_{34}$\\
\hline
$\vcenter{\hbox{\scalebox{.7}{\diagEighteen}}}$
& $\mathfrak{n}^{(2)}_{18}= \frac{1}{2} {\Big(}-2 s_{34} [\![p_2,\ell _6]\!]-[\![\ell _6,p_2,p_4,p_3]\!]+s_{23} \ell _{12}^2{\Big)}$\\
\hline
$\vcenter{\hbox{\scalebox{.7}{\diagNineteen}}}$
& $\mathfrak{n}^{(2)}_{19}= s_{23} s_{45}$\\
\hline
\end{tabular}
\end{center}


\newpage

\bibliographystyle{JHEP}
\bibliography{refs.bib}

\providecommand{\href}[2]{#2}\begingroup\raggedright\begin{thebibliography}{100}

\bibitem{parkeTaylor1985}
S.~J. Parke and T.~R. Taylor, \emph{{Gluonic Two Goes To Four}}, {\emph{Nucl.
  Phys.} {\bfseries B269} (1986) 410}.

\bibitem{Parke:1986gb}
S.~J. Parke and T.~R. Taylor, \emph{{An Amplitude for $n$-Gluon Scattering}},
  \href{https://doi.org/10.1103/PhysRevLett.56.2459}{\emph{Phys. Rev. Lett.}
  {\bfseries 56} (1986) 2459}.

\bibitem{Mangano:1987xk}
M.~L. Mangano, S.~J. Parke and Z.~Xu, \emph{{Duality and Multi-Gluon
  Scattering}}, \href{https://doi.org/10.1016/0550-3213(88)90001-6}{\emph{Nucl.
  Phys.} {\bfseries B298} (1988) 653}.

\bibitem{Mangano:1990by}
M.~L. Mangano and S.~J. Parke, \emph{{Multiparton Amplitudes in Gauge
  Theories}}, \href{https://doi.org/10.1016/0370-1573(91)90091-Y}{\emph{Phys.
  Rept.} {\bfseries 200} (1991) 301}
  [\href{https://arxiv.org/abs/hep-th/0509223}{{\ttfamily hep-th/0509223}}].

\bibitem{Lusztig}
G.~Lusztig, \emph{{Total Positivity in Partial Flag Manifolds}},
  \href{https://doi.org/10.1090/S1088-4165-98-00046-6}{\emph{Represent. Theory}
  {\bfseries 2} (1998) 70}.

\bibitem{Postnikov:2006kva}
A.~Postnikov, \emph{{Total Positivity, Grassmannians, and Networks}},
  \href{https://arxiv.org/abs/math/0609764}{{\ttfamily math/0609764}}.

\bibitem{Postnikov2009matching}
A.~Postnikov, D.~Speyer and L.~Williams, \emph{{Matching Polytopes, Toric
  Geometry, and the Totally Non-Negative Grassmannian}},
  \href{https://doi.org/10.1007/s10801-008-0160-1}{\emph{Journal of Algebraic
  Combinatorics} {\bfseries 30} (2009) 173}
  [\href{https://arxiv.org/abs/0706.2501}{{\ttfamily 0706.2501}}].

\bibitem{Williams:2003a}
L.~K. {Williams}, \emph{{Enumeration of Totally Positive Grassmann Cells}},
  {\emph{{Adv. Math.}} {\bfseries 190} (2005) 319}
  [\href{https://arxiv.org/abs/arXiv:math/0307271}{{\ttfamily
  arXiv:math/0307271}}].

\bibitem{Goncharov:2011hp}
A.~B. Goncharov and R.~Kenyon, \emph{{Dimers and Cluster Integrable Systems}},
  \href{https://arxiv.org/abs/1107.5588}{{\ttfamily 1107.5588}}.

\bibitem{KLS}
A.~Knutson, T.~Lam and D.~Speyer, \emph{{Positroid Varieties: Juggling and
  Geometry}},
  \href{https://doi.org/10.1112/S0010437X13007240}{\emph{{Compositio
  Mathematica}} {\bfseries 149} (2013) 1710}
  [\href{https://arxiv.org/abs/1111.3660}{{\ttfamily 1111.3660}}].

\bibitem{Broadhurst:1996kc}
D.~J. Broadhurst and D.~Kreimer, \emph{{Association of Multiple Zeta Values
  with Positive Knots via Feynman Diagrams up to 9 Loops}},
  \href{https://doi.org/10.1016/S0370-2693(96)01623-1}{\emph{Phys. Lett.}
  {\bfseries B393} (1997) 403}
  [\href{https://arxiv.org/abs/hep-th/9609128}{{\ttfamily hep-th/9609128}}].

\bibitem{Kreimer:1997dp}
D.~Kreimer, \emph{{On the Hopf Algebra Structure of Perturbative Quantum Field
  Theories}}, {\emph{Adv. Theor. Math. Phys.} {\bfseries 2} (1998) 303}
  [\href{https://arxiv.org/abs/q-alg/9707029}{{\ttfamily q-alg/9707029}}].

\bibitem{Bloch:2005bh}
S.~Bloch, H.~Esnault and D.~Kreimer, \emph{{On Motives Associated to Graph
  Polynomials}}, \href{https://doi.org/10.1007/s00220-006-0040-2}{\emph{Commun.
  Math. Phys.} {\bfseries 267} (2006) 181}
  [\href{https://arxiv.org/abs/math/0510011}{{\ttfamily math/0510011}}].

\bibitem{Aluffi:2008sy}
P.~Aluffi and M.~Marcolli, \emph{{Feynman Motives of Banana Graphs}},
  \href{https://doi.org/10.4310/CNTP.2009.v3.n1.a1}{\emph{Commun. Num. Theor.
  Phys.} {\bfseries 3} (2009) 1}
  [\href{https://arxiv.org/abs/0807.1690}{{\ttfamily 0807.1690}}].

\bibitem{Brown:2009ta}
F.~C.~S. Brown, \emph{{On the Periods of Some Feynman Integrals}},
  \href{https://arxiv.org/abs/0910.0114}{{\ttfamily 0910.0114}}.

\bibitem{Marcolli:2009zy}
M.~Marcolli, \emph{{Feynman Integrals and Motives}},
  \href{https://arxiv.org/abs/0907.0321}{{\ttfamily 0907.0321}}.

\bibitem{Brown:2009rc}
F.~Brown and K.~Yeats, \emph{{Spanning Forest Polynomials and the
  Transcendental Weight of Feynman Graphs}},
  \href{https://doi.org/10.1007/s00220-010-1145-1}{\emph{Commun. Math. Phys.}
  {\bfseries 301} (2011) 357}
  [\href{https://arxiv.org/abs/0910.5429}{{\ttfamily 0910.5429}}].

\bibitem{Brown:2010bw}
F.~Brown and O.~Schnetz, \emph{{A K3 in $\phi^4$}},
  \href{https://arxiv.org/abs/1006.4064}{{\ttfamily 1006.4064}}.

\bibitem{Bogner:2014mha}
C.~Bogner and F.~Brown, \emph{{Feynman Integrals and Iterated Integrals on
  Moduli Spaces of Curves of Genus Zero}},
  \href{https://doi.org/10.4310/CNTP.2015.v9.n1.a3}{\emph{Commun. Num. Theor.
  Phys.} {\bfseries 09} (2015) 189}
  [\href{https://arxiv.org/abs/1408.1862}{{\ttfamily 1408.1862}}].

\bibitem{Brown:2015fyf}
F.~Brown, \emph{{Feynman Amplitudes and Cosmic Galois group}},
  \href{https://arxiv.org/abs/1512.06409}{{\ttfamily 1512.06409}}.

\bibitem{Panzer:2015ida}
E.~Panzer, \emph{{Feynman Integrals and Hyperlogarithms}}, Ph.D. thesis,
  Humboldt U., Berlin, Inst. Math., 2015.
\newblock \href{https://arxiv.org/abs/1506.07243}{{\ttfamily 1506.07243}}.

\bibitem{Caron-Huot:2019bsq}
S.~Caron-Huot, L.~J. Dixon, F.~Dulat, M.~Von~Hippel, A.~J. McLeod and
  G.~Papathanasiou, \emph{{The Cosmic Galois Group and Extended Steinmann
  Relations for Planar $\mathcal{N} = 4$ SYM Amplitudes}},
  \href{https://doi.org/10.1007/JHEP09(2019)061}{\emph{JHEP} {\bfseries 09}
  (2019) 061} [\href{https://arxiv.org/abs/1906.07116}{{\ttfamily
  1906.07116}}].

\bibitem{Gurdogan:2020ppd}
O.~G\"urdo\u{g}an, \emph{{From integrability to the Galois coaction on Feynman
  periods}}, \href{https://doi.org/10.1103/PhysRevD.103.L081703}{\emph{Phys.
  Rev. D} {\bfseries 103} (2021) L081703}
  [\href{https://arxiv.org/abs/2011.04781}{{\ttfamily 2011.04781}}].

\bibitem{Bern:1998ug}
Z.~Bern, L.~J. Dixon, D.~C. Dunbar, M.~Perelstein and J.~S. Rozowsky, \emph{{On
  the Relationship Between Yang-Mills Theory and Gravity and its Implication
  for Ultraviolet Divergences}},
  \href{https://doi.org/10.1016/S0550-3213(98)00420-9}{\emph{Nucl. Phys.}
  {\bfseries B530} (1998) 401}
  [\href{https://arxiv.org/abs/hep-th/9802162}{{\ttfamily hep-th/9802162}}].

\bibitem{Bern:2007xj}
Z.~Bern, J.~J. Carrasco, D.~Forde, H.~Ita and H.~Johansson, \emph{{Unexpected
  Cancellations in Gravity Theories}},
  \href{https://doi.org/10.1103/PhysRevD.77.025010}{\emph{Phys. Rev.}
  {\bfseries D77} (2008) 025010}
  [\href{https://arxiv.org/abs/0707.1035}{{\ttfamily 0707.1035}}].

\bibitem{Bern:2012uf}
Z.~Bern, J.~J.~M. Carrasco, L.~J. Dixon, H.~Johansson and R.~Roiban,
  \emph{{Simplifying Multiloop Integrands and Ultraviolet Divergences of Gauge
  Theory and Gravity Amplitudes}},
  \href{https://doi.org/10.1103/PhysRevD.85.105014}{\emph{Phys. Rev.}
  {\bfseries D85} (2012) 105014}
  [\href{https://arxiv.org/abs/1201.5366}{{\ttfamily 1201.5366}}].

\bibitem{Bern:2015xsa}
Z.~Bern, C.~Cheung, H.-H. Chi, S.~Davies, L.~Dixon and J.~Nohle,
  \emph{{Evanescent Effects Can Alter Ultraviolet Divergences in Quantum
  Gravity without Physical Consequences}},
  \href{https://doi.org/10.1103/PhysRevLett.115.211301}{\emph{Phys. Rev. Lett.}
  {\bfseries 115} (2015) 211301}
  [\href{https://arxiv.org/abs/1507.06118}{{\ttfamily 1507.06118}}].

\bibitem{BCF}
R.~Britto, F.~Cachazo and B.~Feng, \emph{{New Recursion Relations for Tree
  Amplitudes of Gluons}},
  \href{https://doi.org/10.1016/j.nuclphysb.2005.02.030}{\emph{Nucl.Phys.}
  {\bfseries B715} (2005) 499}
  [\href{https://arxiv.org/abs/hep-th/0412308}{{\ttfamily hep-th/0412308}}].

\bibitem{BCFW}
R.~Britto, F.~Cachazo, B.~Feng and E.~Witten, \emph{{Direct Proof of Tree-Level
  Recursion Relation in Yang- Mills Theory}},
  \href{https://doi.org/10.1103/PhysRevLett.94.181602}{\emph{Phys. Rev. Lett.}
  {\bfseries 94} (2005) 181602}
  [\href{https://arxiv.org/abs/hep-th/0501052}{{\ttfamily hep-th/0501052}}].

\bibitem{ArkaniHamed:2010kv}
N.~Arkani-Hamed, J.~L. Bourjaily, F.~Cachazo, S.~Caron-Huot and J.~Trnka,
  \emph{{The All-Loop Integrand For Scattering Amplitudes in Planar
  $\mathcal{N}\!=\!4$ SYM}},
  \href{https://doi.org/10.1007/JHEP01(2011)041}{\emph{JHEP} {\bfseries 1101}
  (2011) 041} [\href{https://arxiv.org/abs/1008.2958}{{\ttfamily 1008.2958}}].

\bibitem{Witten:2003nn}
E.~Witten, \emph{{Perturbative Gauge Theory as a String Theory in Twistor
  Space}}, \href{https://doi.org/10.1007/s00220-004-1187-3}{\emph{Commun. Math.
  Phys.} {\bfseries 252} (2004) 189}
  [\href{https://arxiv.org/abs/hep-th/0312171}{{\ttfamily hep-th/0312171}}].

\bibitem{Roiban:2004vt}
R.~Roiban, M.~Spradlin and A.~Volovich, \emph{{A Googly Amplitude from the B
  Model in Twistor Space}},
  \href{https://doi.org/10.1088/1126-6708/2004/04/012}{\emph{JHEP} {\bfseries
  04} (2004) 012} [\href{https://arxiv.org/abs/hep-th/0402016}{{\ttfamily
  hep-th/0402016}}].

\bibitem{Drummond:2006rz}
J.~Drummond, J.~Henn, V.~Smirnov and E.~Sokatchev, \emph{{Magic Identities for
  Conformal Four-Point Integrals}},
  \href{https://doi.org/10.1088/1126-6708/2007/01/064}{\emph{JHEP} {\bfseries
  0701} (2007) 064} [\href{https://arxiv.org/abs/hep-th/0607160}{{\ttfamily
  hep-th/0607160}}].

\bibitem{Alday:2007hr}
L.~F. Alday and J.~M. Maldacena, \emph{{Gluon Scattering Amplitudes at Strong
  Coupling}}, \href{https://doi.org/10.1088/1126-6708/2007/06/064}{\emph{JHEP}
  {\bfseries 06} (2007) 064} [\href{https://arxiv.org/abs/0705.0303}{{\ttfamily
  0705.0303}}].

\bibitem{Drummond:2008vq}
J.~Drummond, J.~Henn, G.~Korchemsky and E.~Sokatchev, \emph{{Dual
  Superconformal Symmetry of Scattering Amplitudes in $\mathcal{N}\!=\!4$ super
  Yang-Mills Theory}},
  \href{https://doi.org/10.1016/j.nuclphysb.2009.11.022}{\emph{Nucl. Phys.}
  {\bfseries B828} (2010) 317}
  [\href{https://arxiv.org/abs/0807.1095}{{\ttfamily 0807.1095}}].

\bibitem{Drummond:2007aua}
J.~M. Drummond, G.~P. Korchemsky and E.~Sokatchev, \emph{{Conformal Properties
  of Four-Gluon Planar Amplitudes and Wilson loops}},
  \href{https://doi.org/10.1016/j.nuclphysb.2007.11.041}{\emph{Nucl. Phys.}
  {\bfseries B795} (2008) 385}
  [\href{https://arxiv.org/abs/0707.0243}{{\ttfamily 0707.0243}}].

\bibitem{Brandhuber:2007yx}
A.~Brandhuber, P.~Heslop and G.~Travaglini, \emph{{MHV Amplitudes in
  $\mathcal{N}\!=\!4$ Super Yang-Mills and Wilson Loops}},
  \href{https://doi.org/10.1016/j.nuclphysb.2007.11.002}{\emph{Nucl. Phys.}
  {\bfseries B794} (2008) 231}
  [\href{https://arxiv.org/abs/0707.1153}{{\ttfamily 0707.1153}}].

\bibitem{Drummond:2007au}
J.~M. Drummond, J.~Henn, G.~P. Korchemsky and E.~Sokatchev, \emph{{Conformal
  Ward Identities for Wilson Loops and a Test of the Duality with Gluon
  Amplitudes}},
  \href{https://doi.org/10.1016/j.nuclphysb.2009.10.013}{\emph{Nucl. Phys.}
  {\bfseries B826} (2010) 337}
  [\href{https://arxiv.org/abs/0712.1223}{{\ttfamily 0712.1223}}].

\bibitem{Mason:2010yk}
L.~Mason and D.~Skinner, \emph{{The Complete Planar $S$-Matrix of
  $\mathcal{N}\!=\!4$ SYM as a Wilson Loop in Twistor Space}},
  \href{https://doi.org/10.1007/JHEP12(2010)018}{\emph{JHEP} {\bfseries 12}
  (2010) 018} [\href{https://arxiv.org/abs/1009.2225}{{\ttfamily 1009.2225}}].

\bibitem{CaronHuot:2010ek}
S.~Caron-Huot, \emph{{Notes on the Scattering Amplitude / Wilson Loop
  Duality}}, \href{https://doi.org/10.1007/JHEP07(2011)058}{\emph{JHEP}
  {\bfseries 1107} (2011) 058}
  [\href{https://arxiv.org/abs/1010.1167}{{\ttfamily 1010.1167}}].

\bibitem{Alday:2010zy}
L.~F. Alday, B.~Eden, G.~P. Korchemsky, J.~Maldacena and E.~Sokatchev,
  \emph{{From Correlation Functions to Wilson Loops}},
  \href{https://doi.org/10.1007/JHEP09(2011)123}{\emph{JHEP} {\bfseries 1109}
  (2011) 123} [\href{https://arxiv.org/abs/1007.3243}{{\ttfamily 1007.3243}}].

\bibitem{Eden:2010zz}
B.~Eden, G.~P. Korchemsky and E.~Sokatchev, \emph{{From Correlation Functions
  to Scattering Amplitudes}},
  \href{https://doi.org/10.1007/JHEP12(2011)002}{\emph{JHEP} {\bfseries 1112}
  (2011) 002} [\href{https://arxiv.org/abs/1007.3246}{{\ttfamily 1007.3246}}].

\bibitem{Eden:2010ce}
B.~Eden, G.~P. Korchemsky and E.~Sokatchev, \emph{{More on the Duality
  Correlators/Amplitudes}},
  \href{https://doi.org/10.1016/j.physletb.2012.02.014}{\emph{Phys. Lett.}
  {\bfseries B709} (2012) 247}
  [\href{https://arxiv.org/abs/1009.2488}{{\ttfamily 1009.2488}}].

\bibitem{ArkaniHamed:2012nw}
N.~Arkani-Hamed, J.~L. Bourjaily, F.~Cachazo, A.~B. Goncharov, A.~Postnikov and
  J.~Trnka, \emph{{Scattering Amplitudes and the Positive Grassmannian}},
  \href{https://arxiv.org/abs/1212.5605}{{\ttfamily 1212.5605}}.

\bibitem{ArkaniHamed:2009dn}
N.~Arkani-Hamed, F.~Cachazo, C.~Cheung and J.~Kaplan, \emph{{A Duality For The
  $S$-Matrix}}, \href{https://doi.org/10.1007/JHEP03(2010)020}{\emph{JHEP}
  {\bfseries 1003} (2010) 020}
  [\href{https://arxiv.org/abs/0907.5418}{{\ttfamily 0907.5418}}].

\bibitem{ArkaniHamed:2009vw}
N.~Arkani-Hamed, F.~Cachazo and C.~Cheung, \emph{{The Grassmannian Origin Of
  Dual Superconformal Invariance}},
  \href{https://doi.org/10.1007/JHEP03(2010)036}{\emph{JHEP} {\bfseries 1003}
  (2010) 036} [\href{https://arxiv.org/abs/0909.0483}{{\ttfamily 0909.0483}}].

\bibitem{Mason:2009qx}
L.~Mason and D.~Skinner, \emph{{Dual Superconformal Invariance, Momentum
  Twistors and Grassmannians}},
  \href{https://doi.org/10.1088/1126-6708/2009/11/045}{\emph{JHEP} {\bfseries
  0911} (2009) 045} [\href{https://arxiv.org/abs/0909.0250}{{\ttfamily
  0909.0250}}].

\bibitem{ArkaniHamed:2009dg}
N.~Arkani-Hamed, J.~Bourjaily, F.~Cachazo and J.~Trnka, \emph{{Unification of
  Residues and Grassmannian Dualities}},
  \href{https://doi.org/10.1007/JHEP01(2011)049}{\emph{JHEP} {\bfseries 1101}
  (2011) 049} [\href{https://arxiv.org/abs/0912.4912}{{\ttfamily 0912.4912}}].

\bibitem{ArkaniHamed:2009sx}
N.~Arkani-Hamed, J.~Bourjaily, F.~Cachazo and J.~Trnka, \emph{{Local Spacetime
  Physics from the Grassmannian}},
  \href{https://doi.org/10.1007/JHEP01(2011)108}{\emph{JHEP} {\bfseries 1101}
  (2011) 108} [\href{https://arxiv.org/abs/0912.3249}{{\ttfamily 0912.3249}}].

\bibitem{Huang:2013owa}
Y.-T. Huang and C.~Wen, \emph{{ABJM Amplitudes and the Positive Orthogonal
  Grassmannian}}, \href{https://doi.org/10.1007/JHEP02(2014)104}{\emph{JHEP}
  {\bfseries 1402} (2014) 104}
  [\href{https://arxiv.org/abs/1309.3252}{{\ttfamily 1309.3252}}].

\bibitem{ArkaniHamed:2010gg}
N.~Arkani-Hamed, J.~L. Bourjaily, F.~Cachazo, A.~Hodges and J.~Trnka, \emph{{A
  {N}ote on {P}olytopes for {S}cattering {A}mplitudes}},
  \href{https://doi.org/10.1007/JHEP04(2012)081}{\emph{JHEP} {\bfseries 1204}
  (2012) 081} [\href{https://arxiv.org/abs/1012.6030}{{\ttfamily 1012.6030}}].

\bibitem{Arkani-Hamed:2013jha}
N.~Arkani-Hamed and J.~Trnka, \emph{{The Amplituhedron}},
  \href{https://doi.org/10.1007/JHEP10(2014)030}{\emph{JHEP} {\bfseries 1410}
  (2014) 30} [\href{https://arxiv.org/abs/1312.2007}{{\ttfamily 1312.2007}}].

\bibitem{Dixon:2011pw}
L.~J. Dixon, J.~M. Drummond and J.~M. Henn, \emph{{Bootstrapping the Three-Loop
  Hexagon}}, \href{https://doi.org/10.1007/JHEP11(2011)023}{\emph{JHEP}
  {\bfseries 1111} (2011) 023}
  [\href{https://arxiv.org/abs/1108.4461}{{\ttfamily 1108.4461}}].

\bibitem{Dixon:2011nj}
L.~J. Dixon, J.~M. Drummond and J.~M. Henn, \emph{{Analytic Result for the
  Two-Loop Six-Point NMHV Amplitude in $\mathcal{N}\!=\!4$ Super Yang-Mills
  Theory}}, \href{https://doi.org/10.1007/JHEP01(2012)024}{\emph{JHEP}
  {\bfseries 1201} (2012) 024}
  [\href{https://arxiv.org/abs/1111.1704}{{\ttfamily 1111.1704}}].

\bibitem{Dixon:2013eka}
L.~J. Dixon, J.~M. Drummond, M.~von Hippel and J.~Pennington, \emph{{Hexagon
  Functions and the Three-Loop Remainder Function}},
  \href{https://doi.org/10.1007/JHEP12(2013)049}{\emph{JHEP} {\bfseries 1312}
  (2013) 049} [\href{https://arxiv.org/abs/1308.2276}{{\ttfamily 1308.2276}}].

\bibitem{Duhr:2011zq}
C.~Duhr, H.~Gangl and J.~R. Rhodes, \emph{{From polygons and symbols to
  polylogarithmic functions}},
  \href{https://doi.org/10.1007/JHEP10(2012)075}{\emph{JHEP} {\bfseries 10}
  (2012) 075} [\href{https://arxiv.org/abs/1110.0458}{{\ttfamily 1110.0458}}].

\bibitem{Bern:2008qj}
Z.~Bern, J.~Carrasco and H.~Johansson, \emph{{New Relations for Gauge-Theory
  Amplitudes}}, \href{https://doi.org/10.1103/PhysRevD.78.085011}{\emph{Phys.
  Rev.} {\bfseries D78} (2008) 085011}
  [\href{https://arxiv.org/abs/0805.3993}{{\ttfamily 0805.3993}}].

\bibitem{Bern:2010ue}
Z.~Bern, J.~J.~M. Carrasco and H.~Johansson, \emph{{Perturbative Quantum
  Gravity as a Double Copy of Gauge Theory}},
  \href{https://doi.org/10.1103/PhysRevLett.105.061602}{\emph{Phys. Rev. Lett.}
  {\bfseries 105} (2010) 061602}
  [\href{https://arxiv.org/abs/1004.0476}{{\ttfamily 1004.0476}}].

\bibitem{Bern:2019prr}
Z.~Bern, J.~J. Carrasco, M.~Chiodaroli, H.~Johansson and R.~Roiban, \emph{{The
  Duality Between Color and Kinematics and its Applications}},
  \href{https://arxiv.org/abs/1909.01358}{{\ttfamily 1909.01358}}.

\bibitem{Bern:2010tq}
Z.~Bern, J.~Carrasco, L.~J. Dixon, H.~Johansson and R.~Roiban, \emph{{The
  Complete Four-Loop Four-Point Amplitude in $\mathcal{N}\!=\!4$
  Super-Yang-Mills Theory}},
  \href{https://doi.org/10.1103/PhysRevD.82.125040}{\emph{Phys. Rev.}
  {\bfseries D82} (2010) 125040}
  [\href{https://arxiv.org/abs/1008.3327}{{\ttfamily 1008.3327}}].

\bibitem{Carrasco:2011mn}
J.~J. Carrasco and H.~Johansson, \emph{{Five-Point Amplitudes in
  $\mathcal{N}\!=\!4$ Super-Yang-Mills Theory and $\mathcal{N}\!=\!8$
  Supergravity}}, \href{https://doi.org/10.1103/PhysRevD.85.025006}{\emph{Phys.
  Rev.} {\bfseries D85} (2012) 025006}
  [\href{https://arxiv.org/abs/1106.4711}{{\ttfamily 1106.4711}}].

\bibitem{Bern:2011rj}
Z.~Bern, C.~Boucher-Veronneau and H.~Johansson, \emph{{N \ensuremath{>}= 4
  Supergravity Amplitudes from Gauge Theory at One Loop}},
  \href{https://doi.org/10.1103/PhysRevD.84.105035}{\emph{Phys. Rev. D}
  {\bfseries 84} (2011) 105035}
  [\href{https://arxiv.org/abs/1107.1935}{{\ttfamily 1107.1935}}].

\bibitem{Boucher-Veronneau:2011rlc}
C.~Boucher-Veronneau and L.~J. Dixon, \emph{{N \ensuremath{>}- 4 Supergravity
  Amplitudes from Gauge Theory at Two Loops}},
  \href{https://doi.org/10.1007/JHEP12(2011)046}{\emph{JHEP} {\bfseries 12}
  (2011) 046} [\href{https://arxiv.org/abs/1110.1132}{{\ttfamily 1110.1132}}].

\bibitem{Bern:2018jmv}
Z.~Bern, J.~J. Carrasco, W.-M. Chen, A.~Edison, H.~Johansson, J.~Parra-Martinez
  et~al., \emph{{Ultraviolet Properties of $\mathcal N = 8$ Supergravity at
  Five Loops}}, \href{https://doi.org/10.1103/PhysRevD.98.086021}{\emph{Phys.
  Rev. D} {\bfseries 98} (2018) 086021}
  [\href{https://arxiv.org/abs/1804.09311}{{\ttfamily 1804.09311}}].

\bibitem{Cheung:2018wkq}
C.~Cheung, I.~Z. Rothstein and M.~P. Solon, \emph{{From Scattering Amplitudes
  to Classical Potentials in the Post-Minkowskian Expansion}},
  \href{https://doi.org/10.1103/PhysRevLett.121.251101}{\emph{Phys. Rev. Lett.}
  {\bfseries 121} (2018) 251101}
  [\href{https://arxiv.org/abs/1808.02489}{{\ttfamily 1808.02489}}].

\bibitem{Kosower:2018adc}
D.~A. Kosower, B.~Maybee and D.~O'Connell, \emph{{Amplitudes, Observables, and
  Classical Scattering}},
  \href{https://doi.org/10.1007/JHEP02(2019)137}{\emph{JHEP} {\bfseries 02}
  (2019) 137} [\href{https://arxiv.org/abs/1811.10950}{{\ttfamily
  1811.10950}}].

\bibitem{Bern:2019nnu}
Z.~Bern, C.~Cheung, R.~Roiban, C.-H. Shen, M.~P. Solon and M.~Zeng,
  \emph{{Scattering Amplitudes and the Conservative Hamiltonian for Binary
  Systems at Third Post-Minkowskian Order}},
  \href{https://doi.org/10.1103/PhysRevLett.122.201603}{\emph{Phys. Rev. Lett.}
  {\bfseries 122} (2019) 201603}
  [\href{https://arxiv.org/abs/1901.04424}{{\ttfamily 1901.04424}}].

\bibitem{Bern:2022wqg}
Z.~Bern, J.~J. Carrasco, M.~Chiodaroli, H.~Johansson and R.~Roiban, \emph{{The
  SAGEX review on scattering amplitudes Chapter 2: An invitation to
  color-kinematics duality and the double copy}},
  \href{https://doi.org/10.1088/1751-8121/ac93cf}{\emph{J. Phys. A} {\bfseries
  55} (2022) 443003} [\href{https://arxiv.org/abs/2203.13013}{{\ttfamily
  2203.13013}}].

\bibitem{Bjerrum-Bohr:2022blt}
N.~E.~J. Bjerrum-Bohr, P.~H. Damgaard, L.~Plante and P.~Vanhove, \emph{{The
  SAGEX review on scattering amplitudes Chapter 13: Post-Minkowskian expansion
  from scattering amplitudes}},
  \href{https://doi.org/10.1088/1751-8121/ac7a78}{\emph{J. Phys. A} {\bfseries
  55} (2022) 443014} [\href{https://arxiv.org/abs/2203.13024}{{\ttfamily
  2203.13024}}].

\bibitem{Kosower:2022yvp}
D.~A. Kosower, R.~Monteiro and D.~O'Connell, \emph{{The SAGEX review on
  scattering amplitudes Chapter 14: Classical gravity from scattering
  amplitudes}}, \href{https://doi.org/10.1088/1751-8121/ac8846}{\emph{J. Phys.
  A} {\bfseries 55} (2022) 443015}
  [\href{https://arxiv.org/abs/2203.13025}{{\ttfamily 2203.13025}}].

\bibitem{Bern:1996je}
Z.~Bern, L.~J. Dixon and D.~A. Kosower, \emph{{Progress in One-Loop QCD
  Computations}},
  \href{https://doi.org/10.1146/annurev.nucl.46.1.109}{\emph{Ann. Rev. Nucl.
  Part. Sci.} {\bfseries 46} (1996) 109}
  [\href{https://arxiv.org/abs/hep-ph/9602280}{{\ttfamily hep-ph/9602280}}].

\bibitem{Cachazo:2005ga}
F.~Cachazo and P.~Svrcek, \emph{{Lectures on Twistor Strings and Perturbative
  Yang-Mills Theory}}, {\emph{PoS} {\bfseries RTN2005} (2005) 004}
  [\href{https://arxiv.org/abs/hep-th/0504194}{{\ttfamily hep-th/0504194}}].

\bibitem{Beisert:2010jr}
N.~Beisert, C.~Ahn, L.~F. Alday, Z.~Bajnok, J.~M. Drummond et~al.,
  \emph{{Review of AdS/CFT Integrability: An Overview}},
  \href{https://doi.org/10.1007/s11005-011-0529-2}{\emph{Lett. Math. Phys.}
  {\bfseries 99} (2012) 3} [\href{https://arxiv.org/abs/1012.3982}{{\ttfamily
  1012.3982}}].

\bibitem{Elvang:2013cua}
H.~Elvang and Y.-t. Huang, \emph{{Scattering Amplitudes}},
  \href{https://arxiv.org/abs/1308.1697}{{\ttfamily 1308.1697}}.

\bibitem{Dixon:2013uaa}
L.~J. Dixon, \emph{{A Brief Introduction to Modern Amplitude Methods}},  in
  \emph{{Proceedings, 2012 European School of High-Energy Physics (ESHEP 2012):
  La Pommeraye, Anjou, France, June 06-19, 2012}}, pp.~31--67, 2014,
  \href{https://arxiv.org/abs/1310.5353}{{\ttfamily 1310.5353}},
  \href{https://doi.org/10.5170/CERN-2014-008.31}{DOI}.

\bibitem{Henn:2014yza}
J.~M. Henn and J.~C. Plefka, \emph{{Scattering Amplitudes in Gauge Theories}},
  \href{https://doi.org/978-3-642-54021-9,
  10.1007/978-3-642-54022-6}{\emph{Lect. Notes Phys.} {\bfseries 883} (2014)
  1}.

\bibitem{Duhr:2014woa}
C.~Duhr, \emph{{Mathematical aspects of scattering amplitudes}},  in
  \emph{{Theoretical Advanced Study Institute in Elementary Particle Physics}:
  {Journeys Through the Precision Frontier: Amplitudes for Colliders}},
  pp.~419--476, 2015, \href{https://arxiv.org/abs/1411.7538}{{\ttfamily
  1411.7538}}, \href{https://doi.org/10.1142/9789814678766_0010}{DOI}.

\bibitem{Weinzierl:2016bus}
S.~Weinzierl, \emph{{Tales of 1001 Gluons}},
  \href{https://arxiv.org/abs/1610.05318}{{\ttfamily 1610.05318}}.

\bibitem{Travaglini:2022uwo}
G.~Travaglini et~al., \emph{{The SAGEX review on scattering amplitudes}},
  \href{https://doi.org/10.1088/1751-8121/ac8380}{\emph{J. Phys. A} {\bfseries
  55} (2022) 443001} [\href{https://arxiv.org/abs/2203.13011}{{\ttfamily
  2203.13011}}].

\bibitem{Brandhuber:2022qbk}
A.~Brandhuber, J.~Plefka and G.~Travaglini, \emph{{The SAGEX Review on
  Scattering Amplitudes Chapter 1: Modern Fundamentals of Amplitudes}},
  \href{https://doi.org/10.1088/1751-8121/ac8254}{\emph{J. Phys. A} {\bfseries
  55} (2022) 443002} [\href{https://arxiv.org/abs/2203.13012}{{\ttfamily
  2203.13012}}].

\bibitem{Simmons-Duffin:2016gjk}
D.~Simmons-Duffin, \emph{{The Conformal Bootstrap}},  in \emph{{Theoretical
  Advanced Study Institute in Elementary Particle Physics}: {New Frontiers in
  Fields and Strings}}, pp.~1--74, 2017,
  \href{https://arxiv.org/abs/1602.07982}{{\ttfamily 1602.07982}},
  \href{https://doi.org/10.1142/9789813149441_0001}{DOI}.

\bibitem{Mueller:1979ih}
A.~H. Mueller, \emph{{On the Asymptotic Behavior of the Sudakov Form-factor}},
  \href{https://doi.org/10.1103/PhysRevD.20.2037}{\emph{Phys. Rev.} {\bfseries
  D20} (1979) 2037}.

\bibitem{Sen:1981sd}
A.~Sen, \emph{{Asymptotic Behavior of the Sudakov Form-Factor in QCD}},
  \href{https://doi.org/10.1103/PhysRevD.24.3281}{\emph{Phys. Rev. D}
  {\bfseries 24} (1981) 3281}.

\bibitem{Collins:1980ih}
J.~C. Collins, \emph{{Algorithm to Compute Corrections to the Sudakov
  Form-Factor}}, \href{https://doi.org/10.1103/PhysRevD.22.1478}{\emph{Phys.
  Rev.} {\bfseries D22} (1980) 1478}.

\bibitem{Becher:2014oda}
T.~Becher, A.~Broggio and A.~Ferroglia, \emph{{Introduction to Soft-Collinear
  Effective Theory}}, vol.~896. Springer, 2015,
  \href{https://doi.org/10.1007/978-3-319-14848-9}{10.1007/978-3-319-14848-9},
  [\href{https://arxiv.org/abs/1410.1892}{{\ttfamily 1410.1892}}].

\bibitem{Lee:2022nhh}
R.~N. Lee, A.~von Manteuffel, R.~M. Schabinger, A.~V. Smirnov, V.~A. Smirnov
  and M.~Steinhauser, \emph{{Quark and Gluon Form Factors in Four-Loop QCD}},
  \href{https://doi.org/10.1103/PhysRevLett.128.212002}{\emph{Phys. Rev. Lett.}
  {\bfseries 128} (2022) 212002}
  [\href{https://arxiv.org/abs/2202.04660}{{\ttfamily 2202.04660}}].

\bibitem{vanNeerven:1985ja}
W.~L. van Neerven, \emph{{Infrared Behavior of On-shell Form-factors in a $N=4$
  Supersymmetric {Yang-Mills} Field Theory}},
  \href{https://doi.org/10.1007/BF01571808}{\emph{Z. Phys. C} {\bfseries 30}
  (1986) 595}.

\bibitem{Brandhuber:2010ad}
A.~Brandhuber, B.~Spence, G.~Travaglini and G.~Yang, \emph{{Form Factors in
  $\mathcal{N}\!=\!4$ Super Yang-Mills and Periodic Wilson Loops}},
  \href{https://doi.org/10.1007/JHEP01(2011)134}{\emph{JHEP} {\bfseries 1101}
  (2011) 134} [\href{https://arxiv.org/abs/1011.1899}{{\ttfamily 1011.1899}}].

\bibitem{Bork:2010wf}
L.~V. Bork, D.~I. Kazakov and G.~S. Vartanov, \emph{{On form factors in N=4
  sym}}, \href{https://doi.org/10.1007/JHEP02(2011)063}{\emph{JHEP} {\bfseries
  02} (2011) 063} [\href{https://arxiv.org/abs/1011.2440}{{\ttfamily
  1011.2440}}].

\bibitem{Brandhuber:2011tv}
A.~Brandhuber, O.~Gurdogan, R.~Mooney, G.~Travaglini and G.~Yang,
  \emph{{Harmony of Super Form Factors}},
  \href{https://doi.org/10.1007/JHEP10(2011)046}{\emph{JHEP} {\bfseries 1110}
  (2011) 046} [\href{https://arxiv.org/abs/1107.5067}{{\ttfamily 1107.5067}}].

\bibitem{Bork:2011cj}
L.~V. Bork, D.~I. Kazakov and G.~S. Vartanov, \emph{{On MHV Form Factors in
  Superspace for $\mathcal{N}=4$ SYM Theory}},
  \href{https://doi.org/10.1007/JHEP10(2011)133}{\emph{JHEP} {\bfseries 10}
  (2011) 133} [\href{https://arxiv.org/abs/1107.5551}{{\ttfamily 1107.5551}}].

\bibitem{Henn:2011by}
J.~M. Henn, S.~Moch and S.~G. Naculich, \emph{{Form factors and scattering
  amplitudes in N=4 SYM in dimensional and massive regularizations}},
  \href{https://doi.org/10.1007/JHEP12(2011)024}{\emph{JHEP} {\bfseries 12}
  (2011) 024} [\href{https://arxiv.org/abs/1109.5057}{{\ttfamily 1109.5057}}].

\bibitem{Gehrmann:2011xn}
T.~Gehrmann, J.~M. Henn and T.~Huber, \emph{{The Three-Loop Form Factor in
  $\mathcal{N}\!=\!4$ Super Yang-Mills}},
  \href{https://doi.org/10.1007/JHEP03(2012)101}{\emph{JHEP} {\bfseries 1203}
  (2012) 101} [\href{https://arxiv.org/abs/1112.4524}{{\ttfamily 1112.4524}}].

\bibitem{Brandhuber:2012vm}
A.~Brandhuber, G.~Travaglini and G.~Yang, \emph{{Analytic Two-Loop Form Factors
  in $\mathcal{N}\!=\!4$ SYM}},
  \href{https://doi.org/10.1007/JHEP05(2012)082}{\emph{JHEP} {\bfseries 1205}
  (2012) 082} [\href{https://arxiv.org/abs/1201.4170}{{\ttfamily 1201.4170}}].

\bibitem{Bork:2012tt}
L.~V. Bork, \emph{{On NMHV form factors in N=4 SYM theory from generalized
  unitarity}}, \href{https://doi.org/10.1007/JHEP01(2013)049}{\emph{JHEP}
  {\bfseries 01} (2013) 049} [\href{https://arxiv.org/abs/1203.2596}{{\ttfamily
  1203.2596}}].

\bibitem{Boels:2012ew}
R.~H. Boels, B.~A. Kniehl, O.~V. Tarasov and G.~Yang, \emph{{Color-Kinematic
  Duality for Form Factors}},
  \href{https://doi.org/10.1007/JHEP02(2013)063}{\emph{JHEP} {\bfseries 02}
  (2013) 063} [\href{https://arxiv.org/abs/1211.7028}{{\ttfamily 1211.7028}}].

\bibitem{Penante:2014sza}
B.~Penante, B.~Spence, G.~Travaglini and C.~Wen, \emph{{On super form factors
  of half-BPS operators in N=4 super Yang-Mills}},
  \href{https://doi.org/10.1007/JHEP04(2014)083}{\emph{JHEP} {\bfseries 04}
  (2014) 083} [\href{https://arxiv.org/abs/1402.1300}{{\ttfamily 1402.1300}}].

\bibitem{Wilhelm:2014qua}
M.~Wilhelm, \emph{{Amplitudes, Form Factors and the Dilatation Operator in
  $\mathcal{N}=4$ SYM Theory}},
  \href{https://doi.org/10.1007/JHEP02(2015)149}{\emph{JHEP} {\bfseries 02}
  (2015) 149} [\href{https://arxiv.org/abs/1410.6309}{{\ttfamily 1410.6309}}].

\bibitem{Nandan:2014oga}
D.~Nandan, C.~Sieg, M.~Wilhelm and G.~Yang, \emph{{Cutting through form factors
  and cross sections of non-protected operators in $ \mathcal{N}=4 $ SYM}},
  \href{https://doi.org/10.1007/JHEP06(2015)156}{\emph{JHEP} {\bfseries 06}
  (2015) 156} [\href{https://arxiv.org/abs/1410.8485}{{\ttfamily 1410.8485}}].

\bibitem{Loebbert:2015ova}
F.~Loebbert, D.~Nandan, C.~Sieg, M.~Wilhelm and G.~Yang, \emph{{On-Shell
  Methods for the Two-Loop Dilatation Operator and Finite Remainders}},
  \href{https://doi.org/10.1007/JHEP10(2015)012}{\emph{JHEP} {\bfseries 10}
  (2015) 012} [\href{https://arxiv.org/abs/1504.06323}{{\ttfamily
  1504.06323}}].

\bibitem{Frassek:2015rka}
R.~Frassek, D.~Meidinger, D.~Nandan and M.~Wilhelm, \emph{{On-Shell Diagrams,
  Gra{\ss}mannians and Integrability for Form factors}},
  \href{https://doi.org/10.1007/JHEP01(2016)182}{\emph{JHEP} {\bfseries 01}
  (2016) 182} [\href{https://arxiv.org/abs/1506.08192}{{\ttfamily
  1506.08192}}].

\bibitem{Brandhuber:2017bkg}
A.~Brandhuber, M.~Kostacinska, B.~Penante and G.~Travaglini, \emph{{Higgs
  amplitudes from $\mathcal{N}=4$ super Yang-Mills theory}},
  \href{https://doi.org/10.1103/PhysRevLett.119.161601}{\emph{Phys. Rev. Lett.}
  {\bfseries 119} (2017) 161601}
  [\href{https://arxiv.org/abs/1707.09897}{{\ttfamily 1707.09897}}].

\bibitem{Bianchi:2018rrj}
L.~Bianchi, A.~Brandhuber, R.~Panerai and G.~Travaglini, \emph{{Dual conformal
  invariance for form factors}},
  \href{https://doi.org/10.1007/JHEP02(2019)134}{\emph{JHEP} {\bfseries 02}
  (2019) 134} [\href{https://arxiv.org/abs/1812.10468}{{\ttfamily
  1812.10468}}].

\bibitem{Brandhuber:2018xzk}
A.~Brandhuber, M.~Kostacinska, B.~Penante and G.~Travaglini,
  \emph{{$\text{Tr}(F^3)$ supersymmetric form factors and maximal
  transcendentality Part I: $\mathcal{N}=4$ super Yang-Mills}},
  \href{https://doi.org/10.1007/JHEP12(2018)076}{\emph{JHEP} {\bfseries 12}
  (2018) 076} [\href{https://arxiv.org/abs/1804.05703}{{\ttfamily
  1804.05703}}].

\bibitem{Yang:2016ear}
G.~Yang, \emph{{Color-kinematics duality and Sudakov form factor at five loops
  for N=4 supersymmetric Yang-Mills theory}},
  \href{https://doi.org/10.1103/PhysRevLett.117.271602}{\emph{Phys. Rev. Lett.}
  {\bfseries 117} (2016) 271602}
  [\href{https://arxiv.org/abs/1610.02394}{{\ttfamily 1610.02394}}].

\bibitem{Sever:2020jjx}
A.~Sever, A.~G. Tumanov and M.~Wilhelm, \emph{{Operator Product Expansion for
  Form Factors}},
  \href{https://doi.org/10.1103/PhysRevLett.126.031602}{\emph{Phys. Rev. Lett.}
  {\bfseries 126} (2021) 031602}
  [\href{https://arxiv.org/abs/2009.11297}{{\ttfamily 2009.11297}}].

\bibitem{Sever:2021nsq}
A.~Sever, A.~G. Tumanov and M.~Wilhelm, \emph{{An Operator Product Expansion
  for Form Factors II. Born level}},
  \href{https://doi.org/10.1007/JHEP10(2021)071}{\emph{JHEP} {\bfseries 10}
  (2021) 071} [\href{https://arxiv.org/abs/2105.13367}{{\ttfamily
  2105.13367}}].

\bibitem{Alday:2007he}
L.~F. Alday and J.~Maldacena, \emph{{Comments on Gluon Scattering Amplitudes
  via AdS/CFT}},
  \href{https://doi.org/10.1088/1126-6708/2007/11/068}{\emph{JHEP} {\bfseries
  11} (2007) 068} [\href{https://arxiv.org/abs/0710.1060}{{\ttfamily
  0710.1060}}].

\bibitem{Maldacena:2010kp}
J.~Maldacena and A.~Zhiboedov, \emph{{Form factors at strong coupling via a
  Y-system}}, \href{https://doi.org/10.1007/JHEP11(2010)104}{\emph{JHEP}
  {\bfseries 11} (2010) 104} [\href{https://arxiv.org/abs/1009.1139}{{\ttfamily
  1009.1139}}].

\bibitem{Gao:2013dza}
Z.~Gao and G.~Yang, \emph{{Y-system for form factors at strong coupling in
  $AdS_5$ and with multi-operator insertions in $AdS_3$}},
  \href{https://doi.org/10.1007/JHEP06(2013)105}{\emph{JHEP} {\bfseries 06}
  (2013) 105} [\href{https://arxiv.org/abs/1303.2668}{{\ttfamily 1303.2668}}].

\bibitem{Sever:2021xga}
A.~Sever, A.~G. Tumanov and M.~Wilhelm, \emph{{An Operator Product Expansion
  for Form Factors III. Finite Coupling and Multi-Particle Contributions}},
  \href{https://doi.org/10.1007/JHEP03(2022)128}{\emph{JHEP} {\bfseries 03}
  (2022) 128} [\href{https://arxiv.org/abs/2112.10569}{{\ttfamily
  2112.10569}}].

\bibitem{Bern:1994zx}
Z.~Bern, L.~J. Dixon, D.~C. Dunbar and D.~A. Kosower, \emph{{One-Loop $n$-Point
  Gauge Theory Amplitudes, Unitarity and Collinear Limits}},
  \href{https://doi.org/10.1016/0550-3213(94)90179-1}{\emph{Nucl. Phys.}
  {\bfseries B425} (1994) 217}
  [\href{https://arxiv.org/abs/hep-ph/9403226}{{\ttfamily hep-ph/9403226}}].

\bibitem{Bern:1994cg}
Z.~Bern, L.~J. Dixon, D.~C. Dunbar and D.~A. Kosower, \emph{{Fusing Gauge
  Theory Tree Amplitudes into Loop Amplitudes}},
  \href{https://doi.org/10.1016/0550-3213(94)00488-Z}{\emph{Nucl. Phys.}
  {\bfseries B435} (1995) 59}
  [\href{https://arxiv.org/abs/hep-ph/9409265}{{\ttfamily hep-ph/9409265}}].

\bibitem{Britto:2004nc}
R.~Britto, F.~Cachazo and B.~Feng, \emph{{Generalized Unitarity and One-Loop
  Amplitudes in $\mathcal{N}\!=\!4$ Super-Yang-Mills}},
  \href{https://doi.org/10.1016/j.nuclphysb.2005.07.014}{\emph{Nucl. Phys.}
  {\bfseries B725} (2005) 275}
  [\href{https://arxiv.org/abs/hep-th/0412103}{{\ttfamily hep-th/0412103}}].

\bibitem{Yang:2019vag}
G.~Yang, \emph{{On-shell methods for form factors in $\mathcal{N}=4$ SYM and
  their applications}},
  \href{https://doi.org/10.1007/s11433-019-1507-0}{\emph{Sci. China Phys. Mech.
  Astron.} {\bfseries 63} (2020) 270001}
  [\href{https://arxiv.org/abs/1912.11454}{{\ttfamily 1912.11454}}].

\bibitem{Dixon:2021tdw}
L.~J. Dixon, O.~Gurdogan, A.~J. McLeod and M.~Wilhelm, \emph{{Folding
  Amplitudes into Form Factors: An Antipodal Duality}},
  \href{https://doi.org/10.1103/PhysRevLett.128.111602}{\emph{Phys. Rev. Lett.}
  {\bfseries 128} (2022) 111602}
  [\href{https://arxiv.org/abs/2112.06243}{{\ttfamily 2112.06243}}].

\bibitem{Liu:2022vck}
Y.-T. Liu, \emph{{Antipodal symmetry of two-loop MHV amplitudes}},
  \href{https://doi.org/10.1007/JHEP09(2022)131}{\emph{JHEP} {\bfseries 09}
  (2022) 131} [\href{https://arxiv.org/abs/2207.11815}{{\ttfamily
  2207.11815}}].

\bibitem{goncharov2001multiple}
A.~B. Goncharov, \emph{Multiple polylogarithms and mixed tate motives},
  \href{https://arxiv.org/abs/math/0103059}{{\ttfamily math/0103059}}.

\bibitem{goncharov2004galois}
A.~B. Goncharov, \emph{Galois symmetries of fundamental groupoids and
  noncommutative geometry},
  \href{https://arxiv.org/abs/math/0208144}{{\ttfamily math/0208144}}.

\bibitem{Goncharov:2009kx}
A.~B. {Goncharov}, \emph{{A Simple Construction of Grassmannian
  Polylogarithms}}, {\emph{{Adv. Math.}} (April, 2013) }
  [\href{https://arxiv.org/abs/0908.2238}{{\ttfamily 0908.2238}}].

\bibitem{Goncharov:2010jf}
A.~B. Goncharov, M.~Spradlin, C.~Vergu and A.~Volovich, \emph{{Classical
  Polylogarithms for Amplitudes and Wilson Loops}},
  \href{https://doi.org/10.1103/PhysRevLett.105.151605}{\emph{Phys. Rev. Lett.}
  {\bfseries 105} (2010) 151605}
  [\href{https://arxiv.org/abs/1006.5703}{{\ttfamily 1006.5703}}].

\bibitem{Brown:2011ik}
F.~Brown, \emph{{On the decomposition of motivic multiple zeta values}},
  \href{https://arxiv.org/abs/1102.1310}{{\ttfamily 1102.1310}}.

\bibitem{Duhr:2012fh}
C.~Duhr, \emph{{Hopf algebras, coproducts and symbols: an application to Higgs
  boson amplitudes}},
  \href{https://doi.org/10.1007/JHEP08(2012)043}{\emph{JHEP} {\bfseries 08}
  (2012) 043} [\href{https://arxiv.org/abs/1203.0454}{{\ttfamily 1203.0454}}].

\bibitem{Dixon:2022xqh}
L.~J. Dixon, O.~G\"urdo\u{g}an, Y.-T. Liu, A.~J. McLeod and M.~Wilhelm,
  \emph{{Antipodal Self-Duality for a Four-Particle Form Factor}},
  \href{https://doi.org/10.1103/PhysRevLett.130.111601}{\emph{Phys. Rev. Lett.}
  {\bfseries 130} (2023) 111601}
  [\href{https://arxiv.org/abs/2212.02410}{{\ttfamily 2212.02410}}].

\bibitem{Guo:2022qgv}
Y.~Guo, L.~Wang and G.~Yang, \emph{{Analytic Four-Point Lightlike Form Factors
  and OPE of Null-Wrapped Polygons}},
  \href{https://arxiv.org/abs/2209.06816}{{\ttfamily 2209.06816}}.

\bibitem{Abreu:2018aqd}
S.~Abreu, L.~J. Dixon, E.~Herrmann, B.~Page and M.~Zeng, \emph{{The two-loop
  five-point amplitude in $\mathcal{N} =4$ super-Yang-Mills theory}},
  \href{https://doi.org/10.1103/PhysRevLett.122.121603}{\emph{Phys. Rev. Lett.}
  {\bfseries 122} (2019) 121603}
  [\href{https://arxiv.org/abs/1812.08941}{{\ttfamily 1812.08941}}].

\bibitem{Chicherin:2018old}
D.~Chicherin, T.~Gehrmann, J.~M. Henn, P.~Wasser, Y.~Zhang and S.~Zoia,
  \emph{{All Master Integrals for Three-Jet Production at
  Next-to-Next-to-Leading Order}},
  \href{https://doi.org/10.1103/PhysRevLett.123.041603}{\emph{Phys. Rev. Lett.}
  {\bfseries 123} (2019) 041603}
  [\href{https://arxiv.org/abs/1812.11160}{{\ttfamily 1812.11160}}].

\bibitem{Chicherin:2020oor}
D.~Chicherin and V.~Sotnikov, \emph{{Pentagon Functions for Scattering of Five
  Massless Particles}},
  \href{https://doi.org/10.1007/JHEP12(2020)167}{\emph{JHEP} {\bfseries 20}
  (2020) 167} [\href{https://arxiv.org/abs/2009.07803}{{\ttfamily
  2009.07803}}].

\bibitem{Bourjaily:2019iqr}
J.~L. Bourjaily, E.~Herrmann, C.~Langer, A.~J. McLeod and J.~Trnka,
  \emph{{Prescriptive Unitarity for Non-Planar Six-Particle Amplitudes at Two
  Loops}}, \href{https://doi.org/10.1007/JHEP12(2019)073}{\emph{JHEP}
  {\bfseries 12} (2019) 073}
  [\href{https://arxiv.org/abs/1909.09131}{{\ttfamily 1909.09131}}].

\bibitem{Bourjaily:2019gqu}
J.~L. Bourjaily, E.~Herrmann, C.~Langer, A.~J. McLeod and J.~Trnka,
  \emph{{All-Multiplicity Nonplanar Amplitude Integrands in Maximally
  Supersymmetric Yang-Mills Theory at Two Loops}},
  \href{https://doi.org/10.1103/PhysRevLett.124.111603}{\emph{Phys. Rev. Lett.}
  {\bfseries 124} (2020) 111603}
  [\href{https://arxiv.org/abs/1911.09106}{{\ttfamily 1911.09106}}].

\bibitem{Bourjaily:2020qca}
J.~L. Bourjaily, E.~Herrmann, C.~Langer and J.~Trnka, \emph{{Building bases of
  loop integrands}}, \href{https://doi.org/10.1007/JHEP11(2020)116}{\emph{JHEP}
  {\bfseries 11} (2020) 116}
  [\href{https://arxiv.org/abs/2007.13905}{{\ttfamily 2007.13905}}].

\bibitem{Bourjaily:2017wjl}
J.~L. Bourjaily, E.~Herrmann and J.~Trnka, \emph{{Prescriptive Unitarity}},
  \href{https://doi.org/10.1007/JHEP06(2017)059}{\emph{JHEP} {\bfseries 06}
  (2017) 059} [\href{https://arxiv.org/abs/1704.05460}{{\ttfamily
  1704.05460}}].

\bibitem{ArkaniHamed:2010gh}
N.~Arkani-Hamed, J.~L. Bourjaily, F.~Cachazo and J.~Trnka, \emph{{Local
  Integrals for Planar Scattering Amplitudes}},
  \href{https://doi.org/10.1007/JHEP06(2012)125}{\emph{JHEP} {\bfseries 1206}
  (2012) 125} [\href{https://arxiv.org/abs/1012.6032}{{\ttfamily 1012.6032}}].

\bibitem{Arkani-Hamed:2014via}
N.~Arkani-Hamed, J.~L. Bourjaily, F.~Cachazo and J.~Trnka, \emph{{Singularity
  Structure of Maximally Supersymmetric Scattering Amplitudes}},
  \href{https://doi.org/10.1103/PhysRevLett.113.261603}{\emph{Phys. Rev. Lett.}
  {\bfseries 113} (2014) 261603}
  [\href{https://arxiv.org/abs/1410.0354}{{\ttfamily 1410.0354}}].

\bibitem{Abreu:2020jxa}
S.~Abreu, H.~Ita, F.~Moriello, B.~Page, W.~Tschernow and M.~Zeng,
  \emph{{Two-Loop Integrals for Planar Five-Point One-Mass Processes}},
  \href{https://doi.org/10.1007/JHEP11(2020)117}{\emph{JHEP} {\bfseries 11}
  (2020) 117} [\href{https://arxiv.org/abs/2005.04195}{{\ttfamily
  2005.04195}}].

\bibitem{Henn:2014qga}
J.~M. Henn, \emph{{Lectures on Differential Equations for Feynman Integrals}},
  \href{https://doi.org/10.1088/1751-8113/48/15/153001}{\emph{J. Phys.}
  {\bfseries A48} (2015) 153001}
  [\href{https://arxiv.org/abs/1412.2296}{{\ttfamily 1412.2296}}].

\bibitem{Badger:2023eqz}
S.~Badger, J.~Henn, J.~Plefka and S.~Zoia, \emph{{Scattering Amplitudes in
  Quantum Field Theory}},  \href{https://arxiv.org/abs/2306.05976}{{\ttfamily
  2306.05976}}.

\bibitem{Smirnov:2004ym}
V.~A. Smirnov, \emph{{Evaluating Feynman Integrals}}, {\emph{Springer Tracts
  Mod. Phys.} {\bfseries 211} (2004) 1}.

\bibitem{Smirnov:2006ry}
V.~A. Smirnov, \emph{{Feynman Integral Calculus}}. {Springer-Verlag}, 2006.

\bibitem{Smirnov:2012gma}
V.~A. Smirnov, \emph{{Analytic Tools for Feynman integrals}},
  \href{https://doi.org/10.1007/978-3-642-34886-0}{\emph{Springer Tracts Mod.
  Phys.} {\bfseries 250} (2012) 1}.

\bibitem{Badger:2021nhg}
S.~Badger, H.~B. Hartanto and S.~Zoia, \emph{{Two-Loop QCD Corrections to Wbb
  Production at Hadron Colliders}},
  \href{https://doi.org/10.1103/PhysRevLett.127.012001}{\emph{Phys. Rev. Lett.}
  {\bfseries 127} (2021) 012001}
  [\href{https://arxiv.org/abs/2102.02516}{{\ttfamily 2102.02516}}].

\bibitem{Ita:2011wn}
Z.~Bern, H.~Ita, L.~Dixon, F.~Ferbers~Cordero, D.~Kosower and D.~Maitre,
  \emph{{Precise Predictions for $Z\!+\!4$ Jets at Hadron Colliders}},
  \href{https://doi.org/10.1103/PhysRevD.85.021301}{\emph{Phys. Rev.}
  {\bfseries D85} (2012) 031501}
  [\href{https://arxiv.org/abs/1108.2229}{{\ttfamily 1108.2229}}].

\bibitem{Passarino:1978jh}
G.~Passarino and M.~Veltman, \emph{{One Loop Corrections for $e^+\,e^-$
  Annihilation Into $\mu^+\,\mu^-$ in the Weinberg Model}},
  \href{https://doi.org/10.1016/0550-3213(79)90234-7}{\emph{Nucl. Phys.}
  {\bfseries B160} (1979) 151}.

\bibitem{Ossola:2006us}
G.~Ossola, C.~G. Papadopoulos and R.~Pittau, \emph{{Reducing Full One-Loop
  Amplitudes to Scalar Integrals at the Integrand Level}},
  \href{https://doi.org/10.1016/j.nuclphysb.2006.11.012}{\emph{Nucl. Phys.}
  {\bfseries B763} (2007) 147}
  [\href{https://arxiv.org/abs/hep-ph/0609007}{{\ttfamily hep-ph/0609007}}].

\bibitem{Mastrolia:2010nb}
P.~Mastrolia, G.~Ossola, T.~Reiter and F.~Tramontano, \emph{{Scattering
  Amplitudes from Unitarity-Based Reduction Algorithm at the Integrand-Level}},
  \href{https://doi.org/10.1007/JHEP08(2010)080}{\emph{JHEP} {\bfseries 08}
  (2010) 080} [\href{https://arxiv.org/abs/1006.0710}{{\ttfamily 1006.0710}}].

\bibitem{Badger:2012dv}
S.~Badger, H.~Frellesvig and Y.~Zhang, \emph{{An Integrand Reconstruction
  Method for Three-Loop Amplitudes}},
  \href{https://doi.org/10.1007/JHEP08(2012)065}{\emph{JHEP} {\bfseries 1208}
  (2012) 065} [\href{https://arxiv.org/abs/1207.2976}{{\ttfamily 1207.2976}}].

\bibitem{Mastrolia:2012an}
P.~Mastrolia, E.~Mirabella, G.~Ossola and T.~Peraro, \emph{{Scattering
  Amplitudes from Multivariate Polynomial Division}},
  \href{https://doi.org/10.1016/j.physletb.2012.09.053}{\emph{Phys. Lett.}
  {\bfseries B718} (2012) 173}
  [\href{https://arxiv.org/abs/1205.7087}{{\ttfamily 1205.7087}}].

\bibitem{Badger:2013sta}
S.~Badger, H.~Frellesvig and Y.~Zhang, \emph{{Multi-loop Integrand Reduction
  with Computational Algebraic Geometry}},
  \href{https://doi.org/10.1088/1742-6596/523/1/012061}{\emph{J. Phys. Conf.
  Ser.} {\bfseries 523} (2014) 012061}
  [\href{https://arxiv.org/abs/1310.4445}{{\ttfamily 1310.4445}}].

\bibitem{Ita:2015tya}
H.~Ita, \emph{{Two-Loop Integrand Decomposition into Master Integrals and
  Surface Terms}},
  \href{https://doi.org/10.1103/PhysRevD.94.116015}{\emph{Phys. Rev.}
  {\bfseries D94} (2016) 116015}
  [\href{https://arxiv.org/abs/1510.05626}{{\ttfamily 1510.05626}}].

\bibitem{Mastrolia:2016dhn}
P.~Mastrolia, T.~Peraro and A.~Primo, \emph{{Adaptive Integrand Decomposition
  in parallel and orthogonal space}},
  \href{https://doi.org/10.1007/JHEP08(2016)164}{\emph{JHEP} {\bfseries 08}
  (2016) 164} [\href{https://arxiv.org/abs/1605.03157}{{\ttfamily
  1605.03157}}].

\bibitem{Cachazo:2008vp}
F.~Cachazo, \emph{{Sharpening The Leading Singularity}},
  \href{https://arxiv.org/abs/0803.1988}{{\ttfamily 0803.1988}}.

\bibitem{Bern:2007ct}
Z.~Bern, J.~Carrasco, H.~Johansson and D.~Kosower, \emph{{Maximally
  Supersymmetric Planar Yang-Mills Amplitudes at Five Loops}},
  \href{https://doi.org/10.1103/PhysRevD.76.125020}{\emph{Phys. Rev.}
  {\bfseries D76} (2007) 125020}
  [\href{https://arxiv.org/abs/0705.1864}{{\ttfamily 0705.1864}}].

\bibitem{Bosma:2017ens}
J.~Bosma, M.~Sogaard and Y.~Zhang, \emph{{Maximal Cuts in Arbitrary
  Dimension}},  \href{https://arxiv.org/abs/1704.04255}{{\ttfamily
  1704.04255}}.

\bibitem{Badger:2013gxa}
S.~Badger, H.~Frellesvig and Y.~Zhang, \emph{{A Two-Loop Five-Gluon Helicity
  Amplitude in QCD}},
  \href{https://doi.org/10.1007/JHEP12(2013)045}{\emph{JHEP} {\bfseries 12}
  (2013) 045} [\href{https://arxiv.org/abs/1310.1051}{{\ttfamily 1310.1051}}].

\bibitem{Badger:2015lda}
S.~Badger, G.~Mogull, A.~Ochirov and D.~O'Connell, \emph{{A Complete Two-Loop,
  Five-Gluon Helicity Amplitude in Yang-Mills Theory}},
  \href{https://doi.org/10.1007/JHEP10(2015)064}{\emph{JHEP} {\bfseries 10}
  (2015) 064} [\href{https://arxiv.org/abs/1507.08797}{{\ttfamily
  1507.08797}}].

\bibitem{Badger:2016ozq}
S.~Badger, G.~Mogull and T.~Peraro, \emph{{Local integrands for two-loop
  all-plus Yang-Mills amplitudes}},
  \href{https://doi.org/10.1007/JHEP08(2016)063}{\emph{JHEP} {\bfseries 08}
  (2016) 063} [\href{https://arxiv.org/abs/1606.02244}{{\ttfamily
  1606.02244}}].

\bibitem{ArkaniHamed:book}
N.~Arkani-Hamed, J.~L. Bourjaily, F.~Cachazo, A.~B. Goncharov, A.~Postnikov and
  J.~Trnka, \emph{{Grassmannian Geometry of Scattering Amplitudes}}. Cambridge
  University Press, 2016.

\bibitem{Drummond:2010mb}
J.~M. Drummond and J.~M. Henn, \emph{{Simple Loop Integrals and Amplitudes in
  $\mathcal{N}\!=\!4$ SYM}},
  \href{https://doi.org/10.1007/JHEP05(2011)105}{\emph{JHEP} {\bfseries 1105}
  (2011) 105} [\href{https://arxiv.org/abs/1008.2965}{{\ttfamily 1008.2965}}].

\bibitem{Arkani-Hamed:2014bca}
N.~Arkani-Hamed, J.~L. Bourjaily, F.~Cachazo, A.~Postnikov and J.~Trnka,
  \emph{{On-Shell Structures of MHV Amplitudes Beyond the Planar Limit}},
  \href{https://doi.org/10.1007/JHEP06(2015)179}{\emph{JHEP} {\bfseries 06}
  (2015) 179} [\href{https://arxiv.org/abs/1412.8475}{{\ttfamily 1412.8475}}].

\bibitem{Franco:2015rma}
S.~Franco, D.~Galloni, B.~Penante and C.~Wen, \emph{{Non-Planar On-Shell
  Diagrams}}, \href{https://doi.org/10.1007/JHEP06(2015)199}{\emph{JHEP}
  {\bfseries 06} (2015) 199}
  [\href{https://arxiv.org/abs/1502.02034}{{\ttfamily 1502.02034}}].

\bibitem{Frassek:2016wlg}
R.~Frassek and D.~Meidinger, \emph{{Yangian-Type Symmetries of Non-Planar
  Leading Singularities}},
  \href{https://doi.org/10.1007/JHEP05(2016)110}{\emph{JHEP} {\bfseries 05}
  (2016) 110} [\href{https://arxiv.org/abs/1603.00088}{{\ttfamily
  1603.00088}}].

\bibitem{Heslop:2016plj}
P.~Heslop and A.~E. Lipstein, \emph{{On-Shell Diagrams for $\mathcal{N}\!=\!8$
  Supergravity Amplitudes}},
  \href{https://arxiv.org/abs/1604.03046}{{\ttfamily 1604.03046}}.

\bibitem{Herrmann:2016qea}
E.~Herrmann and J.~Trnka, \emph{{Gravity On-shell Diagrams}},
  \href{https://doi.org/10.1007/JHEP11(2016)136}{\emph{JHEP} {\bfseries 11}
  (2016) 136} [\href{https://arxiv.org/abs/1604.03479}{{\ttfamily
  1604.03479}}].

\bibitem{Benincasa:2015zna}
P.~Benincasa, \emph{{On-Shell Diagrammatics and the Perturbative Structure of
  Planar Gauge Theories}},  \href{https://arxiv.org/abs/1510.03642}{{\ttfamily
  1510.03642}}.

\bibitem{Benincasa:2016awv}
P.~Benincasa and D.~Gordo, \emph{{On-shell diagrams and the geometry of planar
  N < 4 SYM theories}},  \href{https://arxiv.org/abs/1609.01923}{{\ttfamily
  1609.01923}}.

\bibitem{Bourjaily:2015jna}
J.~L. Bourjaily and J.~Trnka, \emph{{Local Integrand Representations of All
  Two-Loop Amplitudes in Planar SYM}},
  \href{https://doi.org/10.1007/JHEP08(2015)119}{\emph{JHEP} {\bfseries 08}
  (2015) 119} [\href{https://arxiv.org/abs/1505.05886}{{\ttfamily
  1505.05886}}].

\bibitem{Bourjaily:2010wh}
J.~L. Bourjaily, \emph{{Efficient Tree-Amplitudes in $\mathcal{N}\!=\!4$:
  Automatic BCFW Recursion in {\sc Mathematica}}},
  \href{https://arxiv.org/abs/1011.2447}{{\ttfamily 1011.2447}}.

\bibitem{Bourjaily:2012gy}
J.~L. Bourjaily, \emph{{Positroids, Plabic Graphs, and Scattering Amplitudes in
  {\sc Mathematica}}},  \href{https://arxiv.org/abs/1212.6974}{{\ttfamily
  1212.6974}}.

\bibitem{Dixon:2010ik}
L.~J. Dixon, J.~M. Henn, J.~Plefka and T.~Schuster, \emph{{All Tree-Level
  Amplitudes in Massless QCD}},
  \href{https://doi.org/10.1007/JHEP01(2011)035}{\emph{JHEP} {\bfseries 1101}
  (2011) 035} [\href{https://arxiv.org/abs/1010.3991}{{\ttfamily 1010.3991}}].

\bibitem{Nair:1988bq}
V.~P. Nair, \emph{{A Current Algebra for Some Gauge Theory Amplitudes}},
  \href{https://doi.org/10.1016/0370-2693(88)91471-2}{\emph{Phys. Lett.}
  {\bfseries B214} (1988) 215}.

\bibitem{Eden:2011yp}
B.~Eden, P.~Heslop, G.~P. Korchemsky and E.~Sokatchev, \emph{{The
  Super-Correlator/ Super-Amplitude Duality: Part I}},
  \href{https://doi.org/10.1016/j.nuclphysb.2012.12.015}{\emph{Nucl. Phys.}
  {\bfseries B869} (2013) 329}
  [\href{https://arxiv.org/abs/1103.3714}{{\ttfamily 1103.3714}}].

\bibitem{Eden:2011ku}
B.~Eden, P.~Heslop, G.~P. Korchemsky and E.~Sokatchev, \emph{{The
  Super-Correlator/ Super-Amplitude Duality: Part II}},
  \href{https://doi.org/10.1016/j.nuclphysb.2012.12.014}{\emph{Nucl. Phys.}
  {\bfseries B869} (2013) 378}
  [\href{https://arxiv.org/abs/1103.4353}{{\ttfamily 1103.4353}}].

\bibitem{Galperin:2001seg}
A.~S. Galperin, E.~A. Ivanov, V.~I. Ogievetsky and E.~S. Sokatchev,
  \emph{{Harmonic superspace}}, Cambridge Monographs on Mathematical Physics.
  Cambridge University Press, 2007,
  \href{https://doi.org/10.1017/CBO9780511535109}{10.1017/CBO9780511535109}.

\bibitem{Lin:2021pne}
G.~Lin and G.~Yang, \emph{{Double Copy of Form Factors and Higgs Amplitudes: A
  Mechanism for Turning Spurious Poles in Yang-Mills Theory into Physical Poles
  in Gravity}},
  \href{https://doi.org/10.1103/PhysRevLett.129.251601}{\emph{Phys. Rev. Lett.}
  {\bfseries 129} (2022) 251601}
  [\href{https://arxiv.org/abs/2111.12719}{{\ttfamily 2111.12719}}].

\bibitem{Lin:2022jrp}
G.~Lin and G.~Yang, \emph{{Double copy for tree-level form factors I:
  foundations}},  \href{https://arxiv.org/abs/2211.01386}{{\ttfamily
  2211.01386}}.

\bibitem{Lin:2023rwe}
G.~Lin and G.~Yang, \emph{{Double copy for tree-level form factors II:
  generalizations and special topics}},
  \href{https://arxiv.org/abs/2306.04672}{{\ttfamily 2306.04672}}.

\bibitem{Bourjaily:2021iyq}
J.~L. Bourjaily, C.~Langer and Y.~Zhang, \emph{{All two-loop, color-dressed,
  six-point amplitude integrands in supersymmetric Yang-Mills theory}},
  \href{https://doi.org/10.1103/PhysRevD.105.105015}{\emph{Phys. Rev. D}
  {\bfseries 105} (2022) 105015}
  [\href{https://arxiv.org/abs/2112.06934}{{\ttfamily 2112.06934}}].

\end{thebibliography}\endgroup

\end{document}